\documentclass[12pt]{ull_thesis22}
\usepackage[vlined,linesnumbered,ruled,boxed]{algorithm2e}
 \usepackage{amsmath}
 \usepackage{amsfonts}
 \usepackage{amssymb,xcolor}
 \usepackage{mdframed}
 \usepackage{optidef}

 \usepackage{layout}
 \usepackage{mydef}
 \usepackage{url}
\usepackage{bibentry}

\usepackage{tabularx} 
\usepackage{algorithmic}

\nobibliography*

 \usepackage{wrapfig,color,caption,subcaption,balance,multirow}
 \usepackage[vlined,linesnumbered,ruled,boxed]{algorithm2e}
 \usepackage{indentfirst}

\usepackage{etoolbox}
\makeatletter
\patchcmd{\@chapter}{\addtocontents{lof}{\protect\addvspace{10\p@}}}{}{}{}
\patchcmd{\@chapter}{\addtocontents{lot}{\protect\addvspace{10\p@}}}{}{}{}
\makeatother

\makeatletter
\AtBeginEnvironment{thebibliography}{%
   \clubpenalty10000
   \@clubpenalty \clubpenalty
   \widowpenalty10000}
\makeatother

\newcommand{\etal}{{\it et~al.}}
\newcommand{\comments}[1]{}

\usepackage{adjustbox,tabu,tabularx}

\makeatletter
\def\thickhline{%
  \noalign{\ifnum0=`}\fi\hrule \@height \thickarrayrulewidth \futurelet
   \reserved@a\@xthickhline}
\def\@xthickhline{\ifx\reserved@a\thickhline
               \vskip\doublerulesep
               \vskip-\thickarrayrulewidth
             \fi
      \ifnum0=`{\fi}}
\makeatother

\newlength{\boxfigwidth}

\usepackage{titlesec}

 \titleformat{\chapter}[block]
{\singlespacing\bfseries\filcenter}{\chaptertitlename\ \thechapter: \ }{0pt}{}
\titlespacing*{\chapter}
{0pt}{-0.5\baselineskip}{0pt}  

\titleformat{\section}
{\bfseries}{\thesection\ }{1em}{}
\titlespacing*{\section}{0pt}{*0}{0pt}

 \titleformat{\subsection}[runin]
{\bfseries}{\thesubsection}{1em}{}
\titlespacing{\subsection}{0pt}{*0}{0pt}

\titleformat{\subsubsection}[runin]
{\bfseries}{\thesubsubsection}{1em}{}[ ] 
\titlespacing{\subsubsection}
{\parindent}{*0}{0pt}

 \raggedright
\usepackage{graphicx,multirow}
\usepackage[vlined,linesnumbered,ruled,boxed]{algorithm2e}

 \usepackage{layout}
 \usepackage{url}
\usepackage{wrapfig}
\usepackage{bibentry}

\usepackage{tabularx} 
\usepackage{algorithmic}
\usepackage{amssymb}
\usepackage{bbold}

\nobibliography*

\usepackage{wrapfig,color,caption,balance,multirow}

\usepackage{adjustbox,tabu}

\usepackage[vlined,linesnumbered,ruled,boxed]{algorithm2e}
 \usepackage{amsmath}

 \usepackage{caption}
\usepackage{bibentry}

\makeatletter
\def\thickhline{%
  \noalign{\ifnum0=`}\fi\hrule \@height \thickarrayrulewidth \futurelet
   \reserved@a\@xthickhline}
\def\@xthickhline{\ifx\reserved@a\thickhline
               \vskip\doublerulesep
               \vskip-\thickarrayrulewidth
             \fi
      \ifnum0=`{\fi}}
\makeatother

\newlength{\thickarrayrulewidth}
\nobibliography*

 \usepackage{graphicx,multirow}

\usepackage{diagbox}

\usepackage{cleveref}
\newcommand\hl{\bgroup\markoverwith
  {\textcolor{yellow}{\rule[-.5ex]{2pt}{2.5ex}}}\ULon}

\def\BibTeX{{\rm B\kern-.05em{\sc i\kern-.025em b}\kern-.08em
    T\kern-.1667em\lower.7ex\hbox{E}\kern-.125emX}}

\begin{document}

    \title{AI-Driven Confidential Computing across Edge-to-Cloud Continuum}
    
    
     
     \title{Federated Fog Computing for Remote Industry 4.0 Applications}   
    
    \author{Razin Farhan Hussain}
    \convocationdate{Fall}
    \gradyear{2022}
    \degree{Doctor of Philosophy}
    \major{Computer Science}
    \supervisor{Mohsen Amini Salehi}
   \ranksupervisor{Associate Professor of Computer Science \\ The Center for Advanced Computer Studies}
    \deanofgraduateschool{Mary Farmer-Kaiser}
    \secondcommitteemember{Xu Yuan}
  \ranksecondcommitteemember{Assistant Professor of Computer Science \\ The Center for Advanced Computer Studies}
    \thirdcommitteemember{Li Chen}       
  \rankthirdcommitteemember{Assistant Professor of Computer Science \\ The Center for Advanced Computer Studies}
  \firstcommitteemember{Sheng Chen}
  \rankfirstcommitteemember{Associate Professor of Computer Science \\ The Center for Advanced Computer Studies}
 \fileforabstract{B-abstract}
  \filefordedication{B-dedicatory}
    \fileforacknowledgement{B-acknowledgement}





\prefatorypages

    \chapter{Introduction}
Industry 4.0 is revolutionizing the utilization of computing resources across various industries \cite{singh2021dynamic}. With the emergence of the Internet of Things (IoT) and modern computing systems (\eg edge computing, fog computing, and serverless computing), industries are becoming more intelligent with smart sensors and actuators that create a large quantity of data \cite{sensorDataCisco} every day. However, the computational resources required to store and analyze sensor-generated data are expensive and particularly scarce in remote areas \cite{tortonesi2019taming}. In the industrial sector, various sorts of applications (\eg machine learning (ML), reporting, alarm generators, and surveillance) employ sensor-generated data to automate or conduct complex operations. Sometimes these data require real-time feedback to conduct fault-intolerant latency-sensitive activities (\eg drilling operation in an oil rig, workers' safety operation, manufacturing products). Alternatively, certain tasks need large computing capacity and are delay tolerance, necessitating cloud data center assistance. For instance, the ``Fire safety" application \cite{park2019dependable} utilizes a deep neural network (DNN) model that needs extensive training in highly configured cloud data centers. Similarly, ``reservoir simulation," widely used in the petroleum industry to anticipate the field performance under varies producing strategies, requires a large quantity of seismic data and high-performance computing systems \cite{Combier2014Reservoir}.

\begin{figure}[ht]
    \centering
    \includegraphics[width=0.85\textwidth]{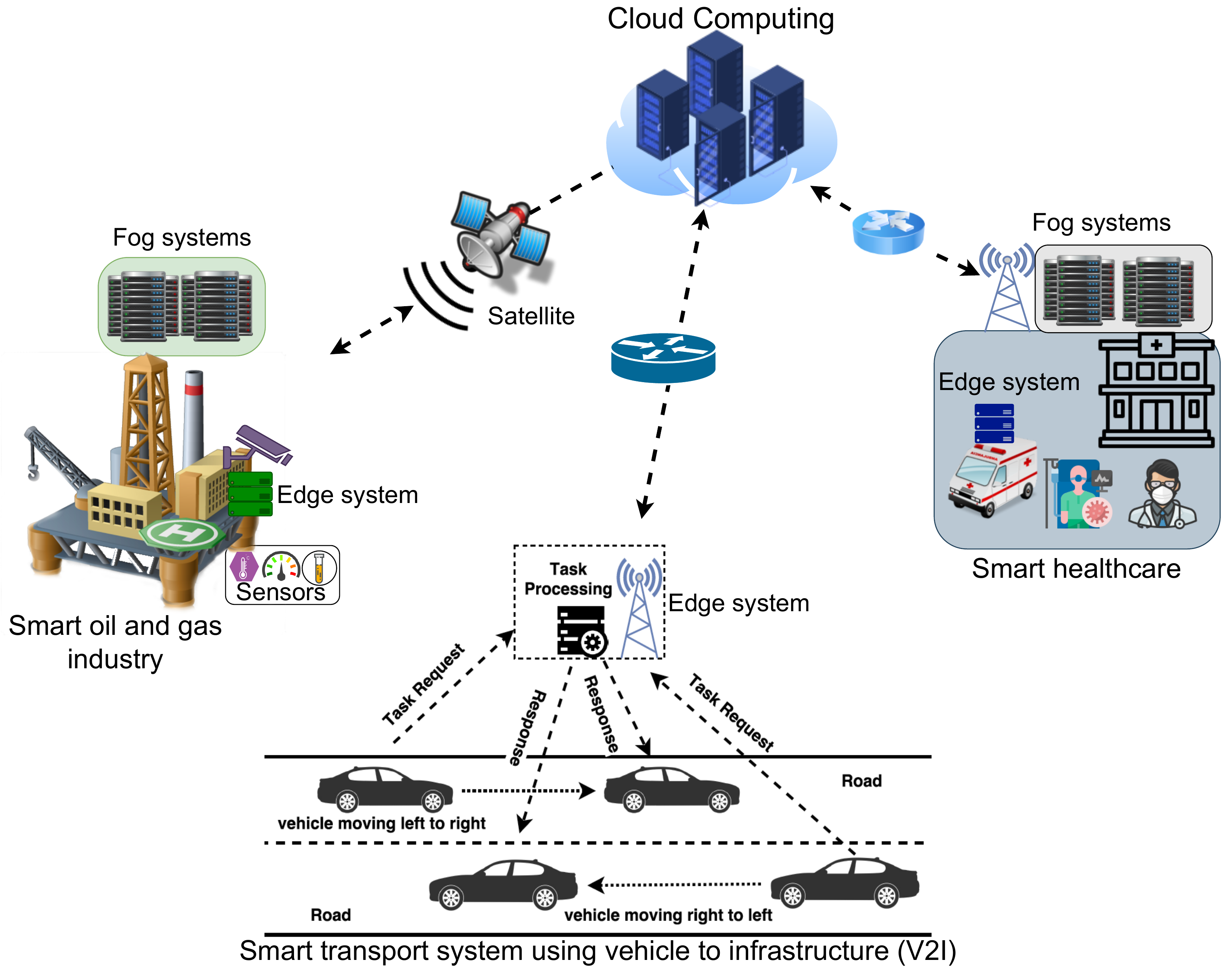}
    \caption{Advanced computing systems in various smart industries (\eg oil and gas, healthcare, transportation) for real-time latency-sensitive tasks}
    \label{fig:C1intro1}
\end{figure}

In remote or distant locations of the industrial sectors (\eg offshore oil extraction sites, solar fields), transferring the sensor-generated data to a cloud data center is expensive and latency intensive, influencing the use of computing near the data sources \cite{trinks2018edge}. Additionally, real-time applications require a faster response time, which is usually not feasible with cloud computing resources. Hence, bringing computational resources to the data sources near the end clients is an essential requirement for remote industries.

One of the solutions for computing near data sources is edge computing \cite{hartmann2022edge} as depicted in figure~\ref{fig:C1intro1}, which brings computational resources closer to the end devices, and data generation sources. As such, \textit{edge computing} can be defined as a distributed computing platform that puts industrial applications closer to data sources like IoT devices or local computer servers. This closeness to data at its source can result in significant business benefits such as faster insights, faster reaction times, and increased bandwidth availability. Although edge computing supports real-time latency-sensitive applications, edge devices are resource constraints that need efficient resource allocation \cite{chen2018thriftyedge} mechanism. Hence, another solution for computing platforms near end users is fog computing systems \cite{rani2022fog} that complement edge computing by having more computing resources, having more comprehensive middleware for managing workload efficiently. 

The main driving force of Industry 4.0 revolution is machine learning (ML) or deep neural network (DNN) applications \cite{jamwal2022deep, batistakis2021ai,webert2022fault} that ensure efficient industrial operations and workplace safety. Hence the ML or DNN-based applications encompass both the training and inference stages \cite{eshratifar2019jointdnn}. The training stage is generally carried out offline due to time and computing resource constraints. Whereas the inference execution can be completed utilizing general-purpose computing systems. The DNN-based applications are mainly trained on cloud data centers or computing servers with high configuration hardware (\eg GPU, TPU, FPGA) support. In contrast, the inference operations are performed on the fog computing systems near the end users. As such, it is essential for a system engineer or system administrator to understand the performance of these DNN-based applications in fog computing systems \cite{hussain2020analyzing}. Especially for fault-intolerance latency-sensitive critical DNN-based applications, the forecast of inference execution time within a computing resource can be significantly vital that sometimes save lives in a disaster situation. As such, we perform a statistical analysis of the inference execution time of various Industry 4.0 applications on the cloud and fog systems. Consequently, we introduce an execution time workload trace that can help system architects to develop a load-balancing solution robust against stochastic execution times of Industry 4.0 smart applications. Therefore, efficiency, productivity, and industrial safety can be ensured by utilizing these robust solutions.

\begin{figure}[!ht]
    \centering
    \includegraphics[width=0.8\textwidth]{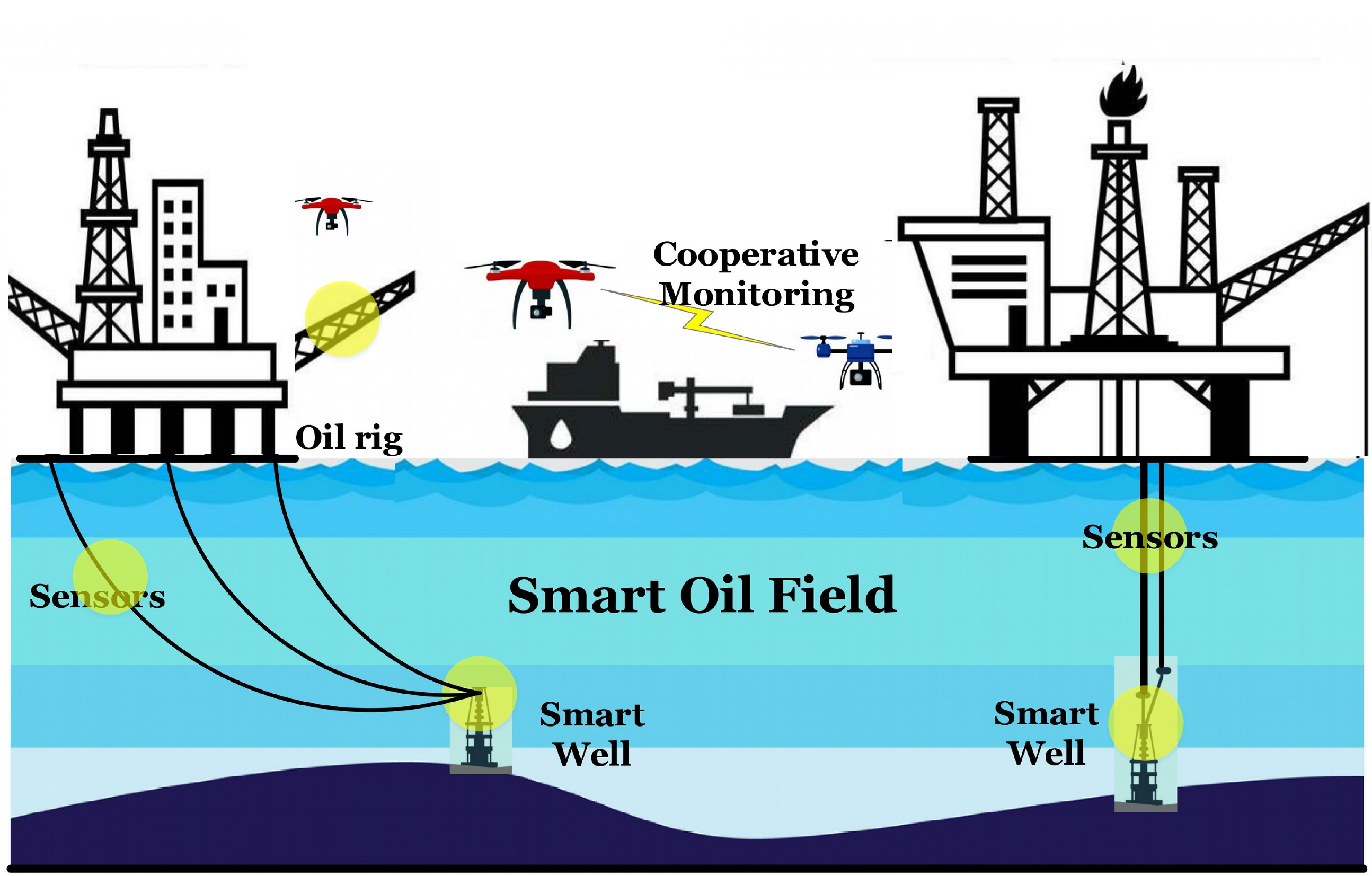}
    \caption{A remote offshore smart oil field consists of multiple oil rigs (oil extraction sites). In this scenario, the oil rigs, drill ships, or even rescue ships have fog computing systems in the form of mobile data centers to support the oil extraction computing demands along with any unpredictable emergencies (\eg oil spill detection, toxic gas detection)}
    \label{fig:C1intro2}
\end{figure}

In remote offshore industry, at times of emergencies (\eg disasters, accidents), the demand for task processing in the edge or even fog computing systems can be significantly high, leading to a drop in some tasks due to not meeting their latency constraints (a.k.a deadlines). As such, we propose to federate nearby privately owned computing resources by forming a federation of fog systems to support the surge of task processing requests in times of emergencies. For instance, in a remote offshore smart oil field, as depicted in Figure~\ref{fig:C1intro2}, multiple oil extraction sites can be built by the respective companies that typically contain privately owned fog computing systems to support their regular computing demands. At a disaster time or other emergencies such as an explosion, the computing demands surge to support multiple recovery procedures to be coordinated. For instance, in a fire breakout event, various activities such as drone-based inspection, fire detection, and alert generation with precise fire locations need real-time coordination to neutralize the emergency. In this scenario, some latency-sensitive tasks can be offloaded \cite{kan2018task} to other fog systems that may have more computational resources or less busy. Hence, the federated fog systems' resilience depends on supporting the surge in computing demands by efficient resource allocation across the federation. Therefore, we propose a probabilistic resource allocation method across fog federation for latency sensitive monolithic tasks to support computing demands in emergency situations.

\begin{figure}[!h]
    \centering
    \includegraphics[width=0.8\textwidth]{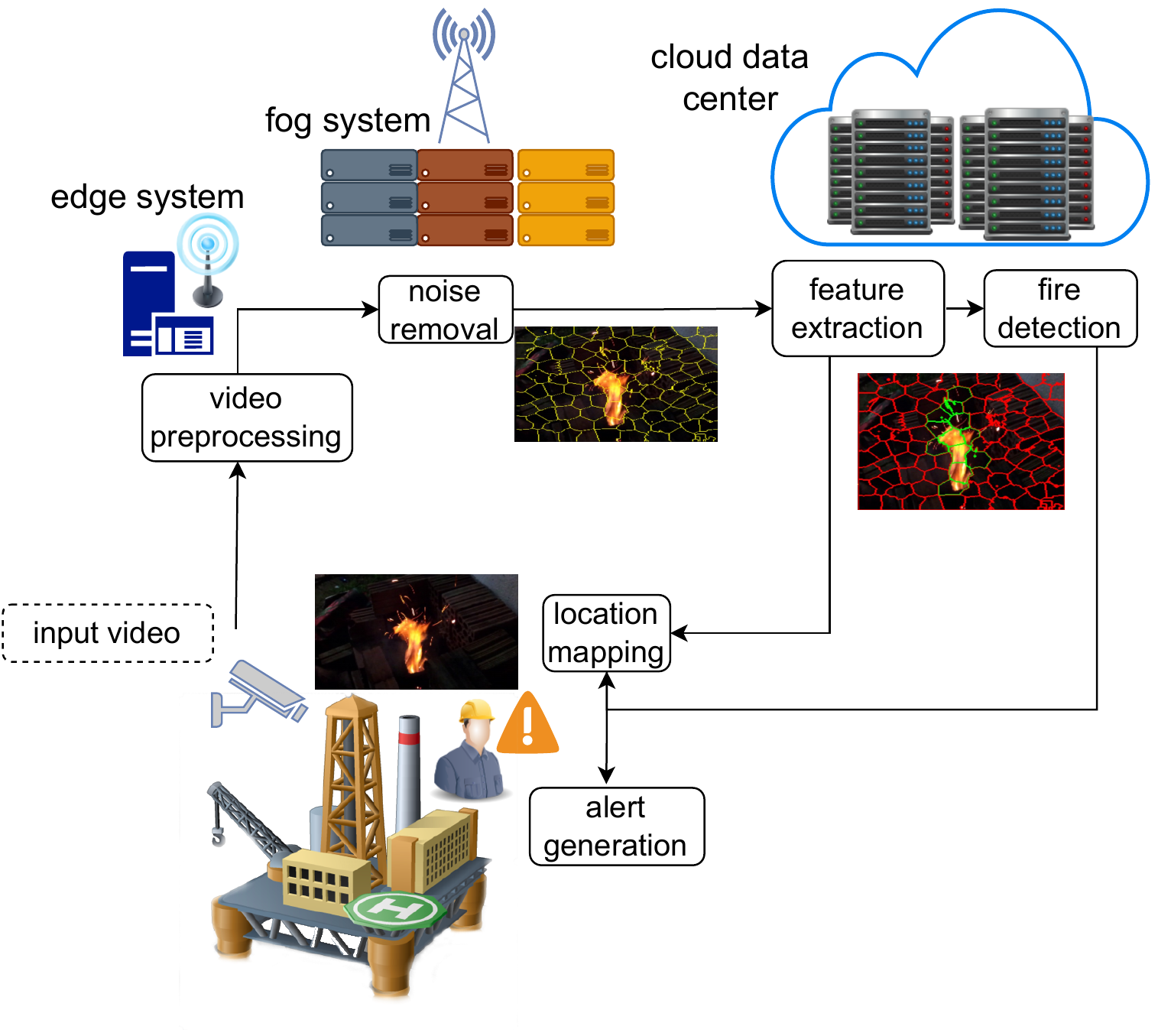}
    \caption{A typical micro-service application, ``fire safety'' execution scenario in edge-fog-cloud paradigm.}
    \label{fig:C1micro}
\end{figure}


Smart applications in Industry 4.0 can have various software architectures (\eg monolithic, micro-service \cite{jwo2022data}) that serve different purposes of industrial operations. Hence, micro-service architecture is one of the widely used software architectures that comprise various micro-level services having immense benefits on development and deployment \cite{dragoni2017microservices}. For instance, as depicted in Figure \ref{fig:C1micro}, the ``fire safety" micro-service-based application comprises video pre-processing, noise removal, feature extraction, fire detection, location mapping, and alert generation micro-services performing different activities. In a typical industrial scenario, these micro-services are supported by various execution platforms that can have stochastic execution latencies. In contrast, a monolithic architecture is the conventional unified paradigm for constructing a software application. Monolithic software is intended to be self-contained, with firmly connected rather than loosely coupled components or services, as in a micro-service architecture. Although the industrial revolution influenced the utilization of micro-service applications, various industries have monolithic legacy applications that are still in operation and need to be supported by existing execution platforms. In an emergency or disaster, various application requests with different latency requirements are generated in the proximity of the disastrous area that needs distinct computational support. Hence, the nearby execution platform gets oversubscribed with the surge in demand for executing numerous applications on time, that can degrade the execution platform's performance. In this case, to support the high computation demands utilizing the proposed federated fog system need an efficient resource allocation method that is aware of receiving applications' internal structure, computation, and communication latencies. Therefore, the reliability of an execution platform in an oversubscribed situation depends on accommodating various computational demands that ensure industrial safety.

Federating computing resources in remote industrial areas imposes security concerns for each participant fog system of the federation, that is owned by private companies. In addition, individual fog systems can have sensitive data that imposes privacy issues for the company owning the computing systems. Hence, considering ML application training across the fog federation suffers from data scarcity, that is an obstacle to building accurate ML models. As such, a secure and privacy-preserving distributed ML training method should be in place to build an accurate ML model that can be crucial in emergency situations in Industry 4.0.


 
\section{Research Problem and Objectives}
The fundamental purpose of this research is to identify, evaluate, and manage robust execution of Industry 4.0 applications in remote areas (\eg offshore oil fields) across modern edge and fog computing systems. Hence, we develop solutions that use fog systems in emergency and oversubscribed circumstances to satisfy the computational demand in remote industrial sectors. This dissertation address the following research challenges to enable a robust and QoS-efficient federated fog system for industry 4.0 applications:
\begin{enumerate}
    \item How to utilize modern distributed computing systems in the context of remote smart industries (\eg oil and gas, energy) considering the industrial revolution, Industry 4.0?

    \item What are the statistical execution behaviors of Industry 4.0 applications in fog systems?
    
    \item How to enable a robust federated fog computing system that can efficiently procure computing demands during a workload surge? 
    
    \item How to support Industry 4.0 applications with modern micro-service architecture along with monolithic legacy applications and maintain the Quality of Service (QoS) of a fog federation?
    
    \item How to utilize federated fog  securely to improve the performance of Industry 4.0 applications?
\end{enumerate}

\section{Contributions}
We identified various obstacles as we investigated many facets of federated fog computing systems in the industrial sector. Therefore, in addition to addressing significant challenges, we present state-of-the-art solutions and give exhaustive performance assessments for recommended methodologies. In light of the research topics outlined in the preceding section, the considerable contributions of this dissertation are as follows:
\begin{itemize}
 \item Identifying the connection of industry 4.0 and modern distributed computing systems (\eg edge, fog) and addressing the scope of utilizing advanced analytics (\eg artificial intelligence, machine learning, deep learning) in the context of the remote smart oil field.

 \item Performance analysis of ML-based Industry 4.0 applications across fog and cloud computing systems?
 
 \item Proposing a real-world workload benchmark of inference execution times for four different ML-based Industry 4.0 applications.

 \item Enabling the notion of federated fog via resource allocation methods operating based on Bayesian probability utilizing fog systems for latency-sensitive tasks in an oversubscribed system that tries to recover from a disaster.

 \item Proposing a statistical resource allocation solution across federated fog systems that is aware of internal software architecture and stochastic latency requirements of Industry 4.0 micro-service workflow applications.



 \item Proposing a data privacy preserving ML-based Industry 4.0 application training solution across federated fog system in remote industrial sites.

 
\end{itemize}


\section{Dissertation Organisation}
\begin{itemize}

\item Chapter~\ref{chap:bg} explores the related research works and provides background for edge \& fog computing, fog federation systems, load balancing \& task allocation techniques, and data privacy aspects of a federated fog system. Hence, the scope of utilizing the modern distributed systems in the remote smart oil fields is addressed with various use case scenarios. 
\begin{itemize}
     \item \textbf{Razin Farhan Hussain}, Ali Mokhtari, Mohsen Amini Salehi, and Ali Ghalambor
     \textit{IoT for Smart Operations in the Oil and Gas Industry} published as a book by Elsevier (ISBN:9780323998444).

\end{itemize}

\item Chapter~\ref{section:performanceAnalysis} studies the performance of ML-based Industry 4.0 applications in heterogeneous cloud computing resources. The statistical analysis of ML-based applications helped to generate a real-world Industry 4.0 application inference execution time workload that can be beneficial for the system architect to develop robust load-balancing solutions.
\begin{itemize}
    \item \textbf{Razin Farhan Hussain}, Alireza Pakravan, and Mohsen Amini Salehi
    \textit{Analyzing the Performance of Smart Industry 4.0 Applications on Cloud Computing Systems} published in Proceedings of the 22nd IEEE International Conference on High-Performance Computing and Communications (HPCC-2020)
\end{itemize}

\item Chapter~\ref{section:loadBalance} explores the possible advantages and practicality of building a fog federation in a distant offshore smart oil field in the event of a disaster. Using probabilistic load balancing heuristics across the fog federation for resource allocation can efficiently assure the system's resiliency. Moreover, compared to baseline approaches, the advantage of employing probabilistic methods is backed by a synthetic workload created in EdgeCloudSim simulation\cite{sonmez2017edgecloudsim}.

\begin{itemize}
    \item \textbf{Razin Farhan Hussain}, Mohsen Amini Salehi, Anna Kovalenko, and Omid Semiari \textit{Federated Edge Computing for Disaster Management in Remote Smart Oil Fields } published in Proceedings of the 21st IEEE International Conference on High Performance Computing and Communications (HPCC-2019)
    
    \item \textbf{Razin Farhan Hussain}, Omid Semiari, and Mohsen Amini Salehi
     \textit{Robust Resource Allocation Using Edge Computing for Vehicle to Infrastructure (V2I) Networks} published in Proceedings of the 3rd IEEE International Conference on Fog and Edge Computing (ICFEC'19)
    
    \item \textbf{Razin Farhan Hussain}, Mohsen Amini Salehi, and Omid Semiari
     \textit{Serverless Edge Computing for Green Oil and Gas Industry} published in Proceedings of IEEE Green Technologies Conference(GreenTech) - 2019
    
\end{itemize}

\item Chapter~\ref{section:microservice} explores the advanced micro-service software architecture for Industry 4.0 applications to enhance the robustness of remote federated fog systems. Hence, the load balancer should be aware of the software architecture of the receiving applications as well as the uncertainties of the execution platform. As a result, the distribution of receiving applications across the fog federation enhances the possibility of applications being completed on time.
\begin{itemize}
    \item \textbf{Razin Farhan Hussain}, Mohsen Amini Salehi
    \textit{Adapting Remote Industry 4.0 Applications to
Federated Fog Computing Systems} prepared for submission to Future Generation Computing System journal in 2022
\end{itemize}

\item Chapter~\ref{section:fedLearn} explores the data privacy aspects of ML-based application training across the federated fog computing systems in Industry 4.0.


\item Chapter~\ref{threatsSideEffect} explores the downsides and side effects of smart solutions for Industry 4.0. This chapter identifies and proposes various cutting-edge solutions for security issues of different industrial sectors.
\begin{itemize}
\item \textbf{Razin Farhan Hussain}, Ali Mokhtari, Mohsen Amini Salehi, and Ali Ghalambor
     \textit{IoT for Smart Operations in the Oil and Gas Industry} published as a book by Elsevier (ISBN:9780323998444).
\end{itemize}

\item Chapter~\ref{section:thesiscon} concludes the dissertation with a discussion of our main findings and future research directions in the area of efficient utilization of fog computing platforms for Industry 4.0.
\end{itemize}
    \chapter{Background and Literature Study} \label{chap:bg}
\section{Computing as a Prominent Aspect of Industry 4.0}
Industrial systems are quickly transitioning from human-controlled processes to closed-loop control services supporting their operations autonomously using extensive sensor and computing infrastructure. This revolutionary change is critical for supporting growing data-intensive and time-sensitive Industry 4.0 applications, particularly at remote locations such as offshore Oil and Gas (O\&G) fields where computer infrastructure is restricted and human resources are limited. Realizing these systems necessitates interdisciplinary research and study at the interplay of Industry 4.0 in remote industry, modern computing infrastructure (such as an Edge and Cloud), and advanced analytics (\eg ML, DNN).

Consequently, this chapter aims to illustrate the challenges, prospects, and solutions for establishing a smart and robust remote industry based on the fundamentals of the Industry 4.0 paradigm. As a result of this study, researchers and practitioners can be more effective in making the remote industry safer, more sustainable, greener, automated, and, subsequently, more cost-efficient. This chapter investigates several computer technologies that support the computing needs of distant industries. Furthermore, it explains how the synergy of cutting-edge computing solutions, such as the Internet of Things (IoT), Machine Learning methodologies, and distributed computing platforms, can be employed to improve industrial processes. As an ideal example of a remote offshore industry, we consider Oil and Gas that has been transforming significantly with the industrial revolution Industry 4.0. On the other hand, the remote offshore O\&G industry has been facing various disasters and catastrophes that raise concerns about production efficiency and safety measures. For instance, the deepwater horizon (2010)\cite{srinivasan2010many}, usumacinta jack-up disaster (2007) \cite{hanlon2013usumacinta}, mumbai high north disaster (2005) \cite{daley2013mumbai}, and the ocean ranger disaster (1982) incidents are significantly connected with safety failures in the industrial sites. Hence these incidents motivated us to improve the computing support in remote industries to ensure safety and productivity. Therefore, this chapter explores various distributed computing technologies, federation-friendly execution platforms, software architecture, and security aspects of Industry 4.0, focusing on the O\&G industry.

\section{Distributed Computing Systems in Industry 4.0}
\subsection{\textit{Cloud Computing}}~\\
Cloud computing is a concept that enables resources (\eg computing, storage, services) to be available as a service, on-demand, configurable, and also shareable \cite{accentureReport}. Modern cloud systems provide diverse services in different levels, such as  infrastructure as a service (IaaS), platform as a service (PaaS), software as a service (SaaS), and function as a service (FaaS).

\begin{figure}[h!]
 \centering
 \includegraphics[width=1\textwidth]{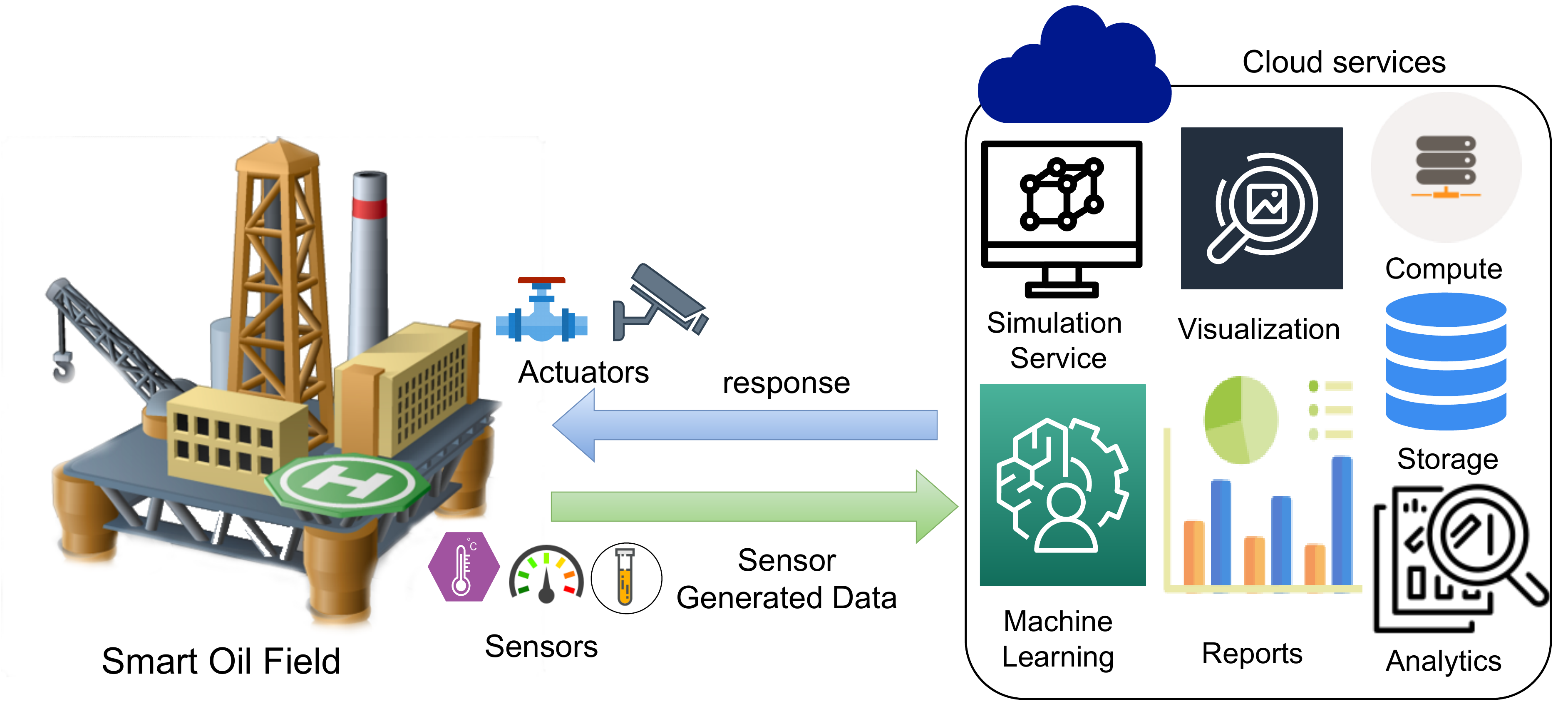}
 \caption{Various cloud services (\eg simulation, analytics, visualization, compute, machine learning, reporting) can be employed to store, process, and analyze sensor-generated data and to control industrial equipment in a smart oil and gas industry.}
 \label{fig:cloudServices}
 \centering
\end{figure}

As presented in Figure \ref{fig:cloudServices}, smart O\&G industry increasingly relies on cloud-based services that are hosted on remote Internet servers (a.k.a. cloud data centers). These data centers are utilized to store and process their data. According to Figure \ref{fig:cloudServices}, various sensor-generated data are sent to cloud providers to avail of different kinds of cloud services. Among these services, some of them send insightful decisions to actuators to close the automation loop in the smart oil field. Cloud technology enables O\&G companies to utilize various data-related and computational services (\eg machine learning and visualization) without the need to maintain any computing infrastructure. However, data privacy and security have remained a concern for such companies to fully embrace the cloud services. These security concerns have caused a small pause and hesitation in adoption cloud services, particularly by major players in this industry. An alternative and more secure approach is to store the data on an on-premise computing facility that is known as a \emph{private cloud} (more recently called \emph{fog computing}). 

On the positive side, cloud systems' performance and ease-of-use are tempting for the O\&G industry. For instance, one of the main users of data-driven cloud services is the North American shale industry that drills thousands of wells every year \cite{2021CloudNorthAmerica}. The scalability feature of cloud services helped the growing amount of data from these wells to be utilized efficiently, allowing the industry to expand remarkably. As such, various modern cloud-based data analytics services have emerged to help O\&G companies to improve their operational workflows and make profitable decisions.

\subsection{\textit{Edge and Fog Computing for Remote Industry 4.0}}~\\
Due to the increasing importance of latency-sensitive applications, real-time operations close to the end user in remote offshore industries, the interest in the notion of edge computing has begun to increase. Additionally, the proliferation of the Internet of Things (IoT) devices and smart sensors in the industrial sector results in a massive amount of data that need to be processed locally. In a typical scenario, the data is transported to cloud data centers \cite{buyya2010energy}, and responses or results are transmitted back to clients through the internet, both of which take time and money. As a result, a distributed computing paradigm has been introduced, which is located close to the end client and processes client data at the network's edge \cite{techTargetWeb}. Researchers call this type of computing ``Edge Computing" since it operates at the network's periphery.

The conventional definition of edge computing is difficult to come by. Different organizations or sources have different definitions, heavily impacted by context. The general perception of edge computing is to provide various computing services (\eg application execution, data pre-processing) through distributed computer systems instead of centralized cloud data centers. Hence, edge computing enables analysis and knowledge collection at the point of information source. In network design, an ``edge computer" is located directly next to or even on top of network endpoints (such as controllers and sensors). The data is then partially or fully processed before being transmitted to the cloud for storage or further processing. Edge computing, on the other hand, may result in the direct transfer of huge volumes of data to the cloud. This might have an impact on system capacity, efficiency, and security. Fog computing \cite{singh2019fog} solves this problem by inserting a processing layer between the edge and the cloud. As a result, `fog computing' collects and analyses data at the edge before it reaches the cloud. Hence, the place from the data source where computing service is offered can be a defining element in distinguishing Fog/Edge computing from cloud computing. For instance, a renewable energy company geared with numerous sensors utilizes fog computing for sensor-data analysis in their operational fields. Accordingly, company`s production efficiency improved by 15\% by reducing data analysis latency from 10 minutes to few seconds. Therefore, fog computing placed near data source in remote industries can enhance efficiency in production.  



The emergence of edge and fog computing does not substitute the cloud computing services; instead, it brings some portion of those services (\eg computing, storage, analytical services) near the end clients. Especially with the ever-growing Internet of Things (IoT) devices, a considerable amount of data is generated \cite{sensorDataCisco} that is significantly valuable for Industry 4.0 ML applications. The generated data sometimes need immediate processing (\ie edge computing support), and alternatively, sometimes need complex processing (\ie cloud computing support). Therefore, a continuous computing platform (\ie Edge-to-Cloud Continuum \cite{balouek2019towards}) is required to support both real-time nature and complex analytical tasks.

\begin{figure}[!h]
    \centering
    \includegraphics[width=0.8\textwidth]{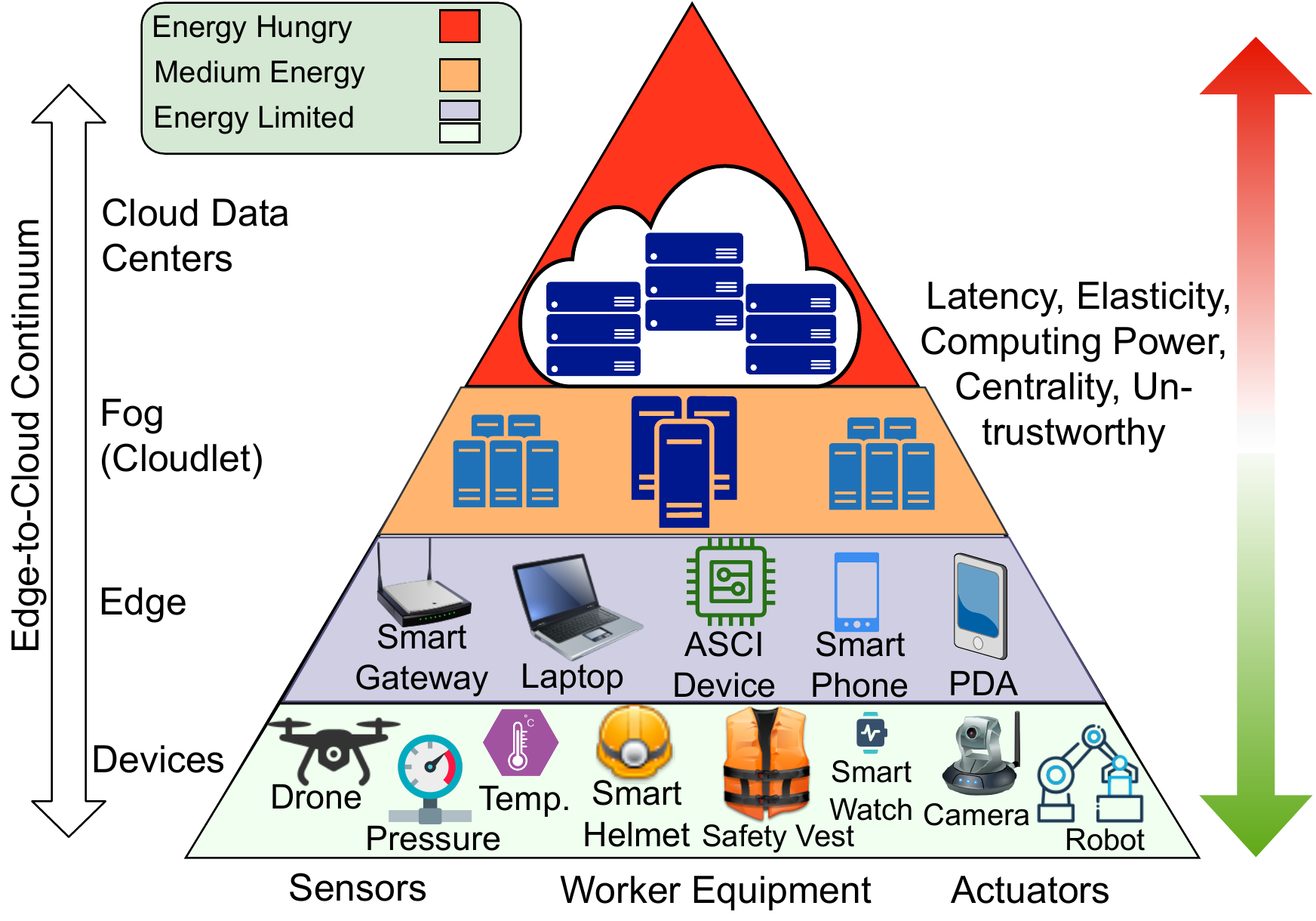}
    \caption{Edge-to-Cloud continuum for oil and gas industry as an example of Industry 4.0. The continuum is mainly divided into four tiers, namely end devices, edge, fog, and cloud data centers. The bottom of the triangle has end devices that are energy limited, whereas traversing to the top, we find more energy-consuming systems.}
    \label{fig:edgeToCloud}
\end{figure}

\subsection{\textit{Edge-to-Cloud Continuum}}~\\
Although edge and cloud computing has a difference in terms of distance and resources, they can be utilized as a complement to each other. For a massive industry such as O\&G, diverse operations and services are needed that require various underlying computing platforms. Hence, the integration of edge or fog computing with cloud computing is a need of time that reflects the usability of the edge-to-cloud continuum\cite{luckow2021pilot}. Accordingly, the Edge-to-Cloud continuum is a service platform that provides various computational resources and infrastructures for supporting different types of services essential for O\&G industry.

Figure \ref{fig:edgeToCloud} demonstrates the Edge-to-Cloud continuum as a triangle where edge devices reside close to end devices (bottom of the triangle) and cloud data centers are the furthest computing entity from end devices. Accordingly, this is a hierarchical arrangement that is distributed vertically. Hence cloud computing has high latency than edge and fog computing. Alternatively, cloud computing has high availability in terms of elasticity and computing power, whereas edge and fog devices are highly secure and privacy-preserving than cloud computing. Therefore, various computing platforms within the continuum serve different purposes for industrial operations.

\subsection{\textit{Use Case of Edge-to-Cloud Continuum in Smart O\&G}}~\\
We investigate drone-based pipeline inspection scenarios in the oil and gas industry to understand how the edge-to-cloud continuum supports computing demands in the industrial sector. Let's consider a scenario where 4K drone-mounted cameras can collect hundreds of gigabytes of data per hour. The current method of analyzing data is to transfer the massive data to the cloud data center, which is cost-prohibitive and impractical, especially if the analysis is real-time. Hence one critical question \emph{Is it feasible or scalable for the future to have any cloud vendor send their container truck with petabytes of storage?}
\begin{figure}[h!]
 \centering
 \includegraphics[width=0.85\textwidth]{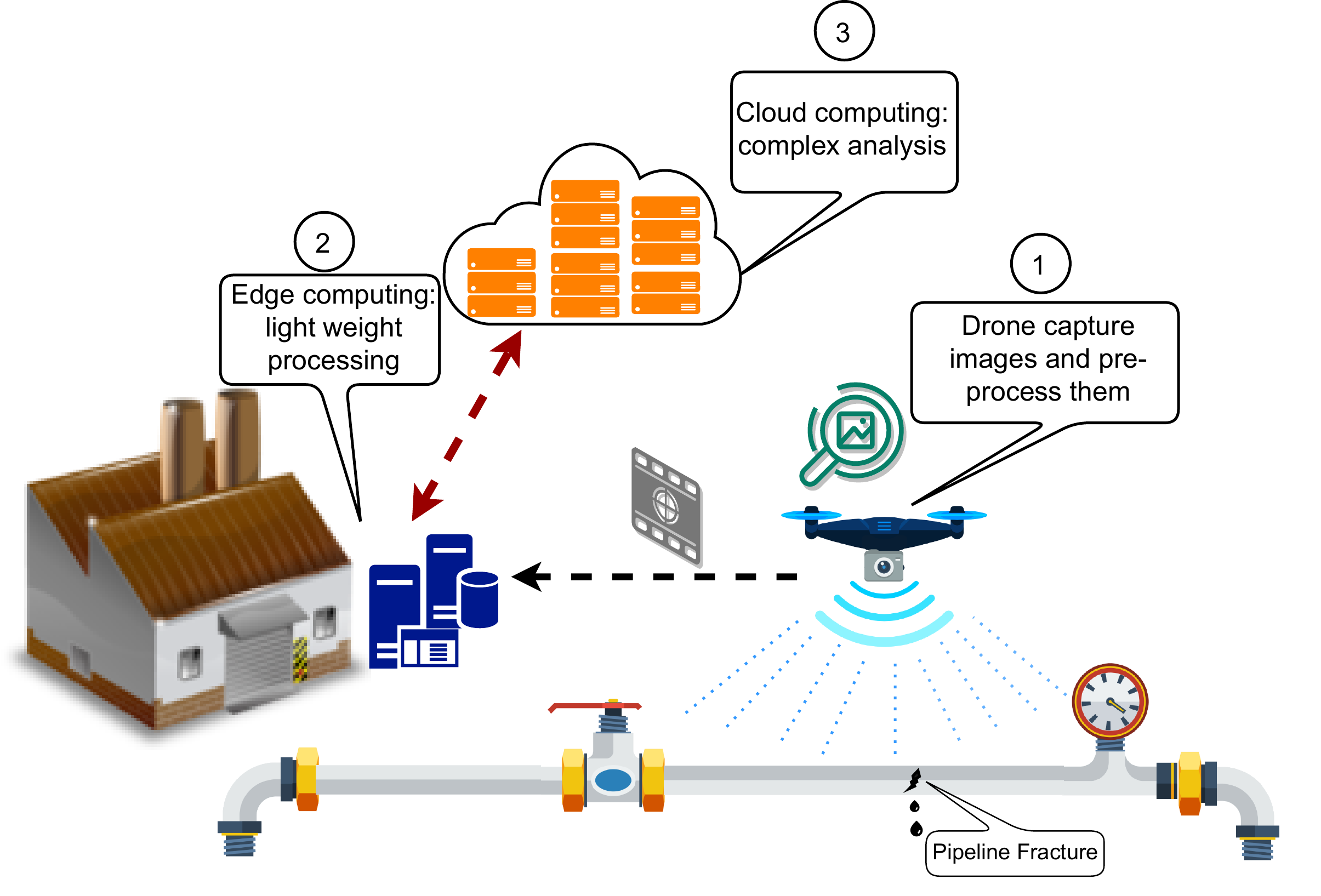}
 \caption{Drone-based inspection scenario where drone captures images and real-time analysis can be performed in edge computing resource whereas long term analysis is performed in distant cloud computing facility.}
 \label{fig:droneInspect}
 \centering
\end{figure}

An example of the same scenario (presented in figure \ref{fig:droneInspect}) from the oil and gas industry perspective is that the drone-based inspection system could use multi-stage value extraction using an edge-to-cloud continuum. The O\&G pipelines can be thousands of miles long and pass through an immense landscape. Pipe sections are generally fitted with analog gauges and smart sensors to measure pressure, flow, and other metrics. By employing an edge AI-enabled surveillance drone to capture these analog gauge images presented in step 1 of figure \ref{fig:droneInspect}, it is possible to separate (step 2) the gauge images and transfer only that critical information to the next compute layer. Here, data pre-processing (image separation) is real-time nature that is performed in the edge computing system. Therefore only the localized necessary data is processed for an accurate reading in the cloud data center (step 3). Then the output of actionable intelligence is sent to the on-site maintenance team to resolve pipeline fracture. Therefore, data pre-processing, lightweight processing, and complex analysis are performed across the edge-to-cloud continuum to conduct efficient drone-based pipeline surveillance.

\subsection{\textit{Landscape of Computing in O\&G}}~\\
Modern computing systems, such as edge, fog, or cloud enable the smooth operation of different fault-intolerant processes across different sectors of the O\&G industry. As a cyber-physical system, the computing technology stack of the O\&G industry is composed of the following components:
\begin{itemize}
\item \emph{Sensors:} Numerous sensors of different types (\eg to gauge pressure, emission of toxic gases, security cameras, etc.) continuously procure multi-modal data in the form of signal, text, images, video, and audio. The data is stored or communicated for offline or online processing to monitor the operation of the oil field or to make management decisions.

\item \emph{Computer networks:} In a smart oil field, short- and long-range wireless and wired computer networks (\eg Bluetooth, CBRS, satellite, etc.) have to be configured for low-latency and high data-rate communication of devices (\eg sensors, servers, and actuators) both for onsite and offsite communication.

\item \emph{Computing systems and middleware:} All the collected data have to be eventually processed to be useful. That is why, in the back-end, smart oil fields are reliant on different forms of computing systems (\eg HPC, cloud, fog, edge, and real-time systems) to perform batch or online data processing for purposes like monitoring, visualization, and human-based or automatic decision making. 

\item \emph{Data processing and software technologies:} The rule of thumb in a smart oil field is that ``the more data can be processed, the more informed decisions can be made''. The large amount of multi-modal data (text, images, video, and signals) continuously generated in a smart oil filed form what is known as \emph{big data}. Such diverse formats of big data have to be processed using various algorithmic techniques, particularly Machine Learning, to provide an insight from the data or to make informed decisions upon them.  

\item \emph{Actuators:} Once a decision is made, it is communicated to an actuator (\eg drilling head and pressure valve) to enact the decision (\eg increase or decrease the pressure).

\end{itemize}

\section{Smart O\&G: Data and Software Aspects }
\subsection{\textit{Big Data in the O\&G industry}}~\\
The oil and gas industry generates a large volume of data on a daily basis, necessitating the need for large-scale computing resources and the cloud. The three key sources of such considerable data in the O\&G industry are as follows::
\begin{itemize}
    \item Hydrocarbon reservoirs are commonly found between 5,000 and 35,000 feet below the Earth's surface. High-resolution images and expensive well logs are the main options for finding and characterizing reservoirs (after the wells are dug).
    
    \item Fluids must pass through complex rock to reach the wellbore, and the fluids themselves are complex, having many different physical properties. Therefore, learning about the unique characteristics of each oil well and evaluating the extracted fluid to treat it properly necessitates collecting vast amounts of data via sensors installed in the oil well and on the drill-head.
    
    \item Oil production entails environmental and human safety hazards, and preventing it requires significant sensor deployment across a large geographical region to gather data regularly and therefore be able to respond rapidly to any ecologically polluting discharge.
\end{itemize}

Big data analytics aids in the automation of critical oil and gas operations, such as exploration, drilling, production, and delivery. The upstream sector, which consists of exploration and drilling, is the most dominant data source among all other sectors, owing to the increasing use of big data analytics for detecting non-conventional shale gas. Furthermore, the oil and gas industry is becoming more volatile due to fluctuating oil prices. As a result, in addition to the engineering team, business teams are increasingly adopting a data-driven strategy to forecast the market and mitigate risks.

\subsection{\textit{Machine Learning as a Data-driven applications in O\&G}}~\\
The smart O\&G industry is a subset of the Industry 4.0 revolution, supported primarily by artificial intelligence (AI), IoT, and cutting-edge computing systems (\eg edge, fog, and cloud computing). An extensive range and volume of relevant data are acquired from many sectors of the O\&G industry due to the widespread adoption of smart sensors and actuators. These data may be evaluated using machine learning models to derive valuable insights and knowledge for the industry and the environment. As a result, in a broad sense, AI is a vital tool for transforming sensor-generated data into new and valuable information and knowledge via Edge-to-Cloud computing.


The term ``data-driven approaches" refers to an arsenal of techniques that can be used to combine different kinds of data, evaluate uncertainties, spot trends, and recover useful facts. Data-dominated software applications running on ML and Deep Neural Network (DNN) models, such as oil production control and emergency surveillance systems ~\cite{eco21,8637768, pei2020towards,mahmoud2019architecture,9144772}, have emerged as the fundamental pillars of the Industry 4.0 revolution \cite{Xuindustry4,LU20171,8231148,8471971}. Especially in remote areas where there is a need for real-time closed-loop automated processes of these applications. The ML-based solutions often take the shape of micro-service processes, each of which may have one or more critical paths that together determine the latency of the whole application \cite{eco21}. These applications require:
\begin{itemize}
    \item A large amount of data to be collected in real-time
    \item Seamless communications of sensing data despite wireless link uncertainties
    \item Dependable execution of ML applications with latency constraints in the face of unexpected load surges (for example, during emergencies)
    \item Transparent deployment and provisioning of applications and resources (also known as ``serverless").
\end{itemize}

Tackling these needs may be difficult, particularly in out-of-the-way places (such as offshore oil fields) with inconsistent connectivity and unstable access to cloud services. These communication and computation constraints become crucial when a remote system must handle massive volumes of data in real-time to manage several facets of an emergency circumstance (for example, an oil spill). While micro datacenters (also known as \emph{fog systems}) are employed to meet the computing demands of such distant systems, their capabilities are sometimes inadequate to deal with the real-time data transport, and processing demands of the load spike \cite{hussain2019federated1}. In the following subsection (ref:edgeAi), we revisit the difficulty posed by limited resources for processing surge in computing demands.

\subsection{\textit{Digital Twin: Another Data-driven Applications in O\&G}}~\\
The term ``digital twin" (DT) refers to a computer simulation of an existing system. Input to the twin may be set from the sensors collecting data from real-world imitation. The twin may then offer real-time feedback to the management about the predicted performance or other repercussions by stimulating the physical object. DT is a data-dominant application that operates based on Machine Learning and the scalability of cloud computing to bring the goal of data integration closer to actuality \cite{parrott2017industry}. The importance of data in a DT system cannot be overstated since it is required for many different types of analysis, prediction, and automation. High-quality, verified, and referenced data is required to produce a practical duplicate. Since the DT operates in real-time, all previously collected data and models must remain accurate, and up-to-date \cite{tao2018digital}. By enabling operators and management in the O\&G sector to transform enormous amounts of data into insights that might make asset failure predictable and hidden revenue opportunities revealed, DT systems can contribute to operational excellence.

\section{Edge-to-Cloud for AI and other Data-driven Applications in Smart O\&G}
\label{subsec:edgeAi}
The wide variety of sensors that communicate through heterogeneous protocols like Modbus, CAN bus, PROFINET, and MQTT \cite{nutanixTechReport} makes it challenging to operationalize an Edge-to-Cloud continuity with local appliances linked to sensors. It is already difficult to implement, with hundreds of agencies and oil rigs involved. In addition, the next generation of cloud-native apps needs different machine learning (ML) frameworks, configurations, and requirements. Furthermore, applications need to be interoperable to function on a variety of devices with diverse processing capabilities (\eg CPU, several kinds of GPU, ASICs, and FPGAs). In addition, the human aspect of IT operational technologies, developers, and data scientists all need to join together to manage the IoT application deployed in the edge-to-cloud continuum. Thus, the primary difficulties throughout the Edge-to-Cloud spectrum might be summed up as follows:

\begin{enumerate}
    \item Connecting a huge number of IoT devices, as well as the edge and the cloud.
    \item Costs associated with wireless communication technologies.
    \item Having access to high-quality computer resources on demand.
    \item Wireless connections that are stagnant, inconsistent, or not operating.
    \item The need for real-time operation of ML-based and other data-driven applications (\eg digital twin).
    \item Data integrity and privacy across Edge-to-Cloud systems.
\end{enumerate}

The Edge-to-Cloud continuum problems for the O\&G industry are broad, complicated, and distinct from traditional solutions. As a result, petroleum professionals and technological specialists are the primary driving forces in developing lucrative eco-friendly solutions for the smart O\&G industry. Meanwhile, academic publications, research papers, and books addressing the junction of petroleum and computer science domains are uncommon. Therefore narrowing the gap between knowing the issue space and providing efficient solutions can help the industry to be more productive and safe.

\section{Federated Fog and It's Challenges in Remote Industry 4.0}~
The earlier sections of this chapter demonstrate the edge-fog-cloud continuum in a hierarchical arrangement where computing resources are distributed vertically. Hence, multiple tiers of execution platforms can be conceptualize where higher tier (\ie cloud) imposes significant latency that may not suitable for latency-sensitive tasks. Accordingly, we investigate the horizontal scalable execution platform, fog systems, in a peer-to-peer arrangement to reduce latency issues. In the industrial sector, fog computing systems are typically located in close proximity that sometimes potential candidates for forming a federation in a peer-to-peer setting to support computing demands in emergencies. For instance, multiple oil rigs with drillships\cite{eakins2015increasing} can be deployed near an offshore hydrocarbon reservoir to extract oil having their private fog systems. Moreover, rescue ships with mobile data centers at disaster time comprise fog systems deployed near disastrous areas whose computing ability can be augmented by forming the fog federation. Hence, it is feasible to assume that some fog systems are underutilized and can support more task processing than their day-to-day requests. Thus, efficient resource allocation across federated computing systems can increase the federated system's quality of Service (QoS). In a related study, \cite{xu2017zenith}, Xu \etal offer a resource allocation instance for edge computing platforms. It uses a decoupled architecture that separates infrastructure management at Edge Computing Infrastructures (ECIs) from service delivery and administration by service providers (SPs). In addition, the authors offer an auction-based resource contract mechanism and a latency-aware scheduling approach that optimizes edge computing systems and service providers' utility. Hence, federating edge computing systems with efficient resource allocation may be used in an emergency to accommodate a spike in task requests. However, several other problems should be addressed to establish a robust and efficient edge federation in an emergency or disaster. The main challenges can be addressed in the following subsections.  

\subsection{\textit{Real-time Services of Industry 4.0}}~\\To improve the response time of latency-intolerant services (\eg sensor data analysis, production monitoring), fog computing systems have been exploited in the literature from the network latency perspective. Lorenzo \etal \cite{lorenzo2017robust} proposed a resource allocation model for wireless edge systems that harvest unused resources of mobile devices to mitigate network congestion. The proposed model utilizes solutions at the physical, access, networking, application, and business layers to reinforce network robustness. This work solely considers networking latency and not end-to-end latency. In \cite{chang2018adaptive}, Chang \etal proposed an optimized resource migration scheme from mobile IoT devices to a heterogeneous Cloud-Fog-Edge computing environment that is aware of the resource-constrained nature of edge devices. It focuses on the performance gain of process migration and assigns tasks based on their run time expectations on the participating systems.

\subsection{\textit{Heterogeneous Fog Systems in Remote Industry 4.0}}~\\ Fog systems in remote industries can be heterogeneous, and the research community addresses two forms of heterogeneity: consistent and inconsistent heterogeneity \cite{salehi2016stochastic}, respectively. Consistent heterogeneity occurs when the same kind of machine has different computational resources. Inconsistent heterogeneity occurs when various types of machines have disparate computational capacities. The requested job may have different execution times depending on the heterogeneity, which substantially impacts the task completion time in an edge system. The problem of heterogeneous data acquisition from sensors in various sectors (e.g., upstream, midstream, downstream) of smart oil fields is addressed in \cite{khan2017reliable} where khan \etal proposed an IoT-based architecture to enable the data acquisition process more simple, secure, robust, reliable and quick. There are several other works (\eg \cite{wunderlich2017network,wang2013nested,hu2016green}) that either do not consider the emergency (oversubscription) or ignore the uncertainties that exist in federated fog environments. In another related work \cite{tareq2018ultra} by the same author, the main focus was on optimizing the wireless network while no resource allocation was performed at the fog system. Therefore, considering both computing and communication latencies is critical to maintaining the QoS of a fog federation.
    
\subsection{\textit{Uncertainty of Task Completion in Fog Systems}}~\\ The primary uncertainty of task completion in a fog federation is influenced by execution and communication latencies. Hence, execution uncertainty mainly refers to the computational resources that execute the assigned task, whereas communication uncertainty is primarily rooted in network systems, especially the upload and download time uncertainty. Both of these uncertainly significantly influence task completion within a fog federation. Xu \etal addressed uncertainty in a similar study \cite{xu2022dynamic}, mentioning that the execution uncertainty caused by performance degradation, service failure, and new service additions remains a significant barrier to the user's service experience. To overcome the uncertainty, this study proposes a software-defined network (SDN)-based fog computing architecture and a dynamic resource provisioning mechanism. Furthermore, the nondominated sorting genetic algorithm-III is used to maximize two objectives, namely energy consumption and completion time, to produce balanced scheduling strategies. In another related paper \cite{li2019resource} on resource allocation and uncertainty, Li \etal suggested a multi-objective optimization problem. Three parallel methods have been developed to increase latency, performance, and resource management. First, a queuing model was investigated in conjunction with task buffering, offloading, and resource allocation algorithms. The authors designed the resource allocation strategy using Lyapunov drift \cite{pejoski2022lyapunov}. An exchange between latency and throughput is found in outcomes for improved system performance.

\begin{figure}[!h]
    \centering
    \includegraphics[width=0.8\textwidth]{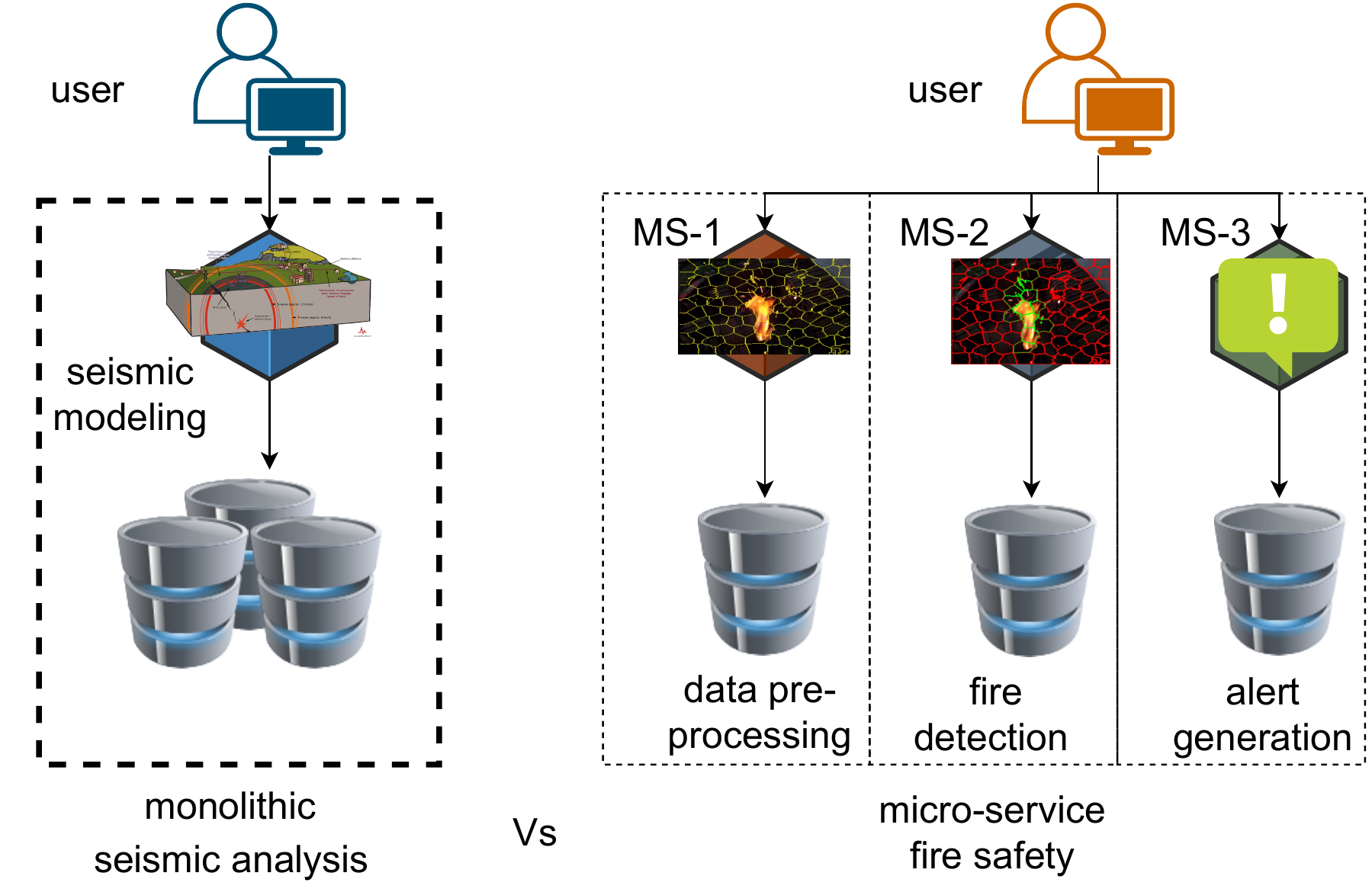}
    \caption{Various software architecture for Industry 4.0 smart applications. Seismic analysis is represented as a monolithic application, whereas fire safety application exhibit micro-service architecture.}
    \label{fig:monvsmicro}
\end{figure}

\subsection{\textit{Software Architecture of Industry 4.0 Applications}}~\\ The industrial revolution has created the demand for emerging smart applications with different software architectures, as depicted in figure \ref{fig:monvsmicro}. Hence, smart applications consist of various micro-services that can be separately deployed with the least amount of administration. For instance, a ``fire safety" application based on micro-service architecture comprised of data pre-processing, fire detection, and alert generation can be deployed in remote industries to ensure the safety of onsite workers from fire hazards. However, the rise of micro-service architecture was mainly introduced to reduce the complexity of large monolithic applications with huge code-base \cite{dragoni2017microservices}. Hence, the research community suggests maintaining the size of micro-service applications optimal \cite{somashekar2021towards,vural2021does,kecskemeti2016entice}, not too large, that can impose complexity in administrating application workflow. In contrast, considering old operational systems, the centralized cloud can only support legacy applications \cite{alam2021cloud,shastry2022approaches} with huge latency-tolerant nature. In contrast, modern Industry 4.0 applications are latency sensitive that need a dynamic execution platform to enable smartness and support swift response time. As such, Rao \etal in \cite{rao2021eco} proposed a dynamic runtime for smart industrial applications that utilize 5G technology with edge-cloud architecture. This work uses application-specific knowledge to map the micro-services into the execution platform. Hence, authors consider only the critical path's latency ignoring various generic micro-services that could play an important role in completing the smart solutions. Additionally, this work considers utilizing cloud data centers to ignore emergency and oversubscribed situations. Similarly, Faticanti \etal in a related study \cite{faticanti2020throughput}, analyzes the throughput needs of micro-service applications while offloading to various fog systems. The authors in this work addressed resource allocation challenges for the fog-native application architectures built on containerized micro-service modules. Two cascading algorithms make up the entirety of the answer. The first one separates fog application components according to throughput, whereas the second governs application orchestration across geographically distributed data centers. 

\section{The Scope of Fog Federation in Industry 4.0}
The smart industry's numerous sensors create massive volumes of data that are often not analyzed due to a lack of storage and processing capabilities \cite{gupta2022wireless}. Alternatively, only some of the data is relevant to any analytical findings. As a result, data pre-processing and filtering of noises and anomalies may be performed in the fog federation \cite{sadri2022data}, leading to effective training of ML models in cloud data centers.

Augmented reality (AR) and real-time video analytics need a quick response and efficient, secure storage systems that fog federation can support \cite{yi2015survey}. For instance, a significant processing delay may confuse a process engineer to perform fault-intolerant work, leading to an accident. Hence AR systems supported by fog computing can maximize throughput and reduce latency in both processing and transmission. Accordingly, K. Ha, \etal in \cite{ha2014towards} design and implement a wearable cognitive assistance spanning backed up by Google Glass and Cloudlet that assists the user by providing hints for social interaction via real-time scene analysis. 

To ensure security and safety, an immense amount of camera sensors are deployed in smart industries (\eg oil and gas, transportation, manufacturing) that perform surveillance 24/7 to detect any anomaly and monitor the hazardous area. Therefore, the captured video needs storage and computational services that can be supported by fog federation. In addition, videos' live streams, transcoding, and ML processing (\eg object detection, classification, object tracking) are more frequent in Industry 4.0 applications. After completing the required services with captured videos, the response can be sent to users in the form of notification, events, description, or video summary. Hence fog federation can be useful for achieving real-time processing (inference) and feedback on a huge amount of video streaming. In addition, scalability can be ensured on low-bandwidth output data. Furthermore, privacy-preserving techniques can also be applied at the fog side to ease the concern of personal privacy leakage in public surveillance systems.

\section{Data Privacy Aspects of a Federated Fog Computing System}
The technological advancement in smart IoT devices and smartphones has increased the possibility of using end devices for various complex ML applications, especially training ML models. The ever-growing power of end devices (\eg mobile phones, PDAs, laptops, wearable) in computing and communication makes the complex Ml model training possible in fog devices. Hence, considering the fog federation, training with various fog systems' local data in a distributed manner can enrich the ML model's accuracy.

\begin{figure}[!h]
    \centering
    \includegraphics[width=0.8\textwidth]{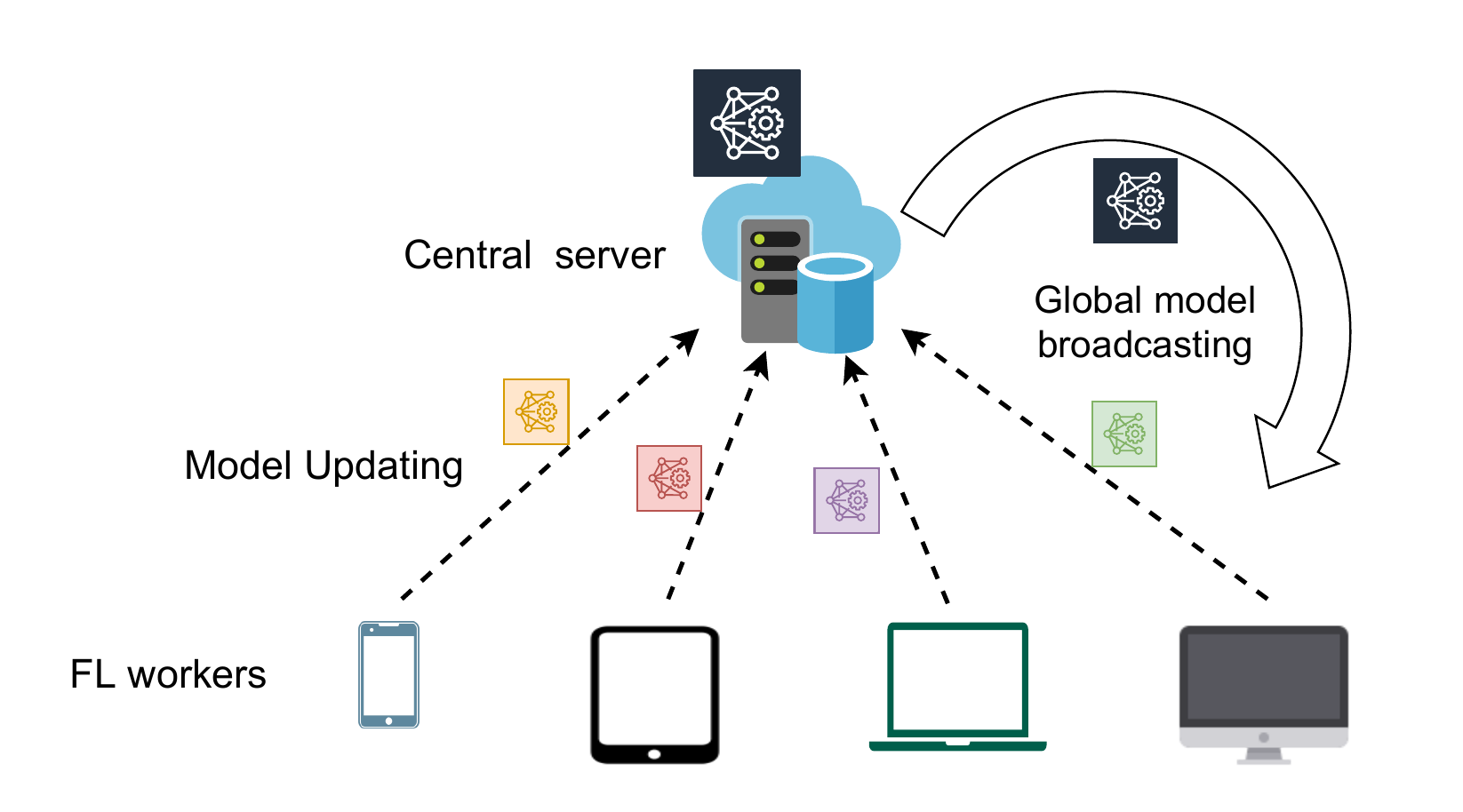}
    \caption{A typical federated learning scenario that consists of FL workers and a central server having the global model. At the beginning of the training, the global model is broadcast to the participating workers to train with their corresponding local training data. After a period of training in FL workers, the updated model is sent back to the server for integration with the global model.}
    \label{fig:typicalFL}
\end{figure}

Although, data security and privacy are the major challenges in this scenario. Hence, federated learning \cite{mcmahan2017communication} is one solution that shares the ML model rather than data that does not leave the owner's fog system. In federated learning, a global model is sent (global model broadcast) to the participating workers' system to train with their local data as presented in figure \ref{fig:typicalFL}. After a certain training period, the updates are sent back to the central server to incorporate the updates into the global model. Then the updated global model is again sent back to the participating FL workers. The process continues until the global model achieves a certain accuracy. Different techniques (\eg fedAvg, fedSGD, fedProx) can be utilized considering the global model's accuracy to incorporate the updates from FL workers.
Furthermore, considering the heterogeneity of FL workers' computation and communication capability, the updates can be generated at different times. Accordingly, two different types of FL techniques are considered in the literature: asynchronous and synchronous FL, respectively \cite{bonawitz2019towards}. Considering the various time to generate updates by the federated worker, some stragglers need to catch up to the certain period of sending the updates to the server. Therefore, asynchronous FL tries to incorporate as many updates as possible, whereas synchronous updates discard the updates that lag behind.

\subsection{\textit{Major Challenges of FL in Fog Federation}}~\\
The federated learning technique in fog federation ensures the ML services while preserving the privacy of the data owners to the end clients. However, due to heterogeneous fog devices and data anomaly, some major challenges need to be addressed that are as following:
\begin{itemize}
    \item \textbf{Class Imbalance Issue in Training Data:}~In FL technique, various FL workers' local data are utilized for ML network model training. Hence, it is possible to have class imbalance issues within some participating workers' local data that can impact the global models' robustness.   
    
    \item \textbf{Communication Cost for Aggregating Updates into Global Model:}~To perform FL training, the global model needs to be transferred to participating workers via the internet. After training, the updates are sent back to the server for synchronization with the global model, and finally updated global model is sent to the FL workers. All the transfer operations utilize internet protocol which can incur a huge amount of communication cost.
    
    \item \textbf{Efficient Management of FL Workers:}~The number of participating FL workers can be huge where unexpected network connectivity and heterogeneous communication protocol make the management scenario nearly impossible\cite{xia2021survey}.
\end{itemize}

\section{Downside of Smart Solutions in Industry 4.0}
Advances in hardware and software technology have evolved the oil and gas sector into a completely automated and machine-dependent industry \cite{lu2020smart}. Although this digital revolution enhances production efficiency, it may produce numerous types of vulnerability and side effects that can lead to catastrophic incidents such as hazardous gas emissions, fire dangers, and oil spills \cite{sakib2021assessment}. Furthermore, the constant advancement of technology opens the potential to hack into information technology (IT) platforms \cite{lu2019oil} that deals with diverse industrial data and communication with the outside network. Another critical technology stack is the operational technology (OT) platform \cite{hahn2016operational}, primarily concerned with direct oil and gas production and processing operations with limited external access. Hence, the bridge between the IT and operational technology (OT) platforms , in particular, raises cyber-threats to oil and gas operations. As a result, while creating smart technology for oil and gas, researchers must study or be cognizant of the drawbacks of smart solutions. As a result, new and current smart solutions should contain better security approaches to ensure the system's reliability. Furthermore, the possible side effects of smart solutions might impede operational efficiencies and become counter-productive.

Accordingly, it is necessary to explore and identify various vulnerable areas of IT and OT platform as well as their interplay aspects in structural categories. Hence, cyber-threats and device incompatibility should be addressed properly to identify various open doors for cyber criminals. One of the issues issue with the oil and gas industry is that it relies on systems that were not designed with network connectivity in consideration. Industrial plants, for example, were never designed to be connected to networks, but with the continuous digital revolution, they are today. This can lead to a risky scenario since a cyber-attack on such a system can impair operations and cause the death of life.

The industrial revolution has increased the utilization of various types of machines that robots or human workers operate. Moreover, these machines sometimes communicate with other machines to complete an industrial operation. Hence, machine-machine and human-machine interactions can go wrong and create opportunities for cyber criminals to sabotage industrial processes. As such, identification of industrial interaction challenges can help to build smart solutions that are safe and secure. Finally, developing any physical or software solution requires human and machine involvement that leads to the engagement of various biases (\eg artificial intelligence, automation, and human-related biases) in smart solutions of Industry 4.0. These biases can lead to unwanted accidents or loopholes for cyber criminals. Therefore, addressing different forms of bias in industrial sectors can help build smart solutions that are resilient to cyber-threats and attacks.

\section{Summary and Positioning of this Dissertation}
This section introduced the Edge-to-Cloud continuum and federation of fog computing paradigms and their goals. First, we discuss various scopes to utilize the Edge-Fog-Cloud continuum for different Industry 4.0 applications, especially real-time nature and machine learning (ML) based applications. Then in chapter 3, we analyze the performance of various Industry 4.0 applications in widely used AWS cloud and Chameleon fog servers. After that, we investigate the challenges of federated fog systems and suggest a statistical resource allocation method across federated fog systems for monolithic workloads in remote industrial sites in chapter 4. Then in chapter 5, we explore the micro-service software architecture of the Industry 4.0 applications and propose a probabilistic partitioning and resource allocation method to improve the robustness of the fog federation. After that, we study the data security and privacy aspects of fog federation by addressing state-of-the-art challenges in privacy-preserving ML-application training for the oil and gas industry in chapter 6. Finally, in chapter 7, we identify the downsides of smart solutions and suggest state-of-the-art solutions for the remote oil and gas industry. In the end, we conclude the dissertation by disclosing a summary of our findings and future avenues to explore in chapter 8.

    \chapter{Performance Analysis of DNN-based Application in Cloud and Fog Systems}
\label{section:performanceAnalysis}
\vspace{14pt}
\section{Overview}
This chapter analyzes the performances of Deep Neural Network (DNN)-based Industry 4.0 applications to study the inference execution times on cloud and fog computing resources. Being an indispensable part of Industry 4.0, DNN-based smart applications make the latency-sensitive inference that needs to be accurate and execute certain application constraints with a specific deadline. The quality of service(QoS) could be compromised due to missing each application's deadline even if the inference accuracy is high. Due to the multi-tenancy and resource heterogeneity inherent to the cloud and fog computing environments, the inference time of DNN-based applications is stochastic. Such stochasticity, if not captured, can potentially lead to a disaster in critical sectors, such as Oil and Gas industry. To make Industry 4.0 robust, solution architects and researchers need to understand the behavior of DNN-based applications and capture the stochasticity that exists in their inference times. Accordingly, in this study, we provide a descriptive analysis of the inference time in the popular cloud platform, Amazon, and in Chameleon as Fog system. 

We employ two statistical methodologies to evaluate DNN-based applications: application-centric and resource-centric. First, we begin with an application-centric analysis in which we statistically model the inference execution time of four categorically unique DNN applications executing on both Amazon and Chameleon. Second, we examine a rate-based indicator known as Million Instruction Per Second (MIPS) for heterogeneous cloud and fog systems using a resource-centric approach. The confidence interval of MIPS for heterogeneous cloud and fog systems is then estimated using non-parametric modeling approaches such as Jackknife and Bootstrap re-sampling. The findings of this work might help academics and cloud solution architects build robust solutions against the stochastic nature of inference time in the cloud, allowing them to deliver higher QoS to their users while avoiding unanticipated repercussions. Furthermore, we provide a DNN-based applications benchmark \footnote{\label{note1}\url{https://github.com/hpcclab/Benchmarking-DNN-applications-industry4.0}} for system architects to employ in building effective resource allocation solutions.

\section{DNN-Based Applications in Industry 4.0}\label{appType}
Among various DNN-based applications utilized in Industry 4.0, we consider four different applications used in the smart O\&G industry. The summary of the chosen applications is demonstrated in table \ref{table:apptype}, which presents the abbreviated name for each application, its DNN (network) model, the type of its input data, the scope of deployment in O\&G Industry \cite{nguyen2020systematic}, and the code base to build the model. The applications' code base, input data, and analysis results are publicly available for reproducibility purposes in the GitHub repository mentioned earlier. In the rest of this section, we explore the characteristics of each application type.

\begin{table}[h!]
\centering
\scalebox{0.85}
{\begin{tabular}{l||l|l|l|l}

\textbf{Application Type} & \textbf{DNN Model} & \textbf{Input Type} & \textbf{Scope} & \textbf{Code Base}\\ \hline \hline
\textit{Fire Detection (Fire)} & Customized Alexnet & Video Segment & \begin{tabular}[c]{@{}l@{}}Control \&\\ Monitoring\end{tabular} & \begin{tabular}[c]{@{}l@{}}Tensorflow \\ (tflearn)\end{tabular}\\ \hline
\textit{\begin{tabular}[c]{@{}l@{}}Human Activity\\ Recognition (HAR)\end{tabular}} & \begin{tabular}[c]{@{}l@{}}Customized Sequential \\ Neural Network\end{tabular} & Motion sensors & \begin{tabular}[c]{@{}l@{}}Workers \\ Safety\end{tabular} & keras\\ \hline
\textit{Oil Spill Detec. (Oil)} & FCN-8 & SAR Images & \begin{tabular}[c]{@{}l@{}}Disaster \\ Management\end{tabular} & keras\\ \hline
\textit{\begin{tabular}[c]{@{}l@{}}Acoustic Impedance \\ Estimation (AIE) \end{tabular}} & \begin{tabular}[c]{@{}l@{}}Temporal Convolutional\\ Network\end{tabular} & Seismic Data & \begin{tabular}[c]{@{}l@{}}Seismic\\ Exploration\end{tabular} & PyTorch\\ 
\end{tabular}}
\caption{\small{DNN-based applications used in O\&G Industry 4.0 along with their network model, input data type, usage scope, and code base.}}
\label{table:apptype}
\end{table}

\subsection{\textit{Fire Detection}}~\\
The fire detection application is an essential component of monitoring systems designed to provide safety and resilience in Industry 4.0. We used a convolutional neural network (CNN) to investigate a fire detection DNN-based application proposed by Dunnings and Breckon \cite{dunnings2018experimentally}. It identifies fire areas (pixels) in real-time in the frames of a monitored video. We use the FireNet model, which correctly identifies and locates fire in each frame of a given video segment, among the several fire detection models offered by the authors. FireNet is a simplified version of the AlexNet model \cite{alom2018history}, with three convolutional layers of sizes 64, 128, and 256. To obtain high-frequency features with a significant response from the preceding layer, each convolutional layer in this model is enhanced with a max-pooling layer and a local response normalization. We created a benchmarking dataset of 240 videos with varied backgrounds to examine the inference time of the fire detection application. All videos are regarded as two seconds long for a fair and accurate appraisal.

\subsection{\textit{Human Activity Recognition}}~\\
Human Activity Recognition (HAR) systems are widely used in Industry 4.0 to ensure workers' safety in hazardous zones. In the HAR system, various sensor data are analyzed that are generated from different sensors used by human workers while performing any physical movement. In this case, motion sensors, such as accelerometers and gyroscopes, that are widely available on handheld PDA devices are utilized to capture human activity-related sensor data. The HAR system we use operates based on the sequential neural network model with four layers to identify the worker's activities (walking, walking upstairs, walking downstairs, sitting). For analysis, we use a dataset of the UCI machine learning repository, known as Human Activity Recognition Using Smartphones \cite{anguita2013public}.

\begin{figure}[h!]
	\centering	
	\includegraphics[scale=0.65]{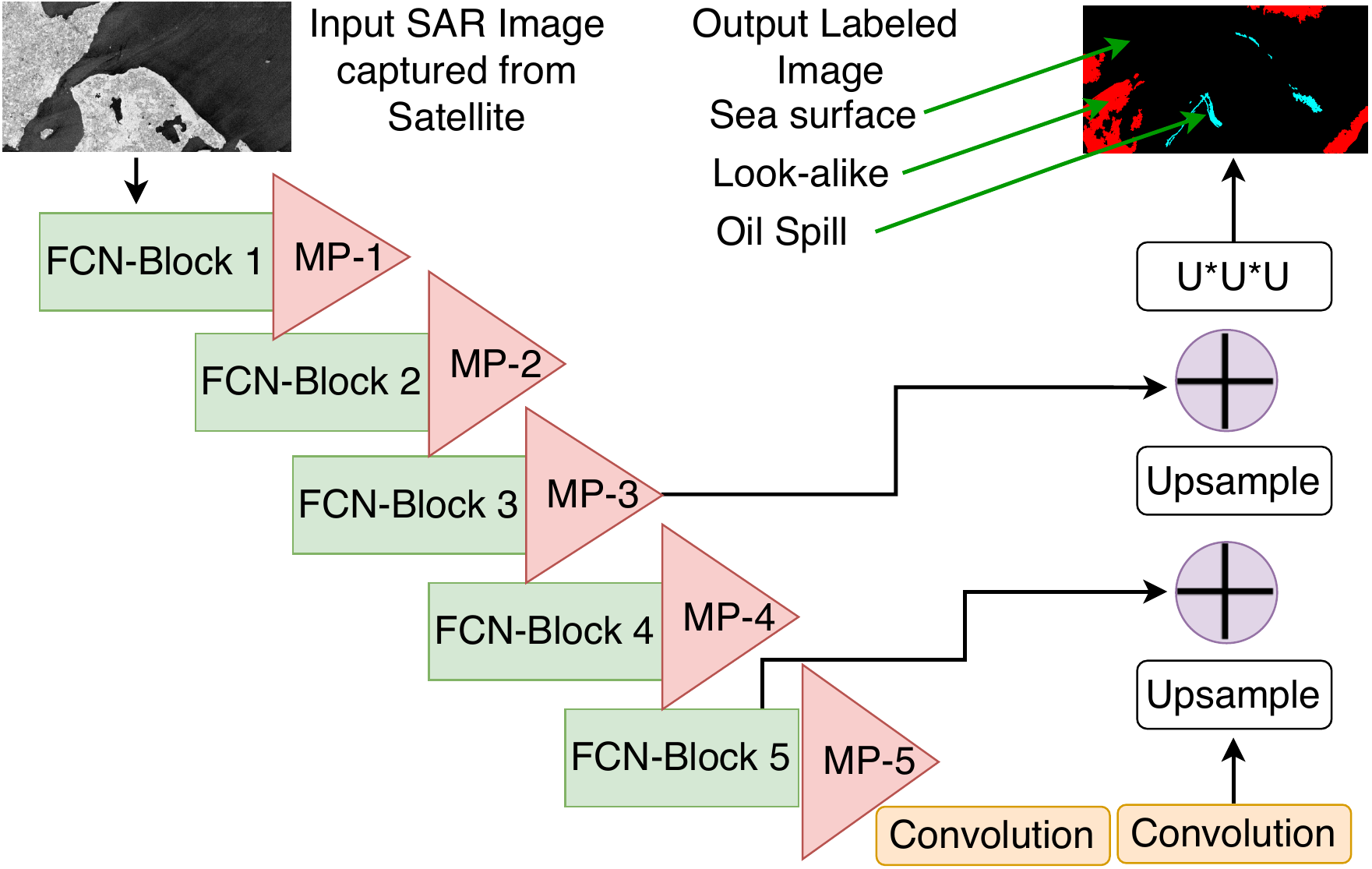}
	\caption{\small{The FCN-8 model is presented in block diagram that consist of 5 fully convolutional network blocks, and 2 up-sampling blocks. The model receives input as a SAR image and perform pixel-wise classification to output a labeled image. }\label{fig:fcn8}}
\end{figure}

\subsection{\textit{Oil Spill Detection}}~\\
Detecting the oil spill is of paramount importance to have a safe and clean O\&G Industry 4.0. The accuracy of DNN-based oil spill detection systems has been promising \cite{krestenitis2019oil}. We utilize a detection system that operates based on the FCN-8 model \cite{long2015fully}, which is depicted in Figure~\ref{fig:fcn8}. As we can see, the model contains five Fully Convolutional Network (FCN) blocks and two up-sampling blocks that collectively perform semantic segmentation (\ie classifying every pixel) of an input image and output a labeled image. The FCN-8 model functions based on the satellite (a.k.a. SAR) \cite{huang2020classification} images. We configure the analysis to obtain the inference time of 110 SAR images collected by MKLab \cite{krestenitis2019oil}.

\begin{figure}[h!]
	\centering	
	\includegraphics[scale=0.66]{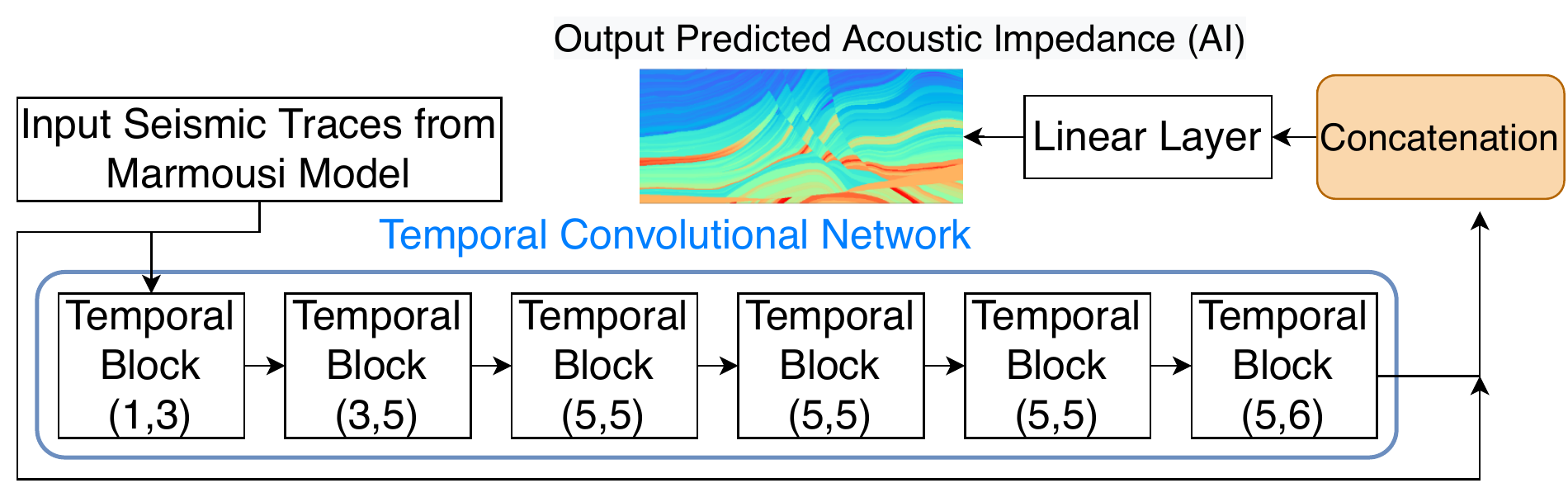}
	\caption{\small{Schematic view of Temporal Convolutional Network (TCN) model that consists of six temporal blocks, the input data, and the output in form of the predicted AI.}}\label{fig:tcn}
\end{figure}

\subsection{\textit{Acoustic Impedance Estimation}}~\\
Estimating acoustic impedance (AI) from seismic data is an important step in O\&G exploration. To estimate AI from seismic data, we utilize a solution functions based on the temporal convolutional network \cite{mustafa2019estimation}, shown in Figure \ref{fig:tcn}. The solution outperforms others in terms of gradient vanishing and overfitting. Marmousi 2 dataset \cite{marmousi} is used to estimate AI.

\section{Computing Platforms for Industry 4.0}\label{section:insType}

\subsection{\textit{Amazon Cloud}}~\\
AWS is a pioneer in the Cloud computing industry and provides more than 175 services, including Amazon EC2 \cite{varia2014overview}, across a large set of distributed data centers. Amazon EC2 provides inconsistently heterogeneous machines (\eg CPU, GPU, and Inferentia) in form of various VM instance types (\eg general purpose, compute-optimized, and machine learning (ML)). Within each VM type, a range of VM configurations (\eg \texttt{large, xlarge, 2xlarge}) are offered that reflect the consistent heterogeneity within that VM type. To realize the impact of machine heterogeneity on the inference time of various applications, we choose one representative VM type of each offered machine type. Table~\ref{awsInstances} represents the type of machines and their specification in terms of number of cores and memory. We note that all the machine types use SSD storage units. Although General Purpose machines are not considered suitable for latency-sensitive DNN-based applications, the reason we study them is their similarity to the specifications of machine types often used in the fog computing platforms. As such, considering these types of machines (and similarly $m1.small$ in the Chameleon cloud) makes the results of this study applicable to cases where fog computing is employed for latency-sensitive applications \cite{icfec19vaughan}.

\begin{table}[h!]
\centering
\scalebox{1.0}
{\begin{tabular}{|l|l|c|c|c|c|c|c|}
\hline
\textbf{Machine Type}           & \textbf{VM Config.} & \multicolumn{1}{l|}{\textbf{vCPU}} & \multicolumn{1}{l|}{\textbf{GPU}} & \multicolumn{1}{r|}{\textbf{Mem. (GB)}}\\ \hline
Mem. Optimized          & \texttt{r5d.xlarge}     & 4                                  & 0                                 & 32                                                                                                                        \\ \hline
ML Optimized & \texttt{inf1.xlarge}             & 4                                  & 0                                 & 8                                                                                                                                                             \\ \hline
GPU                      & \texttt{g4dn.xlarge}             & 4                                  & 1                                 & 16                                                                                                                                         \\ \hline
General Purpose            & \texttt{m5ad.xlarge}             & 4                                  & 0                                 & 16                                                                                                                                                      \\ \hline
Comp. Optimized          & \texttt{c5d.xlarge}              & 4                                  & 0                                 & 8                                                                                                                                        \\ \hline
\end{tabular}}
\caption{Heterogeneous machine types and VM configurations in Amazon EC2 that are considered for performance modeling of DNN-based applications. In this table, ML Optimized represents Inferentia machine type offered by AWS.}
\label{awsInstances}
\end{table}

\begin{table}[h!]
\centering
{\begin{tabular}{l||c|c}
\textbf{VM Config.}       & \textbf{vCPU} & \textbf{Mem. (GB)}   \\ \hline\hline
\texttt{m1.xlarge} & 8 & 16 \\ \hline
\texttt{m1.large}  & 4  & 8  \\ \hline
\texttt{m1.medium}   & 2  & 4 \\ \hline
\texttt{m1.small}  & 1  & 2  \\ \hline
\end{tabular}}
\caption{Various VM flavors in Chameleon cloud are configured to represent a consistently heterogeneous system.}
\label{tab:chameleon}
\end{table}
\vspace{50px}
\subsection{\textit{Chameleon as Fog Computing System}}~\\
Chameleon \cite{chameleon} is a large-scale public computing platform maintained by National Science Foundation (NSF) that usually utilized for academic research purposes. Due to Chameleon's maintenance issues (\eg transient failures, unexpected downtime, resource scarcity), less large scale VM flavors, and distributed nature, we consider Chameleon as Fog computing system. Chameleon supports VM-based heterogeneous computing. It offers the flexibility to manage the compute, memory, and storage capacity of the VM instances. In this study, we use the Chameleon to configure a set of consistently heterogeneous machines (Fog Systems). We configure four VM flavors, namely \texttt{m1.xlarge, m1.large, m1.medium}, and \texttt{m1.small}, as detailed in Table~\ref{tab:chameleon}. We note that VMs offered by Chameleon uses HDD unit as storage.

\section{Environmental Setup for Performance Modeling}
The focus of this study is on latency-sensitive DNN-based applications that are widely used in Industry 4.0. Accordingly, we consider a dynamic (online) system that is already loaded with pre-trained DNN-based applications, explained in the previous section, and executes arriving requests on the pertinent application. This means that we measure the hot start inference time \cite{ogden2019characterizing} in the considered applications. The DNN-based applications mostly use TensorFlow, and the VMs both in AWS and Chameleon are configured to use the framework on top of Ubuntu 18.04.

\begin{figure*}[h!]
\centering
\includegraphics[width=1\textwidth]{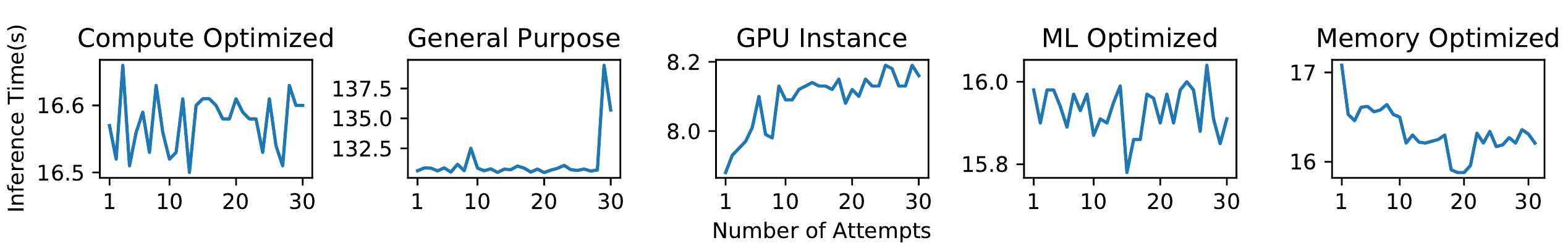}
\caption{\small{The stochastic nature of inference execution time of oil spill application while running on heterogeneous VMs in the AWS. For every VM instance, the oil spill detection application is executed 30 times and those executions are plotted as number of attempts along x-axis. The y-axis represents the inference time for every attempts.}}
\label{fig:stochasticoilaws}
\centering
\end{figure*}

Our initial evaluations in AWS (shown in Figure \ref{fig:stochasticoilaws}) demonstrate that, in different attempts, the inference execution time of an application (Oil Spill) on the same machine type can be highly stochastic. Similar stochasticity is found for chameleon cloud while we run the oil spill detection application 30 times within same VM instance. Hence to capture this randomness (aka consistent heterogeneity) that is caused by several factors, such as transient failures or multi-tenancy \cite{moradi2019adaptive,performanceXiambo}, we base our analysis on 30 times execution of the same request within same VM.

\section{Application-Centric Analysis of Inference Time}\label{section:appCentricAnalysis}
\subsection{\textit{Overview}}~\\
In this part, we capture the inference time of the four DNN applications and try to identify their underlying statistical distributions using various statistical methods. Then, to describe the behavior of inference execution time using a single metric, we explore the central tendency of the distributions.
\subsection{\textit{Statistical Distribution of Inference Execution Time}}~\\
Among various statistical methods, normality tests are widely employed to understand the distribution of the collected samples. Hence, we first use the Shapiro-Wilk test \cite{hanusz2016shapiro} to verify the normality of the inference time distribution on each machine type. Next, we employ the Kolmogorov-Smirnov test \cite{chakravarty1967handbook} to find the best fit distribution based on the sampled inference execution times.

\subsubsection{Shapiro-Wilk test to verify normality of the sampled data. }
The null hypothesis is that the inference execution times are normally distributed. To understand whether a random sample comes from a normal distribution, we perform the Shapiro-Wilk test. The result of this test is considered as $W$, whose low value (lower than $w_\alpha$ threshold) indicates that the sampled data are not normally distributed and vice versa.

The value of $W$ is used to perform the significant testing (\ie calculating P-value). The higher P-value, especially greater than a threshold value (typically 0.05), supports the null hypothesis that the sampled data are normally distributed. 

\begin{table}[h!]
\centering
\scalebox{0.73}
{\begin{tabular}{|c|c|c|c|c|c|}
\hline
\multicolumn{6}{|c|}{\textbf{Execution Time Distribution with Shapiro-Wilk Test in AWS Cloud}} \\ \hline
\textbf{App. Type} & \textbf{\texttt{Mem. Opt.}} & \textbf{\texttt{ML Opt.}} & \textbf{\texttt{GPU}} & \textbf{\texttt{Gen. Pur.}} & \textbf{\texttt{Compt. Opt.}} \\ \hline
\textit{Fire} & \begin{tabular}[c]{@{}c@{}}Not Gaussian\\ (P=2.73$e^{-16}$)\end{tabular} & \begin{tabular}[c]{@{}c@{}}Not Gaussian\\ (P=5.42$e^{-16}$)\end{tabular} & \begin{tabular}[c]{@{}c@{}}Not Gaussian\\ (P=6.59$e^{-16}$)\end{tabular} & \begin{tabular}[c]{@{}c@{}}Not Gaussian\\ (P=2.06$e^{-13}$)\end{tabular} & \begin{tabular}[c]{@{}c@{}}Not Gaussian\\ (P=3.9$e^{-16}$)\end{tabular} \\ \hline
\textit{\begin{tabular}[c]{@{}c@{}}HAR\end{tabular}} & \begin{tabular}[c]{@{}c@{}}Not Gaussian\\ (P=7.12$e^{-8}$)\end{tabular} & \begin{tabular}[c]{@{}c@{}}Not Gaussian\\ (P=1.04$e^{-8}$)\end{tabular} & \begin{tabular}[c]{@{}c@{}}Gaussian\\ (P=0.19)\end{tabular} & \begin{tabular}[c]{@{}c@{}}Not Gaussian\\ (P=1.76$e^{-8}$)\end{tabular} & \begin{tabular}[c]{@{}c@{}}Not Gaussian\\ (P=0.4.62$e^{-5}$)\end{tabular} \\ \hline
\textit{\begin{tabular}[c]{@{}c@{}}Oil\end{tabular}} & \begin{tabular}[c]{@{}c@{}}Not Gaussian\\ (P=8$e^{-4}$)\end{tabular} & \begin{tabular}[c]{@{}c@{}}Not Gaussian\\ (P=2.9$e^{-16}$)\end{tabular} & \begin{tabular}[c]{@{}c@{}}Not Gaussian\\ (P=0.012)\end{tabular} & \begin{tabular}[c]{@{}c@{}}Not Gaussian\\ (P=1.27$e^{-16}$)\end{tabular} & \begin{tabular}[c]{@{}c@{}}Not Gaussian\\ (P=5.86$e^{-14}$)\end{tabular} \\ \hline
\textit{\begin{tabular}[c]{@{}c@{}}AIE\end{tabular}} & \begin{tabular}[c]{@{}c@{}}Gaussian\\ (P=0.46)\end{tabular} & \begin{tabular}[c]{@{}c@{}}Gaussian\\ (P=0.23)\end{tabular} & \begin{tabular}[c]{@{}c@{}}Gaussian\\ (P=0.08)\end{tabular} & \begin{tabular}[c]{@{}c@{}}Not Gaussian\\ (P=1.99$e^{-10}$)\end{tabular} & \begin{tabular}[c]{@{}c@{}}Gaussian\\ (P=0.96)\end{tabular} \\ \hline
\end{tabular}}
\caption{\small{The execution time distributions of DNN-based applications in AWS clouds machines using Shapiro-Wilk test.}}
\label{tab:awsShaprio}
\end{table}

\begin{table}[h!]
\centering
\scalebox{0.78} 
{\begin{tabular}{|c|c|c|c|c|}
\hline
\multicolumn{5}{|c|}{\textbf{Execution Time Distribution with Shapiro-Wilk Test in Chemeleon}} \\ \hline
 \textbf{App. Type} & \textbf{\texttt{m1.xlarge}} & \textbf{\texttt{m1.large}} & \textbf{\texttt{m1.medium}} & \textbf{\texttt{m1.small}} \\ \hline
\textit{Fire} & \begin{tabular}[c]{@{}c@{}}Not Gaussian\\(P=4.05$e^{-5}$) \end{tabular}&\begin{tabular}[c]{@{}c@{}} Not Gaussian\\(P=1.$e^{-4}$) \end{tabular}&\begin{tabular}[c]{@{}c@{}} Not Gaussian\\(P=7.58$e^{-6}$) \end{tabular}&\begin{tabular}[c]{@{}c@{}} Not Gaussian\\(P=1.32$e^{-7}$) \end{tabular}\\ \hline
\textit{\begin{tabular}[c]{@{}c@{}}HAR\end{tabular}} &\begin{tabular}[c]{@{}c@{}} Gaussian \\ (P=0.74) \end{tabular}&\begin{tabular}[c]{@{}c@{}} Not Gaussian \\ (P=0.02) \end{tabular}& \begin{tabular}[c]{@{}c@{}}Gaussian \\ (P=0.18) \end{tabular}&\begin{tabular}[c]{@{}c@{}} Gaussian \\ (P=0.84) \end{tabular}\\ \hline
\textit{\begin{tabular}[c]{@{}c@{}}Oil\end{tabular}} &\begin{tabular}[c]{@{}c@{}} Not Gaussian \\ (P=0.01) \end{tabular}&\begin{tabular}[c]{@{}c@{}} Not Gaussian \\ (P=5.5$e^{-7}$) \end{tabular}& \begin{tabular}[c]{@{}c@{}}Not Gaussian \\ (P=0.01)\end{tabular}& N/A \\ \hline
\textit{\begin{tabular}[c]{@{}c@{}}AIE\end{tabular}} & \begin{tabular}[c]{@{}c@{}}Not Gaussian \\ (P=2.77$e^{-10}$)\end{tabular}& \begin{tabular}[c]{@{}c@{}}Not Gaussian \\ (P= 3.46$e^{-6}$)\end{tabular}& \begin{tabular}[c]{@{}c@{}} Not Gaussian \\ (P= 1.4$e^{-4}$)\end{tabular}& \begin{tabular}[c]{@{}c@{}}Not Gaussian \\ (P=2.46$e^{-6}$)\end{tabular}\\ \hline
\end{tabular}}
\caption{\small{The execution time distributions of DNN applications in Chameleon cloud using Shapiro-Wilk test.}}
\label{tab:chamShap}
\end{table}

The results of Shapiro-Wilk test on the collected inference times for AWS are presented in Table \ref{tab:awsShaprio}, where columns present the various machine types and rows define the application types. The table reflects that our initial assumption is not totally valid. The Shapiro-Wilk test result for the Chameleon cloud, depicted in Table \ref{tab:chamShap}, shows that for only three of the total cases, the normality assumption holds. Considering the lack of normality in several cases, in the next section, we utilize Kolmogorov-Smirnov test to more granularly explore the best fitting distribution for the inference time of each application and also cross validate the prior results we obtained using another statistical method.

\subsubsection{Kolmogorov-Smirnov test to identify the execution time distribution. }
The null hypothesis for the Kolmogorov-Smirnov test is that the inference times of a certain application type on a given machine type follows a statistical distribution. The Kolmogorov-Smirnov Goodness of Fit test (a.k.a. \emph{K-S test}) identifies whether a set of samples derived from a population fits to a specific distribution. Precisely, the test measures the largest vertical distance (called test statistic $D$) between a known hypothetical probability distribution and the distribution generated by the observed inference times (a.k.a. empirical distribution function (EDF)). Then, if $D$ is greater than the critical value obtained from the K-S test P-Value table, then the null hypothesis is rejected.

\begin{table}[h!]
\centering
\scalebox{0.77}
{\begin{tabular}{|c|c|c|c|c|c|}
\hline
\multicolumn{6}{|c|}{\textbf{Execution Time Distribution with Kolmogorov-Smirnov Test in AWS Cloud}} \\ \hline
 \textbf{App. Type} & \textbf{\texttt{Mem. Opt.}} & \textbf{\texttt{ML Opt.}} & \textbf{\texttt{GPU}} & \textbf{\texttt{Gen. Pur.}} & \textbf{\texttt{Compt. Opt.}} \\ \hline
\textit{Fire} & No Distr. & No Distr. & No Distr. & No Distr. & No Distr. \\ \hline
\textit{\begin{tabular}[c]{@{}c@{}}HAR\end{tabular}} & \begin{tabular}[c]{@{}c@{}}Student's t\\ (P=0.08)\end{tabular} & \begin{tabular}[c]{@{}c@{}}Student's t\\ (P=0.77)\end{tabular} & \begin{tabular}[c]{@{}c@{}}Student's t\\ (P=0.99)\end{tabular} & \begin{tabular}[c]{@{}c@{}}Student's t\\ (P=0.57)\end{tabular} & \begin{tabular}[c]{@{}c@{}}Student's t\\ (P=0.95)\end{tabular} \\ \hline
\textit{\begin{tabular}[c]{@{}c@{}}Oil\end{tabular}} & \begin{tabular}[c]{@{}c@{}}Student's t\\ (P=0.44)\end{tabular} & \begin{tabular}[c]{@{}c@{}}Student's t\\ (P=0.96)\end{tabular} & \begin{tabular}[c]{@{}c@{}}Student's t\\ (P=0.5)\end{tabular} & \begin{tabular}[c]{@{}c@{}}Student's t\\ (P=0.20)\end{tabular} & \begin{tabular}[c]{@{}c@{}}Exponential\\ (P=0.21)\end{tabular} \\ \hline
\textit{\begin{tabular}[c]{@{}c@{}}AIE\end{tabular}} & \begin{tabular}[c]{@{}c@{}}Normal\\ (P=0.99)\end{tabular} & \begin{tabular}[c]{@{}c@{}}Normal\\ (P=0.54)\end{tabular} & \begin{tabular}[c]{@{}c@{}}Normal\\ (P=0.47)\end{tabular} & \begin{tabular}[c]{@{}c@{}}Exponential\\ (P=0.16)\end{tabular} & \begin{tabular}[c]{@{}c@{}}Normal\\ (P=0.99)\end{tabular} \\ \hline
\end{tabular}}
\caption{\small{Inference time distributions of DNN-based applications in AWS cloud machines using Kolmogorov-Smirnov test.}} 
\label{tab:kolmoAWS}
\end{table}

\begin{table}[h!]
\centering
\scalebox{0.75}
{\begin{tabular}{|l|l|l|l|l|}
\hline
\multicolumn{5}{|c|}{\textbf{Execution Time Distribution with Kolmogorov-Smirnov test in Chameleon}} \\ \hline
\textbf{App. Type}  & \textbf{\texttt{m1.xlarge}}   & \textbf{\texttt{m1.large}}  & \textbf{\texttt{m1.medium}} & \textbf{\texttt{m1.small}} \\ \hline
\textit{Fire}                   & No Distr   & No Distr & No Distr & Log-normal  \\ \hline
\textit{HAR}       & \begin{tabular}[c]{@{}c@{}}Normal \\ (P=0.98)\end{tabular}  & \begin{tabular}[c]{@{}c@{}}Student's t\\ (P=0.88) \end{tabular}&\begin{tabular}[c]{@{}c@{}} Normal \\ (P=0.66)\end{tabular}& \begin{tabular}[c]{@{}c@{}}Normal \\ (P=0.96)   \end{tabular}  \\ \hline
\textit{Oil}              & \begin{tabular}[c]{@{}c@{}}Log-normal\\(P=0.36) \end{tabular} & \begin{tabular}[c]{@{}c@{}}Log-normal\\(P=0.99) \end{tabular}& \begin{tabular}[c]{@{}c@{}}Log-normal\\(P=0.81)\end{tabular}      & N/A            \\ \hline
\textit{AIE}                & \begin{tabular}[c]{@{}c@{}} Student's t \\(P= 0.47)  \end{tabular} &\begin{tabular}[c]{@{}c@{}} Student's t \\(P=0.12) \end{tabular}   & \begin{tabular}[c]{@{}c@{}}Student's t\\(P=0.25) \end{tabular}  & \begin{tabular}[c]{@{}c@{}}Student's t\\(P=0.83) \end{tabular} \\ \hline
\end{tabular}}
\caption{\small{Inference time distributions of DNN-based applications in Chameleon's machines using the K-S test.}}
\label{tab:chamKol}
\end{table}

The results of the K-S test on the observed inference times in AWS and Chameleon clouds are depicted in Table \ref{tab:kolmoAWS} and \ref{tab:chamKol}, respectively. From Table \ref{tab:kolmoAWS}, we find that, in AWS, majority of the entries either represent Normal distribution or Student's t-distribution that exposes similar properties. However, we observe that the inference time of Fire Detection application does not follow any particular distribution with an acceptable P-Value. We also observe that the inference times of both Oil Spill application on Compute Optimized machine and AIE application on General Purpose machine follow Exponential distribution. However, low P-Value in both of these cases indicate a weak acceptance of the null hypothesis.

On the contrary, Table \ref{tab:chamKol} reflects the dominance of Log-normal (\ie the logarithm of the random variable is normally distributed) and Student's t-distribution over other distributions in the Chameleon cloud. Analyzing the execution traces shows us that the inference times in Chameleon are predominantly larger than the ones in AWS that causes right-skewed property. Hence, the distribution tends to be Log-normal. Consistent with AWS observations, we see that the Fire Detection application does not follow any distribution in most cases. Our further analysis showed that the lack of distribution is due to the input videos' variety (\eg frame rate and resolution). When we reduced the degree of freedom in the input videos and limited them to those with the same configuration (frame rate), we noticed the inference time followed a Log-normal distribution. The observation shows that the characteristics and variation of input data can be decisive in the statistical behavior of inference times (mentioned in highlighted insight). Finally, we note that the Oil Spill application cannot be run on \texttt{m1.small} machine owing to its limited memory.

\noindent\textbf{Insights:}
The key insights of the analysis we conducted on identifying the distribution of inference time are as follows:
\begin{mdframed}[backgroundcolor=blue!20] \begin{itemize}
    \item Shapiro-Wilk test for AWS and Chameleon rejects the null hypothesis that the inference times of DNN-based applications follow the Normal distribution.
    \item The K-S test reflects the dominance of Student's t-distribution of inference time in both AWS (Table \ref{tab:kolmoAWS}), and Chameleon (Table \ref{tab:chamKol}). 
    \item Various configurations of input data is decisive on the statistical distribution of the inference time. 
\end{itemize}
\end{mdframed}

\begin{table}[h!]
\centering
\scalebox{0.72}
{\begin{tabular}{|c|c|c|c|c|c|}
\hline
\multicolumn{6}{|c|}{\textbf{Mean and Standarad Deviation of Inference Execution Times (ms) in AWS}} \\ \hline
\textbf{App. Type} & \textbf{\texttt{Mem. Opt.}} & \textbf{\texttt{ML Opt.}} & \textbf{\texttt{GPU}} & \textbf{\texttt{Gen. Pur.}} & \textbf{\texttt{Compt. Opt.}} \\ \hline
\textit{Fire} & \begin{tabular}[c]{@{}c@{}}$\mu$=1461.8\\ $\sigma$=457.3\end{tabular} & \begin{tabular}[c]{@{}c@{}}$\mu$=1281.7\\ $\sigma$=387.93\end{tabular} & \begin{tabular}[c]{@{}c@{}}$\mu$=1349.5\\ $\sigma$=418.9\end{tabular} & \begin{tabular}[c]{@{}c@{}}$\mu$ =1534.8\\ $\sigma$=494.7\end{tabular} & \begin{tabular}[c]{@{}c@{}}$\mu$=1421.4\\ $\sigma$=441.8\end{tabular} \\ \hline
\textit{\begin{tabular}[c]{@{}c@{}}HAR\end{tabular}} & \begin{tabular}[c]{@{}c@{}}$\mu$=1.27\\ $\sigma$=0.082\end{tabular} & \begin{tabular}[c]{@{}c@{}}$\mu$=0.66\\ $\sigma$=0.006\end{tabular} & \begin{tabular}[c]{@{}c@{}}$\mu$=0.51\\ $\sigma$=0.006\end{tabular} & \begin{tabular}[c]{@{}c@{}}$\mu$ =1.17\\ $\sigma$=0.042\end{tabular} & \begin{tabular}[c]{@{}c@{}}$\mu$=0.66\\ $\sigma$=0.003\end{tabular} \\ \hline
\textit{\begin{tabular}[c]{@{}c@{}}Oil\end{tabular}} & \begin{tabular}[c]{@{}c@{}}$\mu$=269.9\\ $\sigma$=1.01\end{tabular} & \begin{tabular}[c]{@{}c@{}}$\mu$=218.8\\ $\sigma$=0.66\end{tabular} & \begin{tabular}[c]{@{}c@{}}$\mu$=65.98\\ $\sigma$=0.47\end{tabular} & \begin{tabular}[c]{@{}c@{}}$\mu$=667.1\\ $\sigma$=2.26\end{tabular} & \begin{tabular}[c]{@{}c@{}}$\mu$=242.9\\ $\sigma$=0.68\end{tabular} \\ \hline
\textit{\begin{tabular}[c]{@{}c@{}}AIE\end{tabular}} & \begin{tabular}[c]{@{}c@{}}$\mu$=7.02\\ $\sigma$=0.02\end{tabular} & \begin{tabular}[c]{@{}c@{}}$\mu$=6.41\\ $\sigma$=0.03\end{tabular} & \begin{tabular}[c]{@{}c@{}}$\mu$=7.55\\ $\sigma$=0.04\end{tabular} & \begin{tabular}[c]{@{}c@{}}$\mu$=9.35\\ $\sigma$=0.06\end{tabular} & \begin{tabular}[c]{@{}c@{}}$\mu$=7.95\\ $\sigma$=0.02\end{tabular} \\ \hline
\end{tabular}}
\caption{\small{The measurement of central tendency metric ($\mu$), and data dispersion metric ($\sigma$) on the observed inference times in AWS.}}
\label{tab:meanAWS}
\end{table}

\begin{table}[h!]
\centering
\scalebox{0.82}
{\begin{tabular}{|l|l|l|l|l|}
\hline
\multicolumn{5}{|l|}{\textbf{Mean and Std. of Inference Execution Times (ms) in Chameleon}} \\ \hline
\textbf{App. Type} & \textbf{\texttt{m1.xlarge}} & \textbf{\texttt{m1.large}} & \textbf{\texttt{m1.medium}} & \textbf{\texttt{m1.small}} \\ \hline
\textit{Fire} & \begin{tabular}[c]{@{}l@{}}$\mu$=2155.20\\ $\sigma$=725.48\end{tabular} & \begin{tabular}[c]{@{}l@{}}$\mu$=2213.30\\ $\sigma$=731.50\end{tabular} & \begin{tabular}[c]{@{}l@{}}$\mu$=2330.80\\ $\sigma$=742.20\end{tabular} & \begin{tabular}[c]{@{}l@{}}$\mu$=3184.80\\ $\sigma$=1033.30\end{tabular} \\ \hline
\textit{\begin{tabular}[c]{@{}l@{}}HAR\end{tabular}} & \begin{tabular}[c]{@{}l@{}}$\mu$=22.14\\ $\sigma$=0.76\end{tabular} & \begin{tabular}[c]{@{}l@{}}$\mu$=47.69\\ $\sigma$=1.26\end{tabular} & \begin{tabular}[c]{@{}l@{}}$\mu$=49.24\\ $\sigma$=0.57\end{tabular} & \begin{tabular}[c]{@{}l@{}}$\mu$=52.69\\ $\sigma$=0.78\end{tabular} \\ \hline
\textit{\begin{tabular}[c]{@{}l@{}}Oil\end{tabular}} & \begin{tabular}[c]{@{}l@{}}$\mu$=147.16\\ $\sigma$=5.23\end{tabular} & \begin{tabular}[c]{@{}l@{}}$\mu$=222.22\\ $\sigma$=2.89\end{tabular} & \begin{tabular}[c]{@{}l@{}}$\mu$=412.78\\ $\sigma$=4.99\end{tabular} & N/A \\ \hline
\textit{\begin{tabular}[c]{@{}l@{}}AIE\end{tabular}} & \begin{tabular}[c]{@{}l@{}}$\mu$=6.23\\ $\sigma$=0.25\end{tabular} & \begin{tabular}[c]{@{}l@{}}$\mu$=6.23\\ $\sigma$=0.15\end{tabular} & \begin{tabular}[c]{@{}l@{}}$\mu$=6.40\\ $\sigma$=0.13\end{tabular} & \begin{tabular}[c]{@{}l@{}}$\mu$=7.72\\ $\sigma$=0.24\end{tabular} \\ \hline
\end{tabular}}
\caption{\small{Central tendency metric ($\mu$), and data dispersion metric ($\sigma$) of the  inference times in the Chameleon cloud.}}
\label{tab:meanCham}
\end{table}

\subsection{\textit{Analysis of Central Tendency and Dispersion Measures}}~\\
Leveraging the statistical distributions of inference times, in this part, we explore their central tendency metric that summarizes the behavior of collected observations in a single value. In addition, to analyze the statistical dispersion of the observations, we measure the standard deviation of the inference times. In particular, we estimate the arithmetic mean and standard deviation of the inference times. The central tendency metric of inference times for AWS and Chameleon systems are shown in Tables~\ref{tab:meanAWS} and~\ref{tab:meanCham}, respectively. The \textbf{key insights} are as follows:
\begin{mdframed}[backgroundcolor=blue!20] 
\begin{itemize}
    \item Machine Learning Optimized and GPU instances often outperform other AWS machine types.
    \item In both clouds, the inference times of Fire and Oil experience a higher standard deviation in compare with other applications. The high uncertainty is attributed to the characteristics of their input data; Oil Spill input images suffer from class imbalance \cite{krestenitis2019oil}, whereas, Fire input videos are highly variant. 
    \item In Chameleon VMs with a consistent heterogeneity, DNN applications with dense network models (\eg Oil and Fire) can take advantage of powerful machines (\eg \texttt{m1.xlarge}) to significantly reduce their inference times.
    \item Overall, AWS offers a lower inference time than Chameleon. The reason is utilizing SSD units in AWS rather than HDD in Chameleon. In addition, we noticed that Chameleon experiences more transient failures that can slow down the applications.
\end{itemize}
\end{mdframed}

\section{Resource-Centric Analysis of Inference Time}\label{section:resCentric}
In performance analysis of computing systems, a rate-based metric \cite{lilja2005measuring} is defined as the normalization of number of computer instructions executed to a standard time unit. MIPS is a popular rate-based metric that allows comparison of computing speed across two or more computing systems. Given that computing systems (\eg AWS ML Optimized and GPU) increasingly use instruction-level facilities for ML applications, our objective in this part is to analyze the performance of different machine types in processing DNN-based applications. The results of this analysis can be of particular interest to researchers and cloud solution architects whose endeavor is to develop tailored resource allocation solutions for Industry 4.0 use cases.  As for rate-based metrics we do not assume any distribution \cite{patil2010using}, we conduct a non-parametric approach. In addition to MIPS, we provide the range of MIPS in form of \textit{Confidence Intervals} (CI) for each case. \looseness=-1

\begin{table}[h!]
\centering
\scalebox{0.85}
{\begin{tabular}{|c|c|c|c|c|c|}
\hline
\multicolumn{6}{|c|}{\textbf{The MIPS for DNN Applications in AWS Cloud}} \\ \hline
\textbf{App. Type} & \textbf{\texttt{Mem. Opt.}} & \textbf{\texttt{ML Opt.}} & \textbf{\texttt{GPU}} & \textbf{\texttt{Gen. Pur.}} & \textbf{\texttt{Compt. Opt.}} \\ \hline
\textit{Fire} & 1938.63 & 2196.35 & 2092.72 & 1862.04 & 1989.56 \\ \hline
\textit{\begin{tabular}[c]{@{}c@{}}HAR\end{tabular}} & 838640.65 & 1595874.34 & 2040057.33 & 891754.48 & 1581709.12 \\ \hline
\textit{\begin{tabular}[c]{@{}c@{}}Oil\end{tabular}} & 164.54 & 168.58 & 331.98 & 20.46 & 162.01 \\ \hline
\textit{\begin{tabular}[c]{@{}c@{}}AIE\end{tabular}} & 145.58 & 180.28 & 150.25 & 131.25 & 160.32 \\ \hline
\end{tabular}}
\caption{\small{MIPS values of heterogeneous machines in AWS for each DNN-based application.}}
\label{mipsAWS}
\end{table}

\begin{table}[h!]
\centering
\scalebox{0.80}
{\begin{tabular}{|l|l|l|l|l|}
\hline
\multicolumn{5}{|c|}{\textbf{The MIPS for DNN Applications in Chameleon}}                   \\ \hline
\textbf{App. Types} & \textbf{\texttt{m1.xlarge}} & \textbf{\texttt{m1.large}} & \textbf{\texttt{m1.medium}} & \textbf{\texttt{m1.small}} \\ \hline
\textit{Fire}                                   & 1327.81 & 1282.33  & 1249.63 & 871.36\\ \hline
\textit{HAR}                       & 91.78    & 102.51   & 124.76 & 136.62  \\ \hline
\textit{Oil}                              & 18267.35   & 11233.41  & 6243.94  & N/A    \\ \hline
\textit{AIE}                                & 246366.52   & 249551.29 & 236300.93 & 201807.49 \\ \hline
\end{tabular}}
\caption{\small{MIPS vales for heterogeneous machines on Chameleon cloud for each DNN-based application.}}
\label{mipsCham}
\end{table}

Let application $i$ with $n_i$ instructions have $t_{im}$ inference time on machine $m$. Then, MIPS of machine $m$ to execute the application is defined as $MIPS_{mi} = n_i/(t_{im}\times 10^6)$. Hence, before calculating MIPS for any machine, we need to estimate the number of instructions ($n$) of each DNN-based application. For that purpose, we execute each task $t_i$ on a machine whose MIPS is known and estimated $n_i$. Then, for each machine $m$, we measure $t_{im}$ and subsequently calculate $MIPS_{mi}$. Tables~\ref{mipsAWS} and~\ref{mipsCham} show the MIPS values for AWS and Chameleon, respectively.\looseness=-1 

To measure the confidence intervals (CI) of MIPS for each application type in each machine type, we use the non-parametric statistical methods \cite{patil2010using} that perform prediction based on the sample data without making any assumption about their underlying distributions. As we deal with a rate-based metric, we use harmonic mean that offers a precise analysis for this type of metric rather than the arithmetic mean. We utilize Jackknife \cite{patil2010using} re-sampling method and validate it using Bootstrap \cite{patil2010using}, which is another well-known re-sampling method. Both of these methods employ harmonic mean to measure the confidence intervals of MIPS. 

\begin{table}[h!]
\centering
\scalebox{0.83}
{\begin{tabular}{|l|l|l|l|l|l|}
\hline
\multicolumn{6}{|c|}{\textbf{CI of MIPS using Jackknife Method in AWS cloud}}                                                                                                                                                                                                                                                                                     \\ \hline
\textbf{App. Type} & \textbf{\texttt{Mem. Opt.}}                                                    & \textbf{\texttt{ML Opt.}}                                          & \textbf{\texttt{GPU}}                                                                 & \textbf{\texttt{Gen. Pur.}}                                                     & \textbf{\texttt{Compt. Opt.}}                                                   \\ \hline
\textit{Fire} & \begin{tabular}[c]{@{}l@{}}[1549.42,\\ 1975.65]\end{tabular} & \begin{tabular}[c]{@{}l@{}}[1770.81,\\2243.04]\end{tabular} & \begin{tabular}[c]{@{}l@{}}[1671.78,\\2131.66]\end{tabular} & \begin{tabular}[c]{@{}l@{}}[1465.31,\\1889.77]\end{tabular} & \begin{tabular}[c]{@{}l@{}}[1594.78,\\2028.36]\end{tabular} \\ \hline
\textit{HAR}      & \begin{tabular}[c]{@{}l@{}}[812040.26,\\ 856355.96]\end{tabular}   & \begin{tabular}[c]{@{}l@{}}[1592214.75,\\ 1599426.64]\end{tabular}   & \begin{tabular}[c]{@{}l@{}}[2033084.47,\\ 2046727.57]\end{tabular} & \begin{tabular}[c]{@{}l@{}}[880417.69,\\ 901345.49]\end{tabular} & \begin{tabular}[c]{@{}l@{}}[1580275.10,\\ 1585598.85]\end{tabular}   \\ \hline
\textit{Oil}             & \begin{tabular}[c]{@{}l@{}}[163.55,\\165.47]\end{tabular} & \begin{tabular}[c]{@{}l@{}}[168.36,\\168.81]\end{tabular} & \begin{tabular}[c]{@{}l@{}}[330.68,\\333.22]\end{tabular} & \begin{tabular}[c]{@{}l@{}}[20.35,\\20.57]\end{tabular}   & \begin{tabular}[c]{@{}l@{}}[161.86,\\162.17]\end{tabular} \\ \hline
\textit{AIE}               & \begin{tabular}[c]{@{}l@{}}[139.02,\\141.04]\end{tabular} & \begin{tabular}[c]{@{}l@{}}[155.56,\\156.01]\end{tabular} & \begin{tabular}[c]{@{}l@{}}[141.57,\\142.03]\end{tabular} & \begin{tabular}[c]{@{}l@{}}[118.06,\\119.82]\end{tabular} & \begin{tabular}[c]{@{}l@{}}[148.35,\\149.00]\end{tabular} \\ \hline
\end{tabular}}
\caption{\small{The confidence intervals of MIPS values for DNN-based applications in AWS machines, resulted from Jackknife re-sampling method.}}
\label{jackAWS}
\end{table}

\begin{table}[h!]
\centering
\scalebox{0.80}
{\begin{tabular}{|l|l|l|l|l|}
    \hline
    \multicolumn{5}{|c|}{\textbf{CI of MIPS using Jackknife Method in Chameleon Cloud}}                                                                                                                                                                                                                                \\ \hline
    \textbf{App. Type} & \textbf{\texttt{m1.xlarge}}                                                           & \textbf{\texttt{m1.large}}                                                            & \textbf{\texttt{m1.medium}}                                                           & \textbf{\texttt{m1.small}}                                                      \\ \hline
    \textit{Fire}                                   & \begin{tabular}[c]{@{}l@{}}[1032.11,\\1341.75]\end{tabular}   & \begin{tabular}[c]{@{}l@{}}[1010.62,\\1303.02]\end{tabular}   & \begin{tabular}[c]{@{}l@{}}[964.76,\\1259.68]\end{tabular}   & \begin{tabular}[c]{@{}l@{}}[670.82,\\872.85]\end{tabular}   \\ \hline
    \textit{HAR}                       & \begin{tabular}[c]{@{}l@{}}[88.27,\\ 94.20]\end{tabular}   & \begin{tabular}[c]{@{}l@{}}[99.84,\\ 104.49]\end{tabular}   & \begin{tabular}[c]{@{}l@{}}[122.33,\\ 126.67]\end{tabular}   & \begin{tabular}[c]{@{}l@{}}[135.13,\\ 137.92]\end{tabular}   \\ \hline
    \textit{Oil}          & \begin{tabular}[c]{@{}l@{}}[18083.59,\\18628.64]\end{tabular}   & \begin{tabular}[c]{@{}l@{}}[11159.71,\\11662.41]\end{tabular}   & \begin{tabular}[c]{@{}l@{}}[6139.59,\\6262.15]\end{tabular}                                                               & N/A                                                                 \\ \hline
    \textit{AIE}        & \begin{tabular}[c]{@{}l@{}}[237710.12,\\252686.82]\end{tabular} & \begin{tabular}[c]{@{}l@{}}[247166.73,\\251673.68]\end{tabular} & \begin{tabular}[c]{@{}l@{}}[168804.58,\\268273.11]\end{tabular} & \begin{tabular}[c]{@{}l@{}}[199676.71,\\203681.17]\end{tabular} \\ \hline
\end{tabular}}
\caption{\small{Confidence intervals of MIPS values for different DNN-based applications in Chameleon machines, resulted from Jackknife re-sampling method.}}
\label{jackCham}
\end{table}
\subsection{\textit{Estimating Confidence Interval using Jackknife Method}}~\\ Let $p$ be the number of observed inference times. The Jackknife method calculates the harmonic mean in $p$ iterations, each time by eliminating one sample. That is, each time it creates a new sample (re-sample) with size $p-1$. Let $x_j$ be the $j$th observed inference time. Then, the harmonic mean of re-sample $i$ is called the pseudo-harmonic value (denoted as $y_i$) and is calculated based on Equation~\ref{jackEqu}.
\begin{equation}
    y_{i}=\frac{p-1}{\sum\limits_{j=1,j\neq i}^{p}\frac{1}{x_{j}}}
    \label{jackEqu}
\end{equation}
Next, the arithmetic mean (denoted $\bar{y}$) of the $p$ pseudo-harmonic values is computed, and is used to estimate the standard deviation. Finally, the t-distribution table is used to calculate the CI boundaries with a 95\% confidence level. The result of the Jackknife method for AWS machines is shown in Table \ref{jackAWS} that conforms with the MIPS calculation in Table \ref{mipsAWS}. Similarly, the results of analysis for Chameleon cloud using Jackknife method, shown in Table \ref{jackCham}, validate the prior MIPS calculations in Table \ref{mipsCham}. 
However, in the next part, we cross-validate these results using Bootstrap method.
\subsection{\textit{Estimating Confidence Interval using Bootstrap Method}}~\\ Bootstrap repeatedly performs random sampling with a replacement technique \cite{patil2010using} on the observed inference times. The random sampling refers to the selection of a sample with the chance of non-zero probability and the number (represented as $k$) of re-sample data depends on the user's consideration. After re-sampling, the harmonic means of $k$ number of samples are calculated and sorted in ascending order to estimate the confidence intervals. Finally, for a specific confidence level, the ($\alpha / 2 \times k$)th and ($(1 - \alpha / 2) \times k$)th values are selected from the sorted samples as the lower and upper bounds of the CI. We set the $k$ value to 100 and $\alpha$ to 0.05 for 95\% confidence level.\looseness=-1   

For both AWS and Chameleon, the results of CI analysis using the Bootstrap method are similar to, thus validate, the ranges estimated by the Jackknife method. Therefore, due to the shortage of space, we do not report the table of MIPS values for the Bootstrap method. However, we note that the CI ranges provided by the Bootstrap method are shorter (\ie have less uncertainty), regardless of the application type and the cloud platform. The reason for the shorter range is that Bootstrap performs re-sampling with a user-defined number of samples that can be larger than the original sample size.

\begin{figure}[h!]
	\centering	
	\includegraphics[scale=0.80]{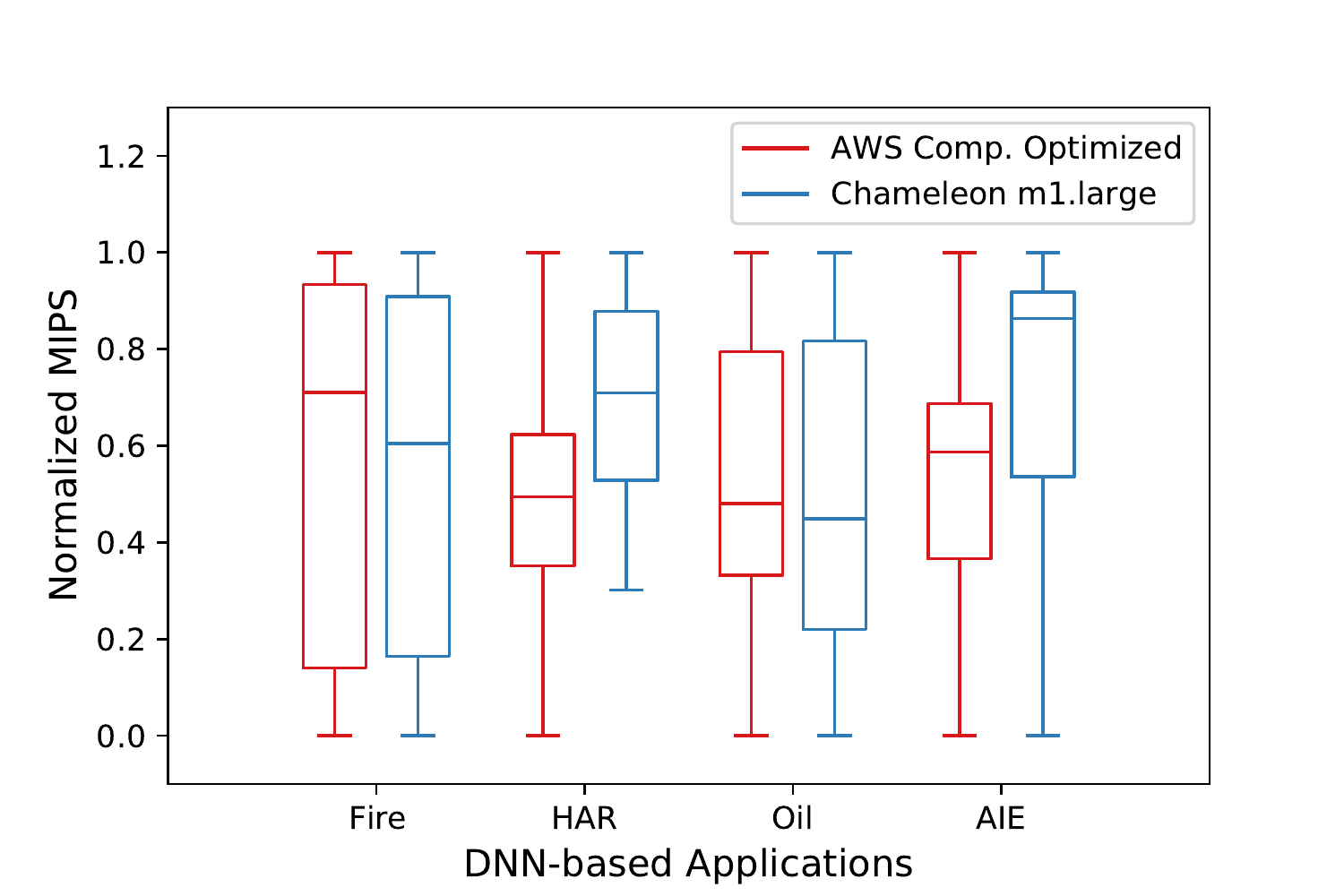}
	\caption{\small{Comparative analysis of the MIPS values of AWS and Chameleon machines for various DNN-based applications. For the sake of presentation, the MIPS values are  normalized between [0,1].}\label{fig:boxplot}}
\end{figure}

To perform a cross-platform analysis of the MIPS values, in Figure~\ref{fig:boxplot}, we compare the range of MIPS values for AWS \texttt{Compute Optimized} against \texttt{m1.large} that is a compatible machine type in Chameleon (see Tables~\ref{awsInstances} and~\ref{tab:chameleon}). The horizontal axis of this figure shows different application types and the vertical axis shows the MIPS values, normalized based on MinMax Scaling  in the range of [0,1], for the sake of better presentation.
Due to high variation in the input videos, we observe a broad CI range for Fire detection across both cloud platforms. 
However, for HAR, Oil Spill, and AIE applications, we observe that the first and third quartiles of the CI range in Chameleon (whose machines are prone to more transient failures \cite{charyyev2019towards}) is larger than those in AWS. This wide range indicates that, apart from variations in the input data, the reliability of underlying resources is also decisive on the stochasticity of the inference times. 

\section{Summary and Discussion}\label{section:concl}
Accurately estimating the inference time of latency-sensitive DNN-based applications plays a critical role in robustness and safety of Industry 4.0. Such accurate estimations enable cloud providers and solution architects to devise resource allocation and load balancing solutions that are robust against uncertainty exists in the execution time of DNN-based applications. In this work, we provide application- and resource-centric analyses on the uncertainty exists in the inference times of several DNN-based applications deployed on heterogeneous machines of two computing platforms, namely AWS and Chameleon. In the first part, we utilized the Shapiro-Wilk test to verify if the assumption of Normal distribution for the inference time holds. We observed that the inference times often do not follow a Normal distribution. Therefore, in the second part, we broaden our distribution testing investigation and utilized the Kolmogorov-Smirnov test to verify the underlying distributions in each case. The analysis showed that inference times across the two computing platforms often follow Student's t-distribution. However, in several cases in Chameleon system we observed the Log-normal distribution that we attribute it to the uncertain performance of VMs in this platform. 
Next, to conduct a resource-centric analysis, we modeled MIPS (as a rate-based performance metric) of the heterogeneous machines for each application type. In the analysis, we took a non-parametric approach, which is suitable for rate-based metrics, and utilized the Jackknife and Bootstrap re-sampling methods with harmonic mean to determine the range of confidence intervals of the MIPS values in each case. The calculated MIPS values and their CI ranges reflect the behavior of different DNN-based applications under various machine types of cloud and fog systems. A comparative analysis of the CI ranges across AWS and Chameleon demonstrate that the uncertainty in the inference time is because of variations in the input data and unreliability of the underlying platforms. 
In the future, we plan to incorporate the findings of this research to devise accurate resource allocation methods in IoT and edge computing systems. In addition, we plan to develop a predictive analysis to determine the execution of each inference task upon arrival.

    \chapter{The Benefits of Federated Fog to Manage Monolithic Workload in Remote Industrial Sites}
\label{section:loadBalance}

\section{Overview}
In the previous chapters, our preliminary research found that fog federation can be a potential computational platform for remote smart industries with stochastic execution behaviors for Industry 4.0 applications. Hence, the stochastic execution of Industry 4.0 applications has an influence on task completion times. In this case, an efficient resource allocation and load balancing technique that is aware of stochastic execution behaviors of Industry 4.0 applications can ensure the system's robustness by enabling the on-time completion of receiving tasks. Accordingly, in this chapter, we \emph{first} strategically develop a load-balancing method for allocating arriving tasks to a fog federation. Hence, our primary goal is to ensure the system's robustness (fog federation) in terms of meeting the deadlines of arriving tasks. To achieve the goal, we estimate the end-to-end latency of a receiving task in a fog system and utilize the latency to predict the task completion time across the fog federation. Hence, we propose a probabilistic task allocation method in the load balancer of each fog system that is aware of the latency constraints of the receiving tasks. Then, in the \emph{second} part, we evaluate our proposed load balancing method using the synthetic workload (customized to industrial tasks workload) of EdgeCloudSim \cite{sonmez2017edgecloudsim} simulator. 

\section{End-to-End Latency in Federated Fog Systems}
\label{sec:latency}
When a task request arrives at a fog system's load balancer, communication and computational latencies combine to generate the end-to-end latency. Furthermore, several factors impact each of these latencies, causing them to behave stochastically. For these reasons, calculating end-to-end latency and capturing its stochastic character in fog computing systems is difficult. In the following sections, we go over the elements that influence communication and processing latencies. In addition, we present a model for estimating end-to-end latency while accounting for its stochastic character.

\subsection{\textit{Estimating Communication Latency}}~\\
The time it takes to process and return a response to a task request is the communication latency. More specifically, communication latency is caused by \emph{transmission latency} and \emph{propagation latency}. The transmission latency between any two points $m$ and $n$ (\eg two fog systems in the fog federation) for task $t$ of type $i$, denoted $\Theta_i(m,n)$, is defined as the sum of uplink transmission latency, denoted $\tau_u (m,n,i)$, and downlink transmission latency, denoted $\tau_d (m,n,i)$. That is,  we have $\Theta_i(m,n)  = \tau_u (m,n,i) + \tau_d (m,n,i)$.
Let $I_u(i)$ be the size of data payload (in bits), originally captured by a sensor, serving as input for task type $i$. Note that, for some sensors (\eg cameras), there can be randomness in the size of captured data, in every sensor reading. Also, let $R_u(m,n)$ represent the uplink bandwidth, through which the data is transmitted. $T$ is the time required to transmit each data packet to the uplink channel (known as Transmission Time Intervals (TTI)). Then, the uplink latency is calculated based on Equation~\ref{eq:2}.
\vspace{-2mm}
\begin{equation}\label{eq:2}
\tau_u (m,n,i)  = \lceil {\frac{I_u(i)}{R_u(m,n)\cdotp T}} \rceil
\end{equation}
Similarly, the downlink latency is defined as Equation~\ref{eq:3}. 

\begin{equation}\label{eq:3}
\tau_d (m,n,i)  = \lceil {\frac{I_d(i)}{R_d(m,n)\cdotp T}} \rceil
\end{equation}
An orthogonal frequency-division multiplexing (OFDM) with total bandwidth $W$ is divided equally into a set of $k$ sub-channels (where $k \in K$) each with bandwidth $w$. Accordingly, the downlink bandwidth is defined based on Equation~\ref{eq:4}.
\begin{equation}\label{eq:4}
R_d(m,n) = w \cdotp \sum_{k \in K} y_{mnk}\log_2(1+\gamma_d(m,n,k))
\end{equation}
where $y_{mnk}$ = 1, if sub-channel $k$ is allocated, otherwise $y_{mnk}$ = 0. As the wireless communication is prone to noise and interference from other fog systems in the federation, the value of $R_d(m,n)$ also depends on downlink \emph{signal to noise plus interference ratio} (also known as $SINR$~\cite{mukherjee2012distribution}). SINR is defined as the power of a particular signal divided by the sum of the interference power (from all the other interfering signals) along with the power of background noise. We note that, details of calculating uplink transmission latency ($\tau_u (m,n,i)$) is similar to those for downlink.

In fog federation, due to the vicinity, the propagation latency between fog systems is negligible. In contrast, the communication between fog systems and cloud datacenters is commonly achieved via satellite that introduces a substantial propagation latency \cite{sof13W}. The propagation latency, denoted $\tau_p$, is calculated based on Equation~\ref{eq:7}.
\vspace{-1mm}
\begin{equation}\label{eq:7}
\tau_p = 2 \cdotp \frac{d(n,st)}{S_l} 
\end{equation}
In the Equation~\ref{eq:7}, $d(n,st)$ is the distance between fog $n$ to satellite $st$ and $S_l$ is the propagation speed in medium or link. To calculate propagation latency in the round trip time, the fraction value should be doubled. Once we know propagation latency, the overall communication latency, denoted $d_{comm}$, to access cloud datacenter is calculated based on Equation~\ref{eq:8}.
\begin{equation}\label{eq:8}
d_{comm} =  \Theta_i(m,n) + \tau_p 
\end{equation}
As we noticed, there are several factors that collectively form the communication latency with stochastic behavior. To capture this stochastic behavior, we treat communication latency as a random variable and model it using statistical distribution. That is, we represent the communication latency between any two points (\eg two fog systems in the federation) using a probability density function (PDF), built upon historical communication information~\cite{salehi2016stochastic}. Based on the central limit theorem, communication latency can be modeled using Normal distribution.

\subsection{\textit{Estimating Computational Latency}}~\\
Once the load balancer assigns arriving task request $t$ to a fog system, the task has to wait in the scheduling queues of the fog system before its execution. For a given task $t$ of type $i$, denoted $t_i$, its completion time (\ie computational latency) is influenced by the waiting time in the queue (queuing latency), plus the task's execution time (execution latency) on the machines of the assigned fog system. Importantly, both of these factors are stochastic, as a result, the task completion time exhibits a stochastic behavior. 

The queuing latency of task $t_i$ is dependent on the number and execution times of tasks ahead of it in the fog system. The stochasticity in execution time can be due to different task types and characteristics of machines in different fog systems. Even the execution time of tasks from the same type on homogeneous machines of the same fog system is stochastic. This can be because of variations in the size of data to be processed and multi-tenancy of tasks in the fog system~\cite{li2016cvss}. Other factors, such as machine failure, can also be reasons for stochastic task execution time.

To capture the stochasticity in computational latency, we consider the task completion time of each task type on each fog system as a random variable. Then, we model the computational latency using statistical distribution. That is, the computational latency is modeled using PDF, built upon historical completion time information of each task type on each fog system. Based on the central limit theorem, the computational latency of each task type on each fog system can be modeled using Normal distribution.

\subsection{\textit{Estimating End-to-End Latency}}~\\
Once we estimate the communication and computational latencies, their compound latency forms the end-to-end latency. More specifically, the compound latency can be obtained by convolving the PDF of communication latency with the PDF of the computational latency. For an arriving task $t_i$ to a load balancer, let $N_i$ be PDF of its communication latency to another fog system in the federation. Also, let $M_i$ be PDF of the computational latency of $t_i$ on the other fog system. Then, the end-to-end latency for $t_i$, denoted $E_i$, is calculated as $E_i=N_i\circledast M_i$.

\section{Robust Resource Allocation in the Federated
Fog Computing System}
\label{sec:resourceAllocation}
\begin{figure}[!h]
    \centering
    \includegraphics[width=0.8\textwidth]{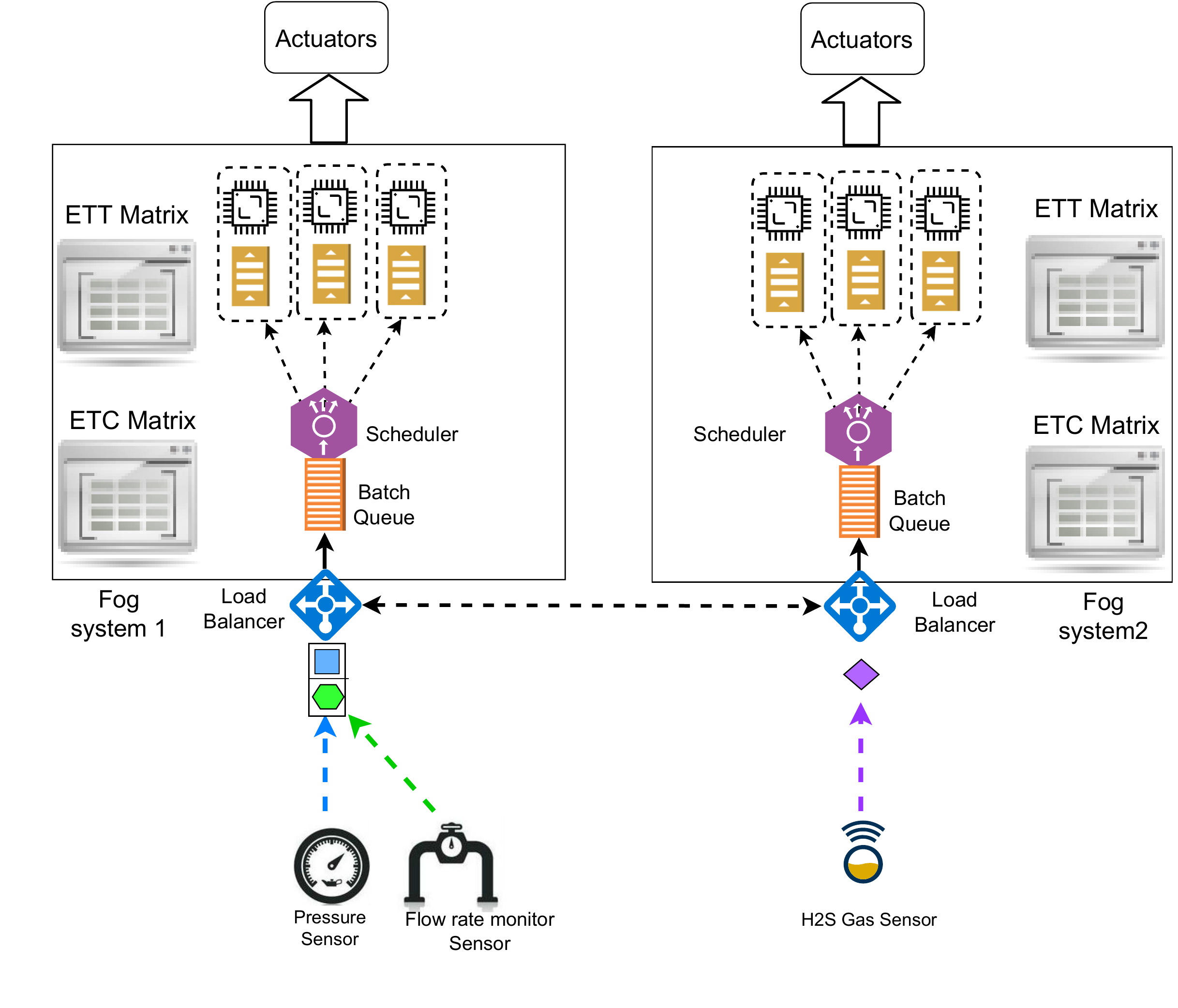}
    \caption{ A Fog system with load balancer module that facilitates fog federation. Task requests generated by sensors are received by the load balancer module and are assigned to the fog system that maximizes the likelihood of success for the task.}
    \label{fig:loadBalancer}
\end{figure}

The synopsis of the proposed resource allocation model in the federated fog computing system is demonstrated in Figure \ref{fig:loadBalancer}. The resource allocation model utilizes a \emph{load balancer} module that is the main enabler of fog federation. Every fog system is equipped with a load balancer that, for each arriving task, it determines the appropriate fog system (either the receiving fog or to a neighboring one) where the task has the highest likelihood of completion before its deadline.

The functionality of load balancer is particularly prominent to cope with the uncertainty exists in task arrivals (\eg during disaster time) and make the fog system robust against it. The load balancer operates in \emph{immediate mode}~\cite{ipdpsw19} and assigns arriving tasks to the appropriate fog system, immediately upon task arrival. The appropriateness is characterized based on the fog system that maximizes the probability of the task meeting its deadline (known as the \emph{probability of success}). The probability of success for task $t_i$ with deadline $\delta_i$ can be calculated for each neighboring fog system, by leveraging the end-to-end latency distribution of executing task $t_i$ on that system. To avoid repetitive task reassignment and compound latency, we determine that once a task assignment decision is made, the task cannot be re-allocated. 

The resource allocation of each fog system leverages the historical information of computational and communication latencies to build PDF of their distributions. For that purpose, each load balancer maintains two matrices, namely Estimated Task Completion (ETC)~\cite{diaz2011energy} and Estimated Task Transfer (ETT), to keep track of computational and communication latencies for each task type on each neighboring fog system. Entry $ETC(i,j)$ keeps the PDF of computational latency for task type $i$ on fog system $j$. Similarly, entry $ETT(i,j)$ keeps the PDF of communication latency for task type $i$ to reach fog system $j$. The entries of ETC and ETT matrices are periodically updated in an offline manner and they do not interfere with the real-time operation of the load balancer. 

Upon arriving task $t_i$, load balancer of the receiving fog can calculate the end-to-end latency distribution of $t_i$ on any neighboring fog $j$, using $ETC(i,j)$ and $ETT(i,j)$. The end-to-end distribution can be used to obtain the probability of completing $t_i$ before its deadline, denoted $p_j(t_i)$, on any of those fog systems. We have: $p_j(t_i)=\mathbb P(E_i\leq \delta_i)$. We note that the probability calculation for task $t_i$ on the receiving fog does not imply further communication latency. As such, for the receiving fog $r$ we have: $p_r(t_i)=\mathbb P(M_i\leq \delta_i)$. In the next step, the fog system that provides the highest probability of success is chosen as a suitable destination to assign task $t_i$. This implies that task $t_i$ is assigned to a neighboring fog system, only if even after considering the communication latency, the neighboring fog provides a higher probability of success.

It is noteworthy that the probability of success on a neighboring fog can be higher than the receiving fog by a non-significant amount. In practice, a task should be assigned to a neighboring fog, only if the neighboring fog system offers a substantially higher probability of success. To understand if the difference between the probabilities is substantial, we leverage confidence intervals (CI) of the underlying end-to-end distributions, from which the probability of success for receiving and remote fogs are calculated. More specifically, we determine a neighboring fog offers a significantly higher probability of success for a given task, only if CI of end-to-end distribution of the neighboring fog does not overlap with the CI of end-to-end distribution of the receiving fog.

\begin{algorithm}[!h]
	\SetAlgoLined\DontPrintSemicolon
	\SetKwInOut{Input}{Input}
	\SetKwInOut{Output}{Output}
	\SetKwFunction{algo}{algo}
	\SetKwFunction{proc}{Procedure}{}{}
	\SetKwFunction{main}{\textbf{TaskAssignment}}
	\Input{Task $t_i$; $ETC$ and $ETT$ matrices; $G$ (set of neighboring fog systems)}
	\Output{Chosen fog $j\in G$ to assign $t_i$}
	$p_r(t_i) \gets$ Probability of success on receiving fog $r$ \;
	\ForEach{fog system $j \in G$} {
	  $p_j(t_i) \gets$ Probability of success on neighbor fog $j$ \;
          \If {$p_j(t_i) > p_r(t_i)$} {	
				      \small{Add $p_j(t_i)$ to $P$, as a potential fog for assignment}\;
       	        	        }
       	}        	        
	Sort elements of set $P$ in descending order\;
        Consider receiving fog $r$ as default assignment for $t_i$ \;
	\ForEach{$p_j \in P$} {
	\If {CI of $E_j$ does not overlap with CI of $N_r$} {	
				      Choose fog $j$ as destination and assign $t_i$ to it\;
				      Exit the loop\;   	
       	        	        }
       	}        	        
	
\caption{Task assignment algorithm for load balancer.}\label{alg:ci}
\end{algorithm}

The pseudo-code provided in Algorithm~\ref{alg:ci} expresses the robust task assignment heuristic that load balancer utilizes to take advantage of federated fog system and increase the robustness of the system. The heuristic is called \emph{Maximum Robustness (MR)} and invoked upon arrival of a new task $t_i$ to the load balancer of a fog system. Based on the deadline of the arriving task ($\delta_i$), the algorithm first calculates the probability of success for $t_i$ on the receiving fog and on its neighboring fog systems (Step 1-7 in Algorithm~\ref{alg:ci}). Then, in Step 8, the calculated probabilities are sorted in the descending order. If the probability of success on the receiving fog is higher, then the task is allocated to the receiving fog system (Step 9). Otherwise, CI of the end-to-end latency distribution for the neighbor with the highest probability of success is compared against receiving fog CI. If the CIs do not overlap, then task $t_i$ is assigned to the neighboring fog (Step 12). Otherwise, the same procedure is performed for the rest of the neighbors of the receiving fog system. If there is no no-overlap neighbor found then, task $t_i$ is assigned to the receiving fog system (default assignment in Step 9).

\section{Performance Evaluation of Federated Fog}
\label{sec:evalLoadBal}
We have used EdgeCloudSim \cite{sonmez2017edgecloudsim}, which is a discrete event simulator for performance evaluation. We simulate five fog systems (micro-datacenters) each one with eight cores and [1500, 2500] Million Instructions Per seconds (MIPs) computational capacity. Cores of each fog system are homogeneous: however, different fog systems have different MIPs that represents the heterogeneity across the fog systems. We also consider a cloud datacenter with 40,000 MIPs to process non-urgent tasks. Task within each fog is mapped in the first come first serve manner. The bandwidth to access cloud is based on satellite communication and set to 200 Mbps, and the propagation delay is 0.57 seconds \cite{skedsmo2016oil}. 

In each workload trial, generated to simulate load of a smart oil field, we consider half of the tasks represent urgent and the other half represent non-urgent tasks. Each task is of a certain type that represents its service type. In each workload trial, urgent tasks are instantiated from two different task types and 
non-urgent tasks are instantiated from two other task types. 
The execution time of each task instantiated from a certain type is sampled from a normal distribution, representing that particular task type. Each task is considered to be sequential (requires one core) and its execution time is simulated in the form of MIPs. Poisson distribution (with different means for different task types) is used to generate the inter-arrival rate of the tasks and simulate task arrival during oversubscription periods. The number of tasks in each workload trial is varied to represent different oversubscription levels.

Deadline for task $i$ in a workload trial is generated as: $\delta_{i}$ = $arr_{i}+\beta\cdotp avg_{comp}^i+\alpha\cdotp avg_{comm}^i+\epsilon$, where $arr_{i}$ is the task arrival time, $avg_{comp}^i$ is average computational latency of the task type across fog systems, and $avg_{comm}^i$ is average communication latency. $\beta$ and $\alpha$ are coefficients, respectively, represent computation and communication uncertainties, and $\epsilon$ is the slack of other uncertainties exist in the system. We consider maintaining ETC and ETT matrices in every fog system and update them in every 10\% of the workload execution. The entries of these matrices are considered as normal distribution as mentioned in the system model. For accuracy, each experiment was conducted 30 times and the mean and 95\% confidence interval of the results are reported.

\subsection{\textit{Baseline Task Assignment Heuristics for Load Balancer}}~\\
\emph{Minimum Expected Completion Time (MECT):}
This heuristic~\cite{salehi2016stochastic} uses the ETC matrix to calculate the average expected completion time for the arriving task on each fog system and selects the fog system with the minimum expected completion time.

\emph{Maximum Computation Certainty (MCC):}
This heuristic (used in \cite{hussainrobust}) utilizes ETC matrix to calculate the difference between the task's deadline and average completion time (called certainty). Then, the task is assigned to the fog that offers the highest certainty. 

\emph{Edge Cloud (EC):}
This heuristic operates based on conventional fog computing model where no federation is recognized. 
Specifically, urgent tasks are assigned to the receiving fog and non-urgent tasks are assigned to the cloud datacenter.

\subsection{\textit{Experimental Results}}

\subsubsection{Analyzing the impact of oversubscription. }
\begin{figure}[!h]
    \centering
    \includegraphics[width=0.8\textwidth]{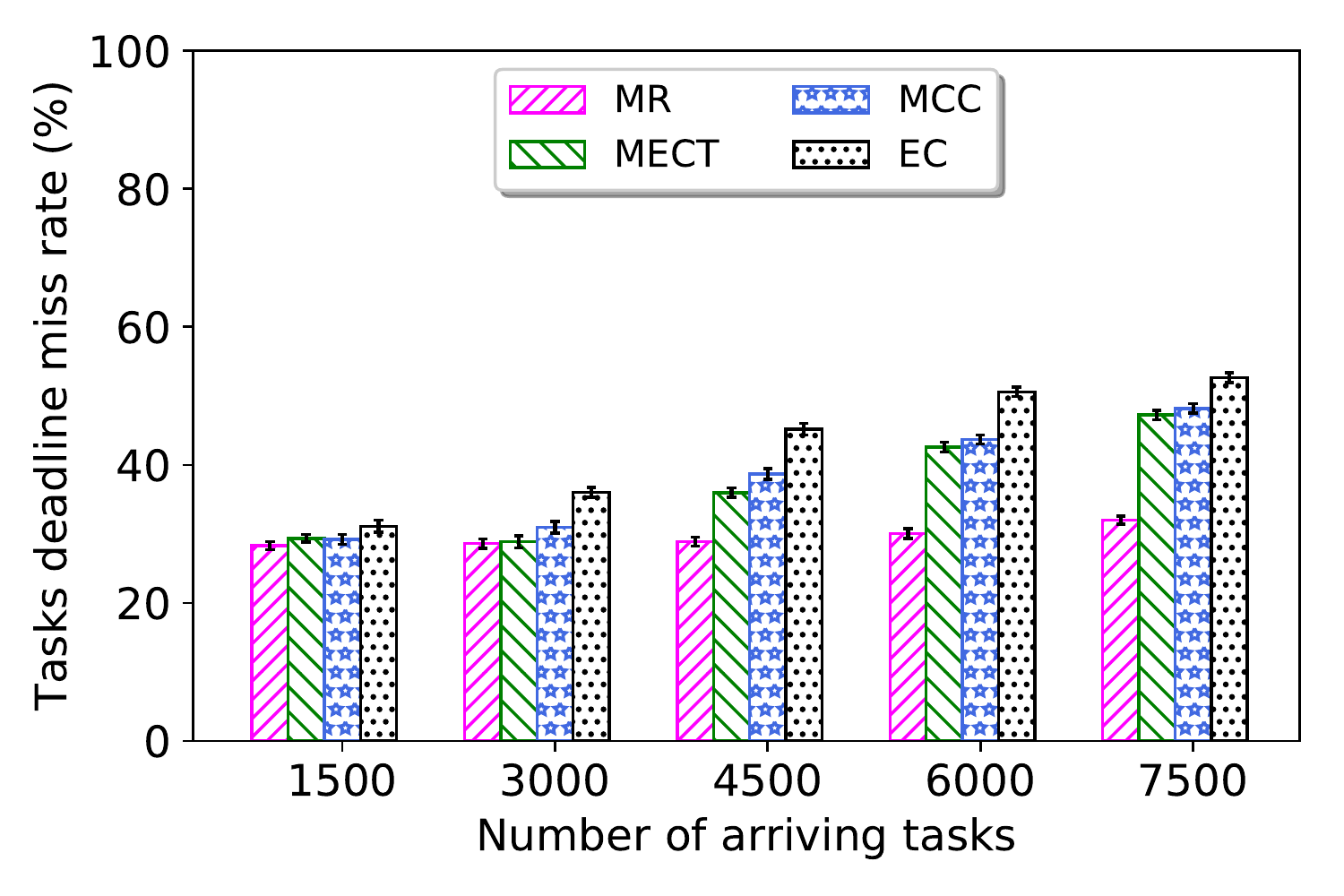}
    \caption{The impact of increasing oversubscription level (number of arriving tasks) on deadline miss rate using different task assignment heuristics in the load balancer.}
    \label{fig:resultDeadlineMissRate}
\end{figure}

The main metric to measure the robustness of an oversubscribed fog system in a smart oil field is the deadline miss rate of tasks. In this experiment, we study the performance of our system by increasing the number of tasks sensors generate (\ie oversubscription level). Figure~\ref{fig:resultDeadlineMissRate} shows the results of varying the number of arriving tasks (from 1,500 to 7,500 in the horizontal axis) on deadline miss rate (vertical axis) when different task assignment heuristics is applied.

In Figure~\ref{fig:resultDeadlineMissRate}, it is visible that as the number of tasks increases, the deadline miss rate grows for all of the heuristics. Under low oversubscription level (1,500 tasks), MR, MECT, and MCC perform similarly. However, as the system gets more oversubscribed (4,500 tasks) the difference becomes substantial. With 7,500 tasks, MR offers around 16\% lower deadline miss rate than MECT and MCC and approximately 21\% better than EC. The reason is that MR captures end-to-end latency and proactively utilizes federation, only if it has a remarkable impact on the probability of success. Nonetheless, EC does not consider federation, and other baseline heuristics only consider the computational latency. We can conclude that considering end-to-end latency and capturing its underlying uncertainties can remarkably improve the robustness, particularly, when the system is oversubscribed (\eg at a disaster time).

\subsubsection{Analyzing communication overhead of fog federation. } 
\begin{figure}[!h]
    \centering
    \includegraphics[width=0.8\textwidth]{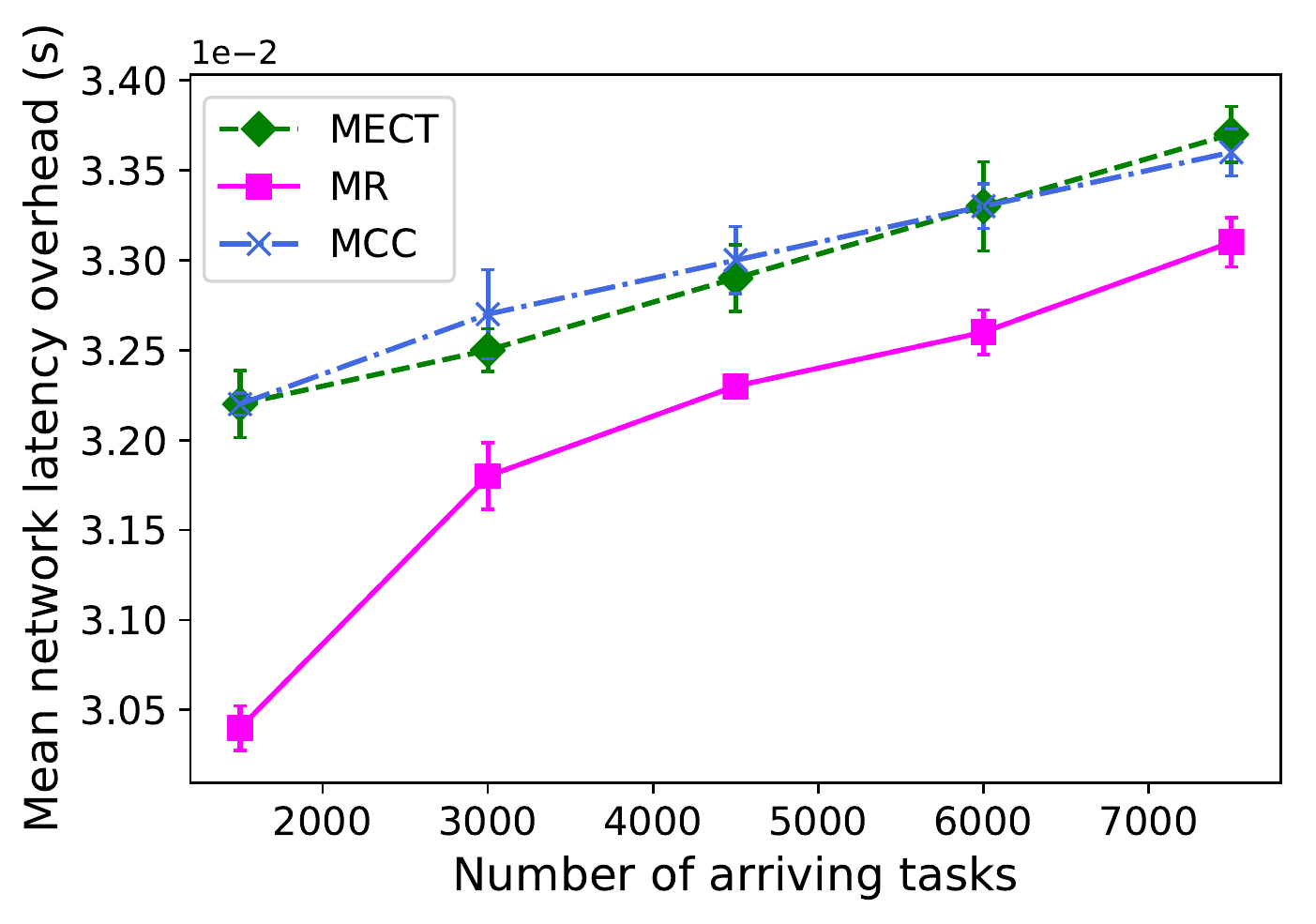}
    \caption{Mean communication latency overhead introduced to each task in fog federation by different heuristics.}
    \label{fig:resultComm}
\end{figure}

Although we showed in the previous experiment that using federation improves system robustness, we are unaware of the communication overhead of task assignment in the federated environment. Therefore, in this experiment, we evaluate the communication latency imposed as a result of applying different task assignment heuristics.
Specifically, we measure the mean communication latency overhead (vertical axis in Figure~\ref{fig:resultComm}) induced to each task, for the various number of arriving tasks (horizontal axis in Figure~\ref{fig:resultComm}).

Figure~\ref{fig:resultComm} shows that MECT and MCC cause higher average communication latency. The reason is that these heuristics do not consider the communication latency and aggressively redirect tasks to the same fog system, making that particular network link (between receiving fog and redirected fog system) congested. In contrast, MR that considers communication latency and redirect tasks more conservatively, only if the improvement in the probability of success is substantial. 

\subsubsection{Analyzing average makespan of tasks. }   
Different task assignment heuristics cause various computational latencies for the tasks. To understand the computational latency, we measure the average makespan of tasks, resulted by applying various task assignment heuristics.

\begin{figure}[!h]
    \centering
    \includegraphics[width=0.8\textwidth]{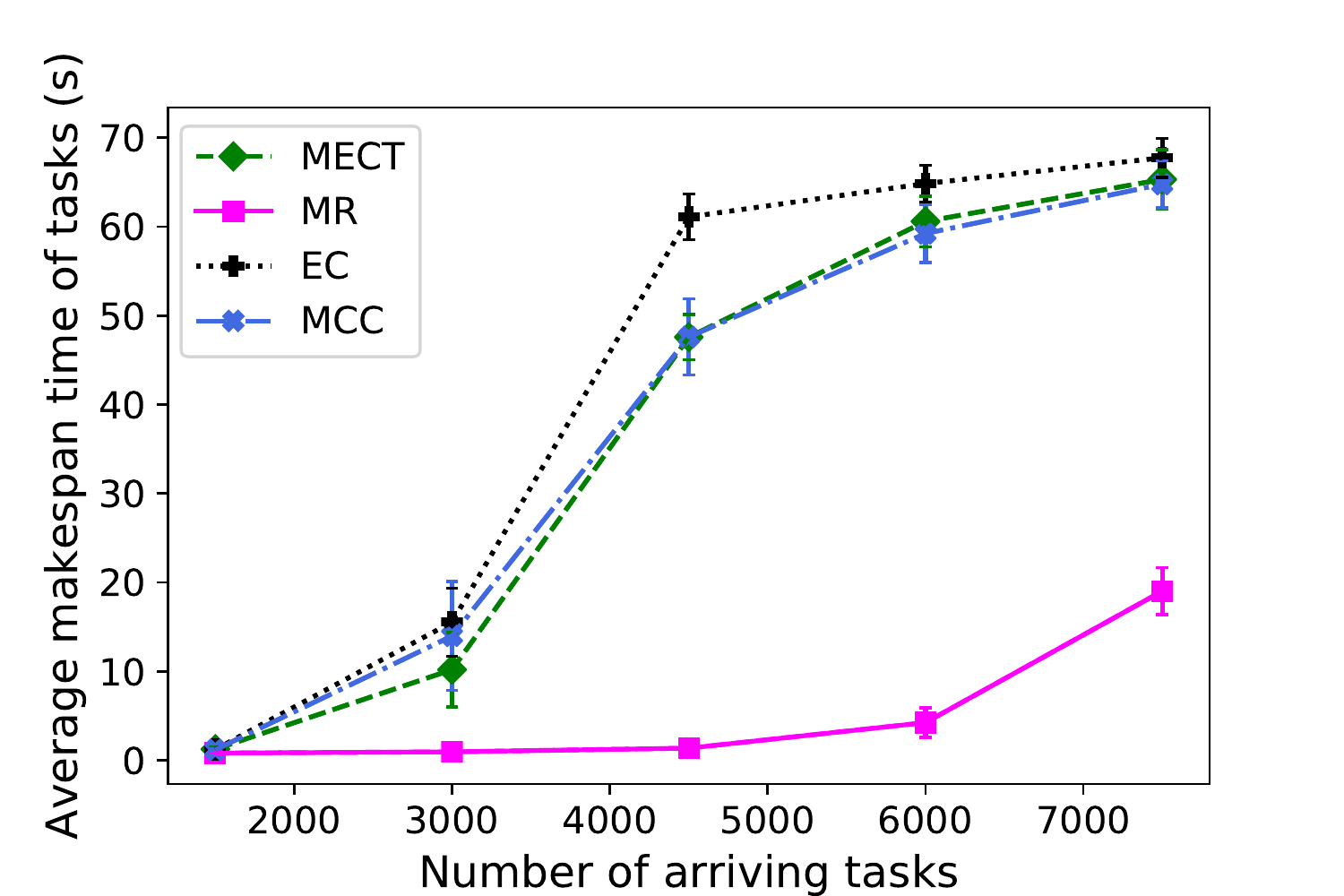}
    \caption{Average makespan time(seconds) of tasks using various task assignment heuristics.}
    \label{fig:resultComp}
\end{figure}

Figure~\ref{fig:resultComp} demonstrates that EC leads to the maximum average makespan time. The reason is that EC does not utilize federation, making the receiving fog system highly oversubscribed while other neighboring fog systems are underutilized. Hence, average makespan time rapidly rises after the receiving fog is saturated with 3,000 tasks. MECT and MCC do not consider the stochastic nature of task completion time; hence, they can potentially assign arriving tasks to one fog and oversubscribe that. As a result, the average makespan of tasks rises. In contrast, MR considers stochastic nature of end-to-end latency and calculates the probability of success on neighboring fog systems. Besides, it assigns tasks to a neighboring fog system, only if it offers a sufficiently higher probability of success.  Hence, MR offers the lowest average makespan time than other heuristics.

\section{Summary}
\label{C3Summary}
In this chapter, we explored the usability of a fog federation for a smart Industry (Oil and Gas) in a disastrous situation. To support the computational demands in an emergency situation allocating various tasks in suitable fog system is challenging due to heterogeneity across fog systems. Hence, maintaining the robustness of the system in terms of every real-time urgent tasks deadline can be difficult unless any efficient load balancing technique adopted by the system. To achieve that, we presented dynamic federation of fog computing systems, exist in nearby industries. Within the federated environment, we captured two sources of uncertainty, namely communication and computation, that are otherwise detrimental to the real-time services. The federation is achieved by a load-balancer module in each fog system that is aware of the end-to-end latency between fog systems and can capture the stochasticity in it. The load balancer leverages this awareness to find the fog system that can substantially improve the probability of success for each arriving task. Experimental results demonstrate that our proposed federated system can enhance the robustness of fog computing systems against uncertainties in arrival time, communication, and computational latencies. We concluded that the load balancer could be particularly useful (by up to 27\%) for higher levels of oversubscription. Even for na\"ive load balancing methods (MCC and MECT) in the federation, the performance improvement is approximately 13\%. 




    \chapter{Adapting Remote Industry 4.0 Smart Micro-Service Applications to  Federated Fog Computing Systems}\label{section:microservice}
\section{Overview}
The advancement of IoT technologies with smart applications drives the wheel of Industry 4.0 \cite{lu2019oil} revolution. Various smart sensors, actuators, and smart devices are deployed in different industries (\eg manufacturing, food processing, oil \& gas) to control the operational technology platform \cite{bramantyo2022data,titu2020merging}. Accordingly, sensors utilized in industrial operations frequently produce tons of data every day \cite{daily2017predictive}. The oil and gas industry is an example of generating enormous amounts of sensor data and the necessity for processing close to the data source. For instance, a typical offshore oil rig produces 1 to 2 TB of data daily \cite{sof2}. The majority of this data is fed to advanced computing applications (\eg machine learning, report generation, automation) that can make smart latency-sensitive decisions to improve energy efficiency, production, and safety measures. For example, applications like workplace air quality estimation \cite{kristiani2021implementation} for workers' safety utilize environmental sensors that measure the quality (\ie the existence of harmful particles in the air) of breathable air in the surrounding of the workplace. Hence, the air quality estimation must be fast to avoid potential occurrences. 


In the remote offshore industry, several services (\eg data acquisition, alert generation, object tracking) are critical for complex or safety-related operations that need to be performed synchronously. The situation can worsen when any unwanted emergency brings many more computational activities completed within limited time frames. In this case, our motivation is the smart Oil and Gas industry that has been facing various disasters and catastrophes (\eg the deepwater horizon (2010)\cite{srinivasan2010many}, usumacinta jack-up disaster (2007) \cite{hanlon2013usumacinta}, mumbai high north disaster (2005) \cite{daley2013mumbai}, the ocean ranger disaster (1982) \cite{heising1989ocean} ) due to complex fault intolerant industrial processes in exploration, drilling, and production operations. Therefore, remote offshore industries need latency-aware support \cite{deng2021intelligent} that can not be feasible with typical cloud data centers due to the remote locations of the industrial operation sites. The current solution utilizes fluctuating satellite communication \cite{cricelli2022market} for sending data to mainland cloud data centers reducing the quality of service (QoS) and increasing the industrial safety risk. Hence, the high-level challenge is the lack of computational resources to support over-subscribed situations in remote industries.

\begin{figure}[h!]
	\centering	
	\includegraphics[scale=0.85]{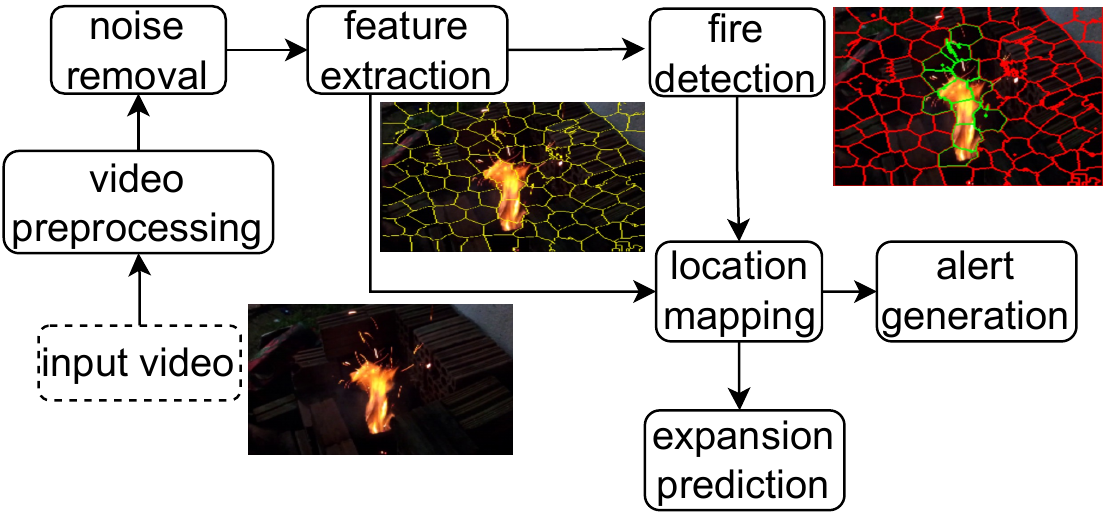}
	\caption{The structure of a microservice-based workflow is presented in a block diagram. Every microservice need to be processed to complete the fire safety application.\label{fig:workflowMicro}}
\end{figure}

\subsection{\textit{Smart Micro-Service Applications for Industry 4.0}}~\\
Industry 4.0 smart applications typically follow modern software architecture \cite{jwo2022data,aberle2022microservice} where various micro-services \cite{dragoni2017microservices} need to be executed in order. However, micro-services can be separately deployed using an automated deployment process, require the least amount of administration, can be developed using a variety of programming languages and data storage techniques, and can each be independently updated, changed, and scaled. Thus, we concentrate on micro-services applications frequently used in remote industries. For instance, as depicted in figure \ref{fig:workflowMicro}, a ``fire safety" application can include micro-services for capturing video surveillance data, pre-processing captured video, noise removal, feature extraction, fire detection, location mapping, alert generation, and expansion prediction. In contrast, many industries have previously deployed legacy applications \cite{calderon2018integration} with inflexible software architecture. Hence, the execution platform should support both monolithic legacy applications and modern micro-services to ensure industrial safety and fault-tolerant operations. However, modern industry 4.0 applications are comprised of micro-services that pose new challenges for the execution platforms. Under this arrangement, an application`s latency constraint is subject to the completion time of the micro-services defined by the underlying software architecture. Therefore, to develop a robust execution platform for industry 4.0, system architects need to understand the software architecture of the receiving applications. 

\subsection{\textit{Federated Fog Systems for Industry 4.0 Micro-service Applications}}~
\begin{figure}[h!]
	\centering	
	\includegraphics[scale=0.60]{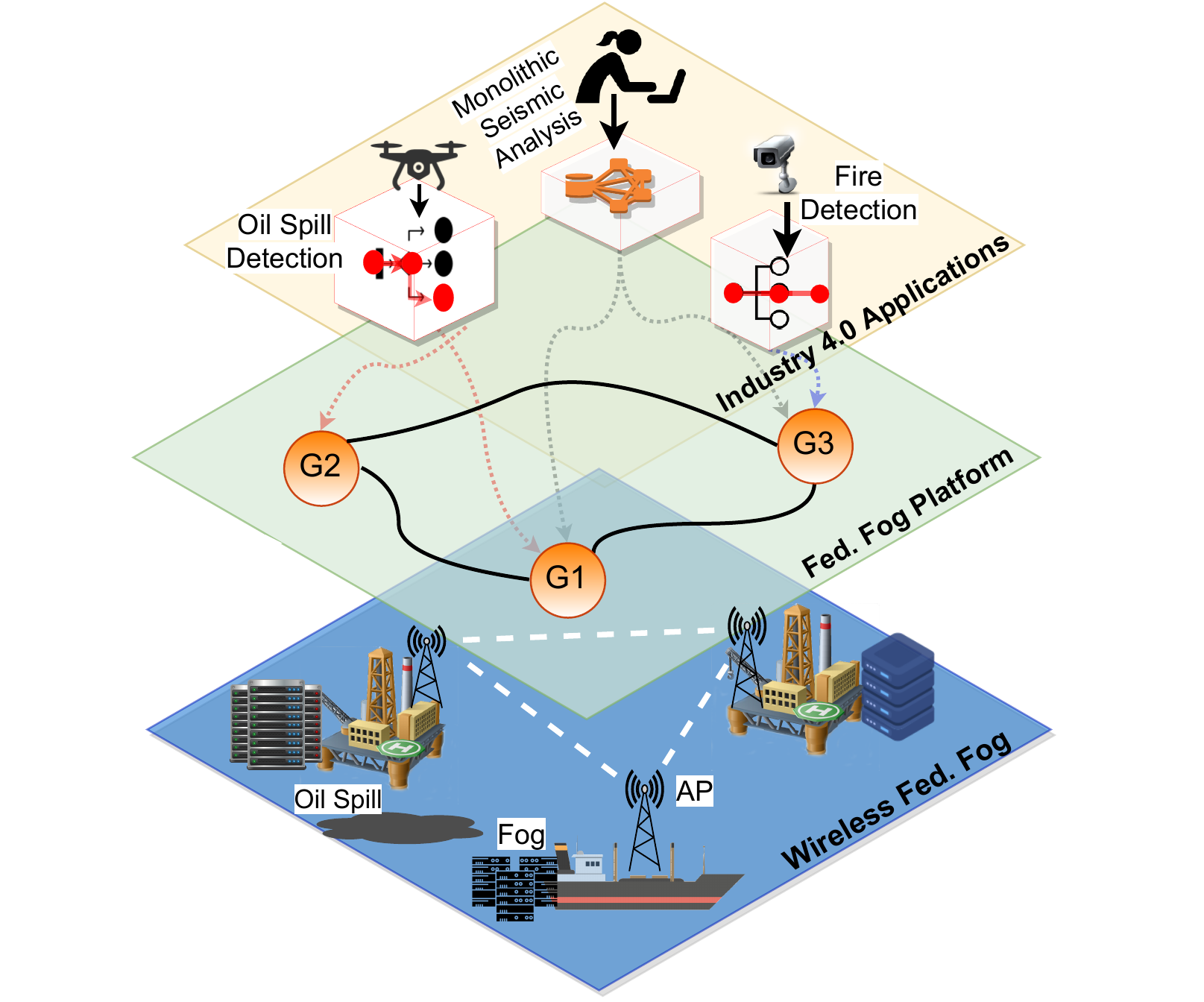}
	\caption{Offshore oil and gas industry has the fog federation infrastructure that can support smart microservice-based applications. \label{fig:layerFog}}
\end{figure}

The emerging industrial IoT and advancements in communication technology have brought computational resources near the data sources and end devices. For instance, nowadays, fog computing systems \cite{verba2019modeling} in remote industries typically execute the industrial computational process to enable smooth production and workplace safety. However, as depicted in figure \ref{fig:layerFog}, a federated fog platform can be conceptualized from chapter~\ref{section:loadBalance} that can form by connecting through wireless gateways denoted as $G_i$. Hence, various applications with heterogeneous latency constraints require computational support from federated fog computing systems. Accordingly, the federation should be cognizant of communications and computing uncertainties, as well as the applications' software structure and latency requirements. Thus, an application execution plan needs to perform for monolithic and micro-service software structures, considering the stochastic execution times and uncertainties that derive from the execution platform and communication technology.

Consequently, in our previous work \cite{hussain2019federated1} presented in chapter~\ref{section:loadBalance}, 
the resource allocation methods are explored intensively for monolithic applications. Furthermore, considering a complex operational process performed by various micro-services, one of the main problems is ensuring the completion of the whole application workflow within the time limit known as the deadline. Hence, it is crucial to know the optimal point to partition the application workflow so that it can be completed on time. Accordingly, the question that needs to be addressed is ``How to distribute Industry 4.0 applications (\eg monolithic, micro-service) across fog federation so that the application workflow can be completed within the given time frame?''. Hence, from a system administration perspective executing the smart micro-service applications raises two more questions, and they are  1) How to partition the micro-service workflows so that its deadline constraint can be realized? 2) How do we allocate partitioned micro-services across fog federation so that it has the highest likelihood of completing on time?

Our prior work \cite{hussain2019federated1} suggests that federating nearby resources is one solution to the resource restrictions (\ie oversubscribe) encountered by edge computing systems in distant sectors like Oil and Gas. Furthermore, we explore new challenges imposed by smart software architecture, a.k.a micro-service workflows. Therefore, to address the difficulties faced by the offshore O\&G sector at large, we propose a resource provisioning method for Industry 4.0 applications across the federated fog system that is aware of both the software architecture and the underlying execution platform's structure. More so, the solution maintains the deadline limitations of the micro-service workflow, which in turn makes the execution platform more reliable. As a result, our approach consists of two stages: understanding the software architecture of the receiving applications and allocating computational resources for the successful completion of these applications. Therefore, the following are the contribution of this research:
\begin{itemize}
    \item Proposing a probabilistic partitioning method that is aware of the underlying software architecture of Industry 4.0 applications.
    
    \item Proposing a statistical resource allocation heuristic considering the time constraints of the application. 
    
    \item Providing extensive evaluation of partitioning technique along with resource allocation across fog federation.
\end{itemize}

The suggested solution can serve as a foundation upon which system architects or industry-focused research associates might construct more elaborate solutions referring to distant offshore sectors at peak demand. In addition, the solution is compatible with monolithic legacy applications, which may aid conventional industries in transitioning to and adapting to the changes brought about by Industry 4.0.

\section{Partitioning Method for Micro-service Application Workflow}~
Maintaining latency constraints of a smart application comprising multiple micro-services depends on underlying software architecture and mapping of computational resources. For example, executing a micro-service application into a single fog system may not be possible or may not maintain its deadline constraint. On the other hand, a monolithic application can not be partitioned and can be considered an application with a single micro-service. For micro-service software architecture, partitioning the application into multiple partitions and allocating them across fog federation can increase the likelihood of its completion within the latency constraint. Furthermore, allocating appropriate computational resources to the partitioned micro-services also ensures the completion of the whole application workflow. 

\begin{figure}[h!]
	\centering	
	\includegraphics[scale=0.55]{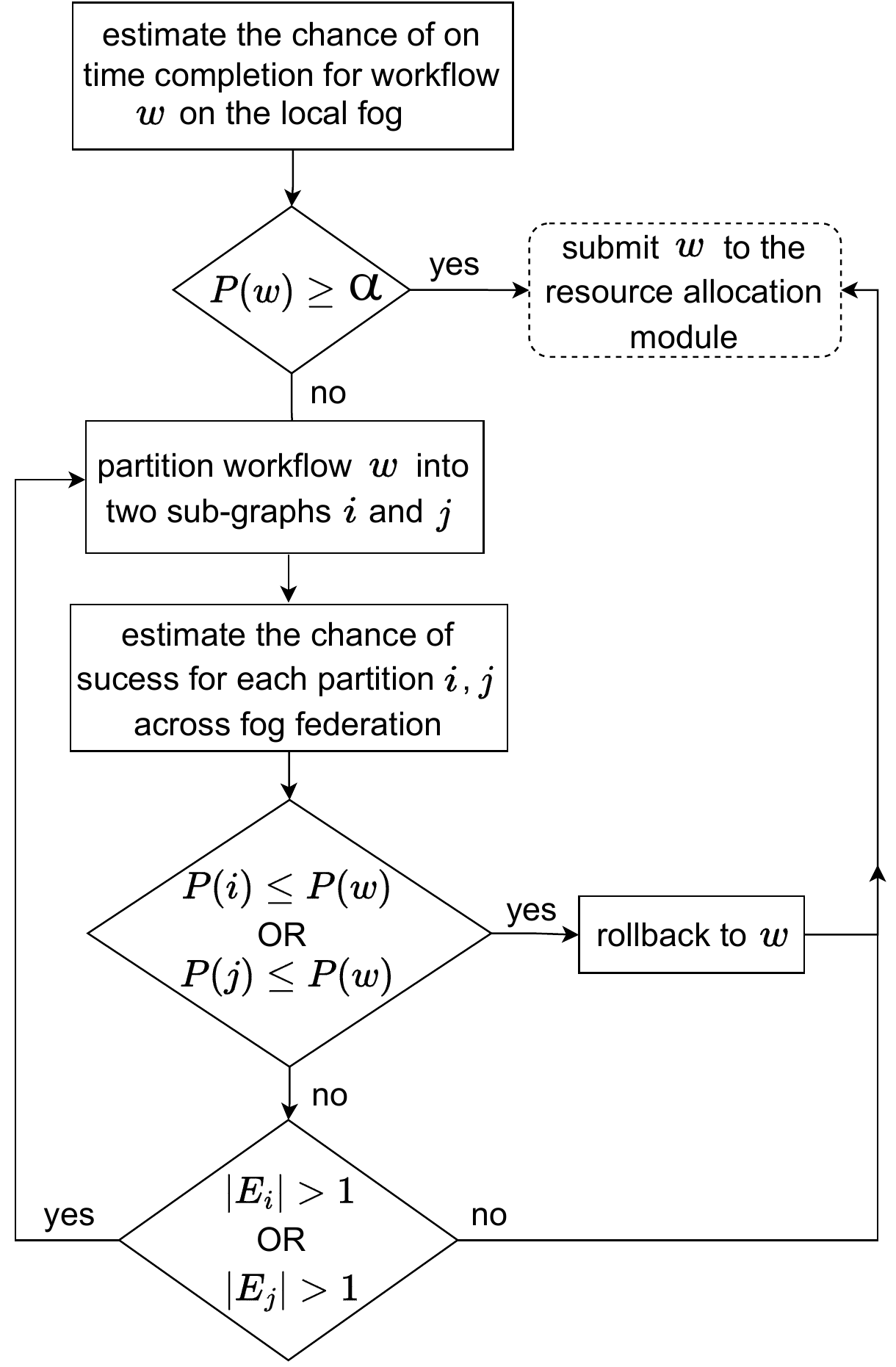}
	\caption{The flowchart of the workflow partitioning method. The partitioned workflow is sent to the resource allocation module, which is denoted as the end box for this flow chart.\label{fig:partitionTech}}
\end{figure}  

The main goal of the partitioning method is to partition the micro-service application in a way such that the application can meet its deadline. Hence, we considered an application having micro-service architecture as a set of micro-services that are connected together in some manner to form a graph $G = (V,E)$, where the set of vertices $V = (m_1,m_2,..m_n)$ denotes the micro-services and edge $e(m_i,m_j) \in E$ represents the communication between micro-service $m_i$ and $m_j$. As the first step of the partitioning method, we consider executing the whole micro-service workflow into the local fog system without partitioning. As such, the partitioning method estimates the chance of on-time completion for workflow $w$ on the local fog, which is the first processing box of flowchart~\ref{fig:partitionTech}. To estimate the deadline for the whole application workflow $w$, we perform a summation of the deadlines for the micro-services that can be defined as $ \delta_w = m_1^\delta + m_2^\delta + ...+m_n^\delta$. 
Hence, each micro-service has a deadline ($m_i^\delta$) known in advance to the load balancer. Furthermore, for each micro-service type, we have computational latency distribution ($m_i^d$) that represents the execution times across fog federation. Hence, to estimate the probability of success for the entire $w$, we convolve the computational latency distributions of the application's micro-services that can be defined as 
\begin{equation}\label{eq:convolve}
D_w = m_1^d \circledast m_2^d \circledast ....m_n^d    
\end{equation}
Finally, using the convolved distribution $D_A$, we measure the probability of success as follows, 
\begin{equation}\label{eq:prob}
P(w) =\mathbb P(D_A \leq \delta_A)
\end{equation}
The output of equation~\ref{eq:prob} is compared with a conditional variable $\alpha$ as depicted in first condition of flowchart~\ref{fig:partitionTech}. We choose an average success rate (\ie 50\%) for $\alpha$ as our experimental evaluation scenario. When the likelihood of completing the workflow is less than $\alpha$, the partitioning service takes place using the min-cut \cite{lakhan2022deep} graph partitioning algorithm, which is the \textit{partition workflow $w$ into two sub-graphs $i$ and $j$} process box in figure \ref{fig:partitionTech}. Hence, considering the flow of actions within the application, we employ one of the widely utilized graph theorems, max-flow min-cute \cite{lochbihler2022mechanized} in our proposed solution.  

Due to finding the minimum number of partitions which is an np-hard problem, we developed our customized solution for Industry 4.0 micro-service applications. Thus, the partitions resulting from the min-cut are estimated for the chance of success across fog federation using equation~\ref{eq:convolve} and \ref{eq:prob} in the third process box of flowchart~\ref{fig:partitionTech}. As we utilize probability to determine the partitions, we named the proposed partitioning method as \textit{Probabilistic Paritioning} (ProPart). If the new sub-graphs completion success is less than the prior success rate (2nd condition of the flowchart), we consider earlier partitions as optimal (rollback to $w$ process box). Accordingly, the resource allocation methods for those partitions are started, which is the end process box of the flowchart.

On the other hand, if the latest partition's chance of on-time completion is greater than the prior success rate, then we evaluate each partition's micro-service architecture, which is the third condition of the flowchart. If the condition fails (``no" line from the third condition), the partitioning process is halted for partitions with only one micro-service, and the resource allocation service takes place. In contrast, for partitions with more than one micro-service (``yes" line from the third condition), the partitioning process is repeated until each has a single micro-service. Therefore, the partitioning method is a repeated process where the output is the optimal number of partitions that are submitted to the resource allocation module.

\section{Resource Allocation Method for Partitioned Micro-service Applications Across Fog Federation}
Resource allocation occurs when the partitioning is completed with an optimum number of partitions. The partitioning method returns the whole application as one part to the resource allocation module for a monolithic application that is considered a single micro-service workflow. The efficacy of the resource allocation approach is significant in dealing with the unpredictability that occurs in applications' arrival (\eg during disaster time) and making the fog system resilient. The resource allocation module runs in \emph{immediate mode}~\cite{ipdpsw19} and quickly allocates incoming applications or micro-service partitions to the relevant fog system. The relevance is defined by the fog system, which increases the likelihood of the micro-services achieving their deadlines (a.k.a the \emph{probability of success}). Hence, the likelihood of on-time completion for a micro-service $m_i$ on a particular fog system can be estimated using the historical end-to-end latency distribution. Furthermore, to minimize frequent application reassignment and compound delay, we have decided that the micro-service cannot be relocated once an assignment choice is made.

Each fog system's resource allocation module uses prior data on computational and communication latencies of various micro-service types across fog federation to generate PDFs of their distributions. To that end, each load balancer keeps track of computational and communication latencies for each micro-service type on each nearby fog using two matrices: Estimated Task Completion (ETC) \cite{diaz2011energy} and Estimated Task Transfer (ETT). The PDF of computational delay for micro-service type $i$ on fog system $j$ is stored in entry $ETC(i,j)$ that is previously used in the partitioning method. Similarly, the item $ETT(i,j)$ maintains the PDF of communication delay for micro-service type $i$ to reach fog system $j$. Hence, the resource allocation module is aware of communication latencies as well, whereas the partitioning method is only aware of computation latencies. The entries of the ETC and ETT matrices are regularly updated offline and do not interfere with the load balancer's real-time functionality.

The resource allocation module can compute the end-to-end latency distribution across fog federation upon the arrival of a partition of micro-services using convolution of $ETC(i,j)$ and $ETT(i,j)$. On any fog system $j$, the end-to-end distribution can be used to calculate the probability of completing each micro-service partition ${mp}_i$ before its deadline, denoted $p_j(mp_i)$. Hence, we estimate the deadline $\delta_i$ for the given partition $mp_i$ by adding each micro-service's deadline within that partition. Then we convolve each micro-service's computational latency distribution $d_{comp}$ with communication distribution $d_{comm}$ to measure the completion time $e_i$ in a particular fog system. To estimate the completion time of the partition $mp_i$ denoted as $E_i$, we convolve the completion time distribution for each micro-service within a given partition. We have: $p_j(mp_i)=\mathbb P(E_i\leq \delta_i)$. We see that the probability of $mp_i$ on the receiving fog does not entail any additional communication delay. Consequently, for receiving fog system, we don't convolve communication latency distribution to completion time estimation. In the subsequent stage, the fog system with the greatest likelihood of completion is selected as a viable destination to allocate $mp_i$. This assignment entails that the micro-service partition $mp_i$ is only given to an adjacent fog system if the surrounding fog offers a greater chance of on-time completion after accounting for communication delay.

It`s important to note that the success rate on a neighboring fog could be greater than on the receiving fog. This is because assigning a micro-service partition to a fog system in close proximity should only be done if doing so significantly increases the likelihood of the partition being completed successfully. Hence, we use confidence intervals (CI) of the underlying end-to-end distributions, from which we derive the likelihood of success for receiving and distant fogs, to assess the significance of the discrepancy. In particular, we find that the CI of the end-to-end distribution of the nearby fog does not overlap with the CI of the receiving fog only if the neighboring fog gives a much better likelihood of success for a given micro-service partition.

\begin{algorithm}[!h]
	\SetAlgoLined\DontPrintSemicolon
	\SetKwInOut{Input}{Input}
	\SetKwInOut{Output}{Output}
	\SetKwFunction{algo}{algo}
	\SetKwFunction{proc}{Procedure}{}{}
	\SetKwFunction{main}{\textbf{TaskAssignment}}
	\Input{Micro-service partition set $M$; $ETC$ and $ETT$ matrices; $G$ (set of neighboring fog systems)}
	\Output{Chosen fog $f\in G$ to assign micro-service partitions $mp_n \in M$}
        \ForEach{micro-service partition $mp_i \in M$}{
	$p_r(mp_i) \gets$ Probability of success on receiving fog $r$ \;
	\ForEach{fog system $f \in G$} {
	  $p_f(mp_i) \gets$ Probability of success on neighbor fog $f$ \;
          \If {$p_f(mp_i) > p_r(mp_i)$} {	
				      \small{Add $p_f(mp_i)$ to $P$, as a potential fog for assignment}\;
       	        	        }
       	}        	        
	Sort elements of set $P$ in descending order\;
        Consider receiving fog $r$ as default assignment for $mp_i$ \;
	\ForEach{$p_f \in P$} {
	\If {CI of $E_f$ does not overlap with CI of $N_r$} {	
				      Choose fog $f$ as destination and assign $mp_i$ to it\;
				      Exit the loop\;   	
       	        	        }
       	}        	        
	}
\caption{Resource allocation algorithm}\label{alg:confidence}
\end{algorithm}

The pseudo-code provided in Algorithm~\ref{alg:confidence} expresses the resource allocation method that the load balancer utilizes to take advantage of the federated fog system and increase the system's robustness. The method is called \emph{Maximum Robustness (MR)} and is invoked when the partitioning method sends micro-service partitions $M$ for resource allocation. At first, the micro-service partitions are separated for further processing. Then based on the deadline ($\delta_i$) of each micro-service partition $mp_i$, the algorithm calculates the probability of success on the receiving fog and on its neighboring fog systems (Step 2-8 in Algorithm~\ref{alg:confidence}). Next, step 9 sorts the calculated probabilities in descending order. If the probability of success on the receiving fog is higher, then the micro-service partition $mp_i$ is considered for allocation to the receiving fog system (Step 10). Otherwise, the CI of the end-to-end latency distribution for the neighbor with the highest probability of success is compared against receiving fog's CI. If the CIs do not overlap, then partition $mp_i$ is assigned to the neighboring fog (Step 13). Otherwise, the same procedure is performed for the rest of the neighbors of the receiving fog system. If no non-overlap neighbor is found, then partition $mp_i$ is assigned to the receiving fog system (default assignment in Step 10).

\section{Performance Evaluation of Software Architecture-Aware Federated Fog Systems}
The partitioning and resource allocation components of the proposed technique occur one after the other within the load balancer module. As a consequence, we evaluate each component separately in various experiments. The recommended partitioning approach is compared to different baselines in the first experiment to examine how the deadline constraints for workflow applications based on microservices have improved. Following partitioning, the allocation of resources to those partitions must be evaluated. We execute the second category of trials to assess the system's efficacy, which compares our proposed resource allocation techniques to alternative baselines. The third experiment is then performed to determine how scaling the fog federation impacts the suggested solution. Finally, for microservice and monolithic applications, we examine the computational latencies resulting from partitioning and resource allocation approaches. The experiments are thoroughly described in the subsections that follow.

\subsection{\textit{Comparison of Micro-service Workflow Partitioning Methods}}
\begin{figure}[h!]
	\centering	
	\includegraphics[scale=0.70]{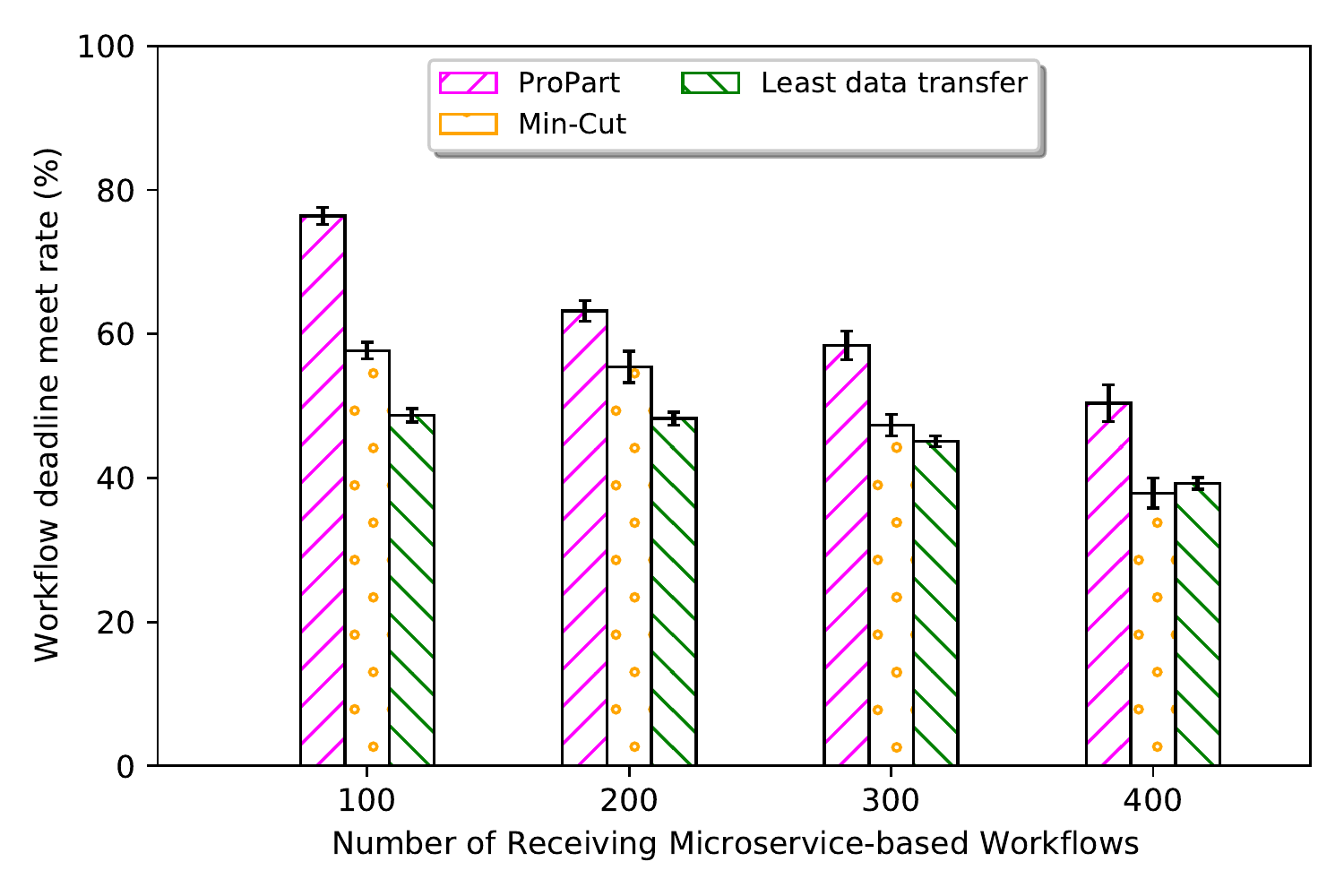}
	\caption{Comparison of the partitioning techniques in terms of workflow deadline meet rate while utilizing proposed probabilistic partitioning technique. The x-axis represents the increasing number of arriving workflow execution requests, whereas the y-axis represents the workflow deadline meet rate.\label{fig:partitionComparison}}
\end{figure}

~\\
In this experiment, we use the suggested partitioning technique (Probabilistic Partitioning, defined as ProPart) for accepting microservice-based workflow applications and compare it to the other two baselines (Min-cut, Least data transfer, for example). In this experiment, we increase the number of microservice applications submitted to the system to generate oversubscribed conditions and record the applications' deadline meet rate in each round of request submission, shown as a bar chart in figure \ref{fig:partitionComparison}. The figure's x-axis indicates the number of microservice-based applications received by the system, while the y-axis reflects the rate at which application deadlines have been met.

The results of this experiment, shown in figure \ref{fig:partitionComparison}, indicate that the deadline meet rate decreases as the number of workflow requests to the system increases for all partitioning techniques. However, in every round of submissions, ProPart surpasses other baselines. For less overloaded scenarios (\eg 100 \& 200 requests), the performance gap between the least efficient strategy (least data transfer) and the suggested technique ProPart is greater than for completely oversubscribed conditions (\eg 300 \& 400 requests). The primary reason for ProPart's superior performance is its statistical assessment of each partition's success likelihood. For up to 200 application requests, the min-cut strategy performed better than the least data transfer. In contrast, in totally overloaded scenarios, the least data transfer performed marginally better than the min-cut because it considers the connection that generates the least output data for splitting. Min-cut, in contrast, examines the smallest communication channel when partitioning. Finally, due to the repeated probabilistic calculation of deadline fulfillment for all microservices, ProPart performed better in totally oversubscribed conditions.

\subsection{\textit{Comparison of Resource Allocation Methods}}~\\
The load balancer in every fog system utilizes a resource allocation technique after the partitioning steps for microservice-based workflow applications. In contrast, for monolithic applications, as soon as load balancer receives a request, it performs resource allocation using probabilistic estimation across fog federation. As such, to compare the proposed resource allocation technique, we performed the following experiments with three different resource allocation methods for microservice and monolithic applications respectively.

\begin{figure}[h!]
    \centering
    \includegraphics[scale=0.65]{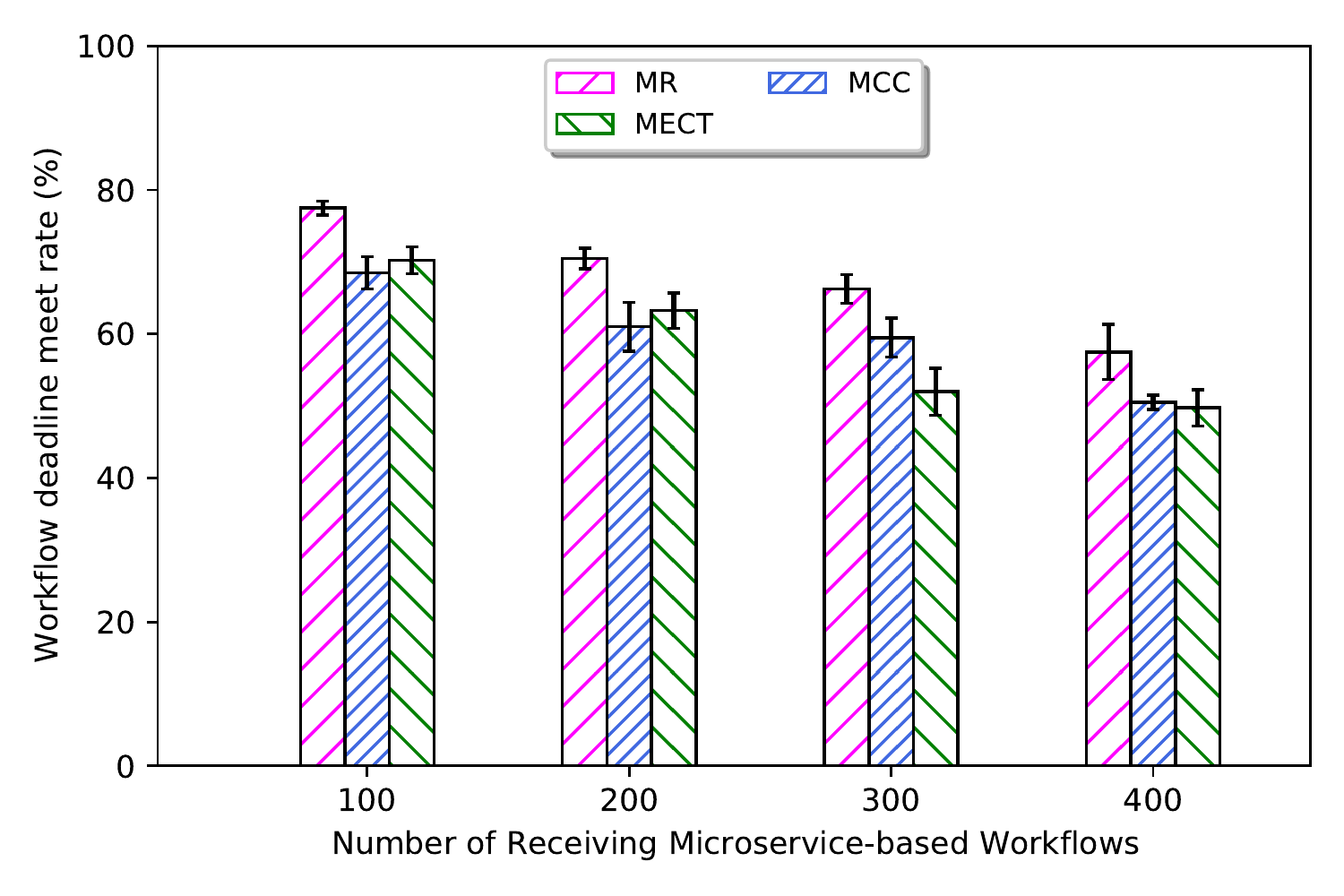}
    \caption{Comparison of resource allocation techniques while utilizing proposed workflow partitioning technique for microservice-based workflow applications.}
    \label{fig:resourcAllocationComp}
\end{figure}

\emph{Microservice-based Workflow Applications: }  Similar to the previous experiment, the number of receiving microservice-based application is incremented to create more oversubscribed situations(\ie the x-axis of the graph). To visualize the performance of the resource allocation techniques, the deadline meet rates of receiving applications are captured and plotted in figure \ref{fig:resourcAllocationComp}.

The result represents a downward trend for all the resource allocation techniques with increasing oversubscribed situations. Hence, it is visible that the proposed resource allocation technique, $MR$ outperforms other baselines in every oversubscribed situation. This is because $MR$ is aware of uncertainty in computation and communication of receiving microservices. In contrast, MECT is only aware of computation, and Certainty utilizes deadlines in its resource allocation technique which lacks communication information.

\begin{figure}[h!]
    \centering
    \includegraphics[scale=0.60]{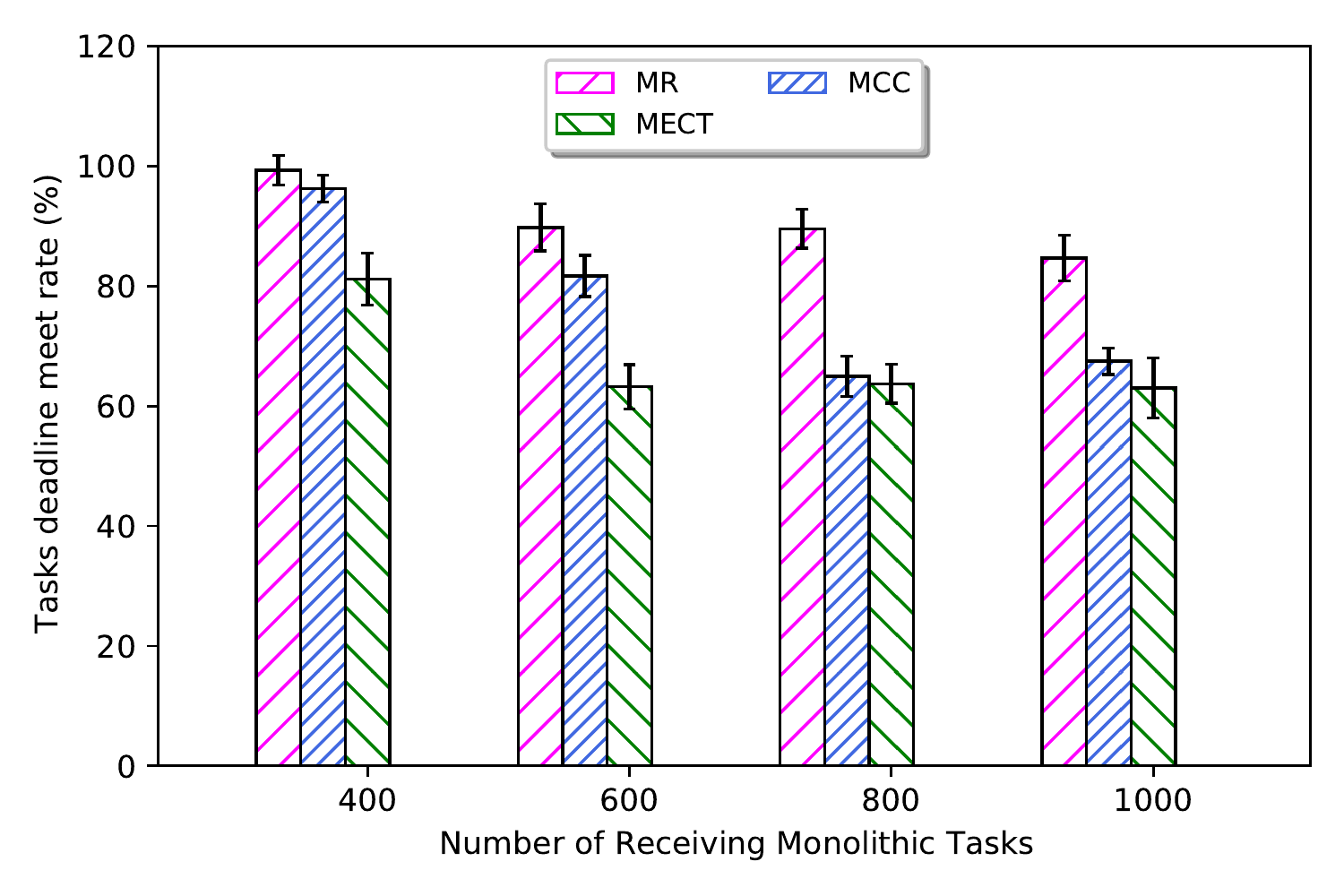}
    \caption{Comparison of resource allocation techniques for monolithic applications. The proposed resource allocation technique $MR$ outperforms other baselines in every application arrival trial.}
    \label{fig:resourcAllocationComp2}
\end{figure}

\emph{Monolithic Independent Applications: } 
In this experiment, we investigate the performance of our system by increasing the number of monolithic applications generated by sensors (\ie the oversubscription level). Figure \ref{fig:resourcAllocationComp2} shows the effects of altering the number of incoming applications (from 400 to 1000 on the horizontal axis) on the deadline meet rate (vertical axis) when various resource allocation heuristics are employed.

In figure~\ref{fig:resourcAllocationComp2}, it is visible that as the number of applications increases, the deadline meets rate decreases for all of the heuristics. Under low oversubscription levels (400 tasks), MR, MECT, and MCC perform similarly. However, the difference becomes substantial as the system gets more oversubscribed (800 applications). With 1000 applications, MR offers around 18-20\% higher deadline-meeting rates than MECT and MCC. The reason is that MR captures end-to-end latency and proactively utilizes federation only if it remarkably impacts the probability of success. Nonetheless, other baseline heuristics only consider computational latency. Therefore, we can conclude that for monolithic applications considering end-to-end latency and capturing its underlying uncertainties can remarkably improve the robustness, particularly when the system is oversubscribed (\eg at a disaster time).

\subsection{\textit{Fog Federation Scaling Impact}}
\begin{figure}[h!]
    \centering
    \includegraphics[scale=0.60]{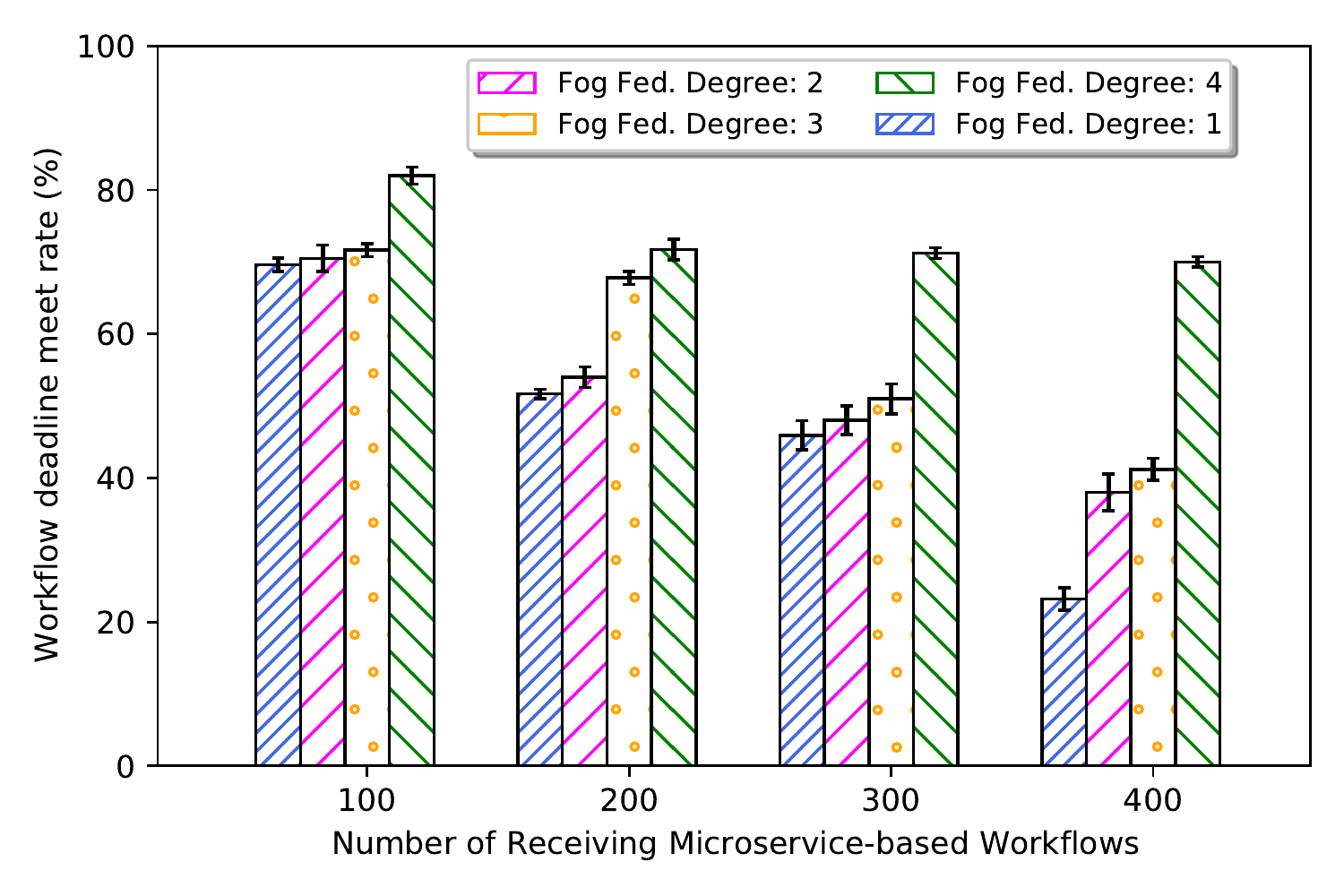}
    \caption{Impact of scaling the fog federation for proposed partitioning and resource allocation techniques in increasing oversubscribed situations considering microservice applications. The degree represents the number of neighbors each fog system has for executing the Industry 4.0 applications.}
    \label{fig:fedScaling}
\end{figure}

\begin{figure}[h!]
    \centering
    \includegraphics[scale=0.60]{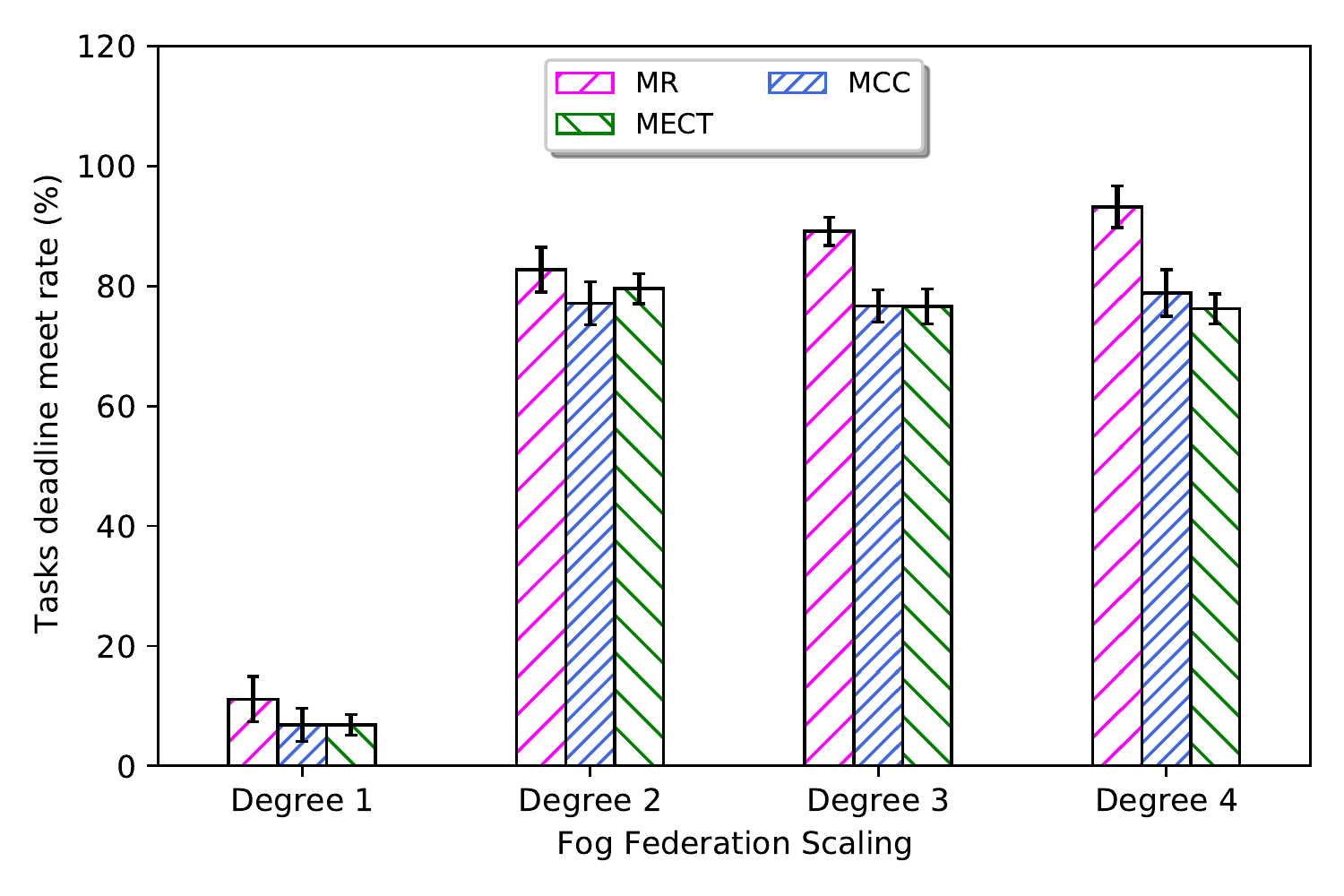}
    \caption{Impact of scaling the fog federation for proposed resource allocation techniques on monolithic applications. The degree represents the number of neighbors each fog system has for executing the Industry 4.0 applications.}
    \label{fig:fedScalingMon}
\end{figure}

~\\
Fog federation in remote offshore areas can be scaled up in times of emergencies by utilizing mobile fog systems mounted on a boat or other vehicles. In contrast, a scaled-down fog federation can decrease the system's robustness. Hence, to understand the impact of federation scaling over the proposed solution, we increase the fog federation degree that represents the number of neighbors and captures the deadline meet rates of the received applications within the increasing oversubscribed situations. The result of this experiment is presented in figure \ref{fig:fedScaling}. In addition, we performed a similar experiment for monolithic applications, where we fixed the number of receiving tasks and incremented the fog federation degree. The result for monolithic applications is presented in figure \ref{fig:fedScalingMon}.

\emph{Microservice-based Application: }
The result shown in \ref{fig:fedScaling} demonstrates the advantages of scaling up the fog federation. As a result, in any overcrowded circumstance, the federation with the greatest number of neighbors (\ie fog fed. degree 4) excels. Despite this, considerable performance improvements are seen in most oversubscribed circumstances (\ie a system processing 400 microservice-based workflows). For less overloaded scenarios (for example, a system with 100 - 200 receiving microservice-based workflows), the performance difference for minor scale-up fog federation is negligible. This is due to the suggested method, particularly the partitioning technique, attempting to put the whole application into a fog system rather than partitioning and distributing them around the federation in less oversubscribed conditions. As a result, the performance increase is substantial in the fully oversubscribed scenario with the most neighbors. 

\emph{Monolithic Applications: }
In this experiment, we compare the resource allocation techniques for monolithic applications while scaling up the fog federation. Similar to microservice-based workflows, the monolithic applications positively impact federation scaling, which is visible from figure \ref{fig:fedScalingMon}. The result reflected a significant performance improvement when the federation scaled up from degree 1 to degree 2 for all heuristics. Hence, degree 1 defines only one neighbor, and the federation is formed with two fog systems. Therefore, none of the heuristics performed well. Even though the proposed method MR, performed better than baselines. Whereas for the highest degree of the federation, the proposed MR heuristic performed approximately 18-20\% better than MECT and MCC. However, for all of the federation scales up, the proposed MR heuristic outperforms others. The main reason is that MR, efficiently utilizes fog federation resources, considering the communication and computation latencies to complete every monolithic application on time.

\section{Summary}
The advancement in software and hardware stack has brought the industrial revolution, Industry 4.0, that changed many legacy system architectures and imposed latency constraints. As such, complex industrial processes are adopting smart solutions every day. Hence, computation near the data source supports smart microservice-based solutions that significantly face resource scarcity and latency constraints challenges. Especially in remote offshore Industries (\eg Oil and Gas, mining ), the latency issue can be critical for complex fault-intolerant industrial processes (\eg hydrocarbon exploration, drilling). Moreover, in emergency situations, the computational execution platform gets oversubscribed with various types of microservices. To overcome challenges enforced by the smart microservice-based solutions, a robust task allocation scheme proposed in this research work that is aware of the software architecture of the solution as well as uncertainties imposed by fog federation. Hence, the proposed solution works on two levels within a load balancer module that exists in every fog system of the federation. The first level considers the software architecture of the receiving application and performs partitioning if necessary, utilizing the probabilistic success rate to complete the applications. Then in the second level, the received applications (\eg monolithic applications or partitioned microservices) are mapped across fog federation, considering the computation and communication constraints. The evaluation results reflect the benefits of using the proposed solution in oversubscribed situations that are approximately 15$\sim$20\% better than the baseline partitioning and resource allocation techniques. In the future, we plan to incorporate an ML-based resource provisioning method to improve the robustness of the federated fog system.
    \chapter{Data Security \& Privacy Aspects in Federated Fog Computing System}\label{section:fedLearn}

\section{Overview}
\begin{figure}[h!]
\centering
\includegraphics[width=0.9\textwidth]{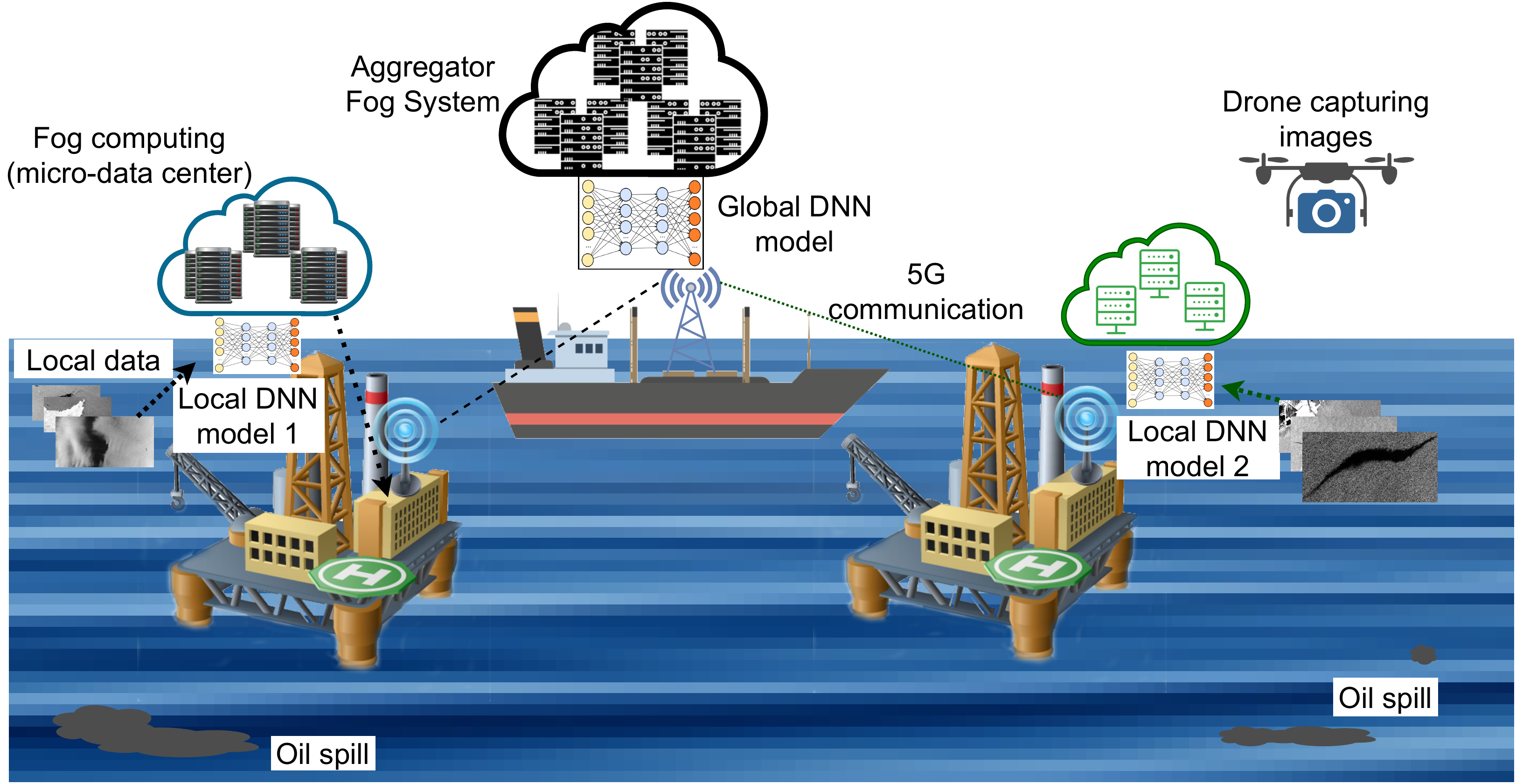}
\caption{A federated learning setup in fog federation. Multiple company share their fog systems to train oil spill detection DNN model where data security is preserved by federated learning.}
\label{fig:fogFLintro}
\centering
\end{figure}

~
The rise of Industry 4.0 \cite{ccinar2020machine} elevates the utilization of IoT devices (\eg sensors, actuators) and fog computing for developing deep neural network (DNN) applications in various industrial sectors (\eg smart oil field, smart farms, smart factory). The DNN-based applications mainly backed up by ML network models that are supposed to train with huge amount of data for achieving relatively high accuracy. Although the training data could be privacy preserving (\ie sensitive to any company), and sometimes data acquisition (\eg Satellite image data, high resolution camera data) is expensive, and time consuming. The expense of developing these DNN-based applications could also increase with data transfer to cloud datacenter using internet for training operation. Hence fog federation (formed by multiple private companies fog systems) can be a potential candidate for supporting computational demand of ML-model training where data security and privacy of the participant private fog systems in the federation need to be addressed to efficient ML training. In this case, federated Learning (FL) techniques \cite{mcmahan2017communication} that brings ML-model to participant end user without leaving the data their source device, can be applied to overcome the privacy constrains of the fog systems owned by private companies. Although the privacy preservation constraints are mitigated by FL as depicted in Figure \ref{fig:fogFLintro}, it can impose some new challenges for the ML models training operation. Here, the problem is that data are coming from various sources, and it is feasible to assume data distribution tends to be non-identical and independent distribution (non-IID). As such, lack of any priority class (\ie consider \textit{oil spill} class in oil spill detection problem) that is termed as \emph{class imbalance} \cite{duan2020self} can reduce the performance of the global DNN model in a FL setup as presented in Figure \ref{fig:fogFLintro}. Hence, ignoring the class imbalance issue, current federated learning methods \cite{mcmahan2017communication} are providing less robust DNN model for oil spill detection. In this case, an object detection model can show misleading high accuracy for all other classes while providing low performance for the desired class (oil spill).

Federated learning is a special branch of distributed machine learning where the global model needs to be converged at a constraint rate. Hence, the convergence of FL mainly depends on the local workers' aggregation that affects the global model's performance. Among two types of federated learning (synchronize and asynchronous), we propose to utilize the synchronize FL method as it is a proven model, especially for class imbalance issue \cite{duan2019astraea,mhaisen2020analysis}. As such to overcome the challenges of FL for oil spill detection, we have adopted an objective function (loss function) to train the local model considering the class imbalance problem. Considering the priority class (\ie oil spill), we introduce a weight for each participating worker that intensifies or attenuates its influence over the global model. The relevant worker selection based on the worker weights is verified by empirical evaluation in section \ref{section:performanceEval}. ~Finally, a dynamic threshold mechanism has been proposed to select relevant workers efficiently considering the global model's performance and fast convergence.

\section{Problem Formulation for Federated Learning}\label{section:problemForm}
The oil spill detection problem can be well defined in semantic segmentation domain of deep learning where various classes are identified in pixel level from original source image. In oil spill detection training various classes (\eg oil-spill, look-alike, land, ship, sea-surface) are found in real world satellite image data set \cite{krestenitis2019oil}. Here, each class is labeled as an individual color in ground truth image. For training a deep neural network (DNN) model (\eg Unet) with federated learning settings a set of workers (\ie fog systems of a fog federation) $S= {1,2,3,...,S}$ are considered with its own local data set $D^L$ where $L \in S$ with $n_L$ samples. Here, $D =$ $\bigcup_{L \in S} D^L$ is the full training data set. The total size of these workers' data set for a random set of workers $S^\prime$ is $N(S^\prime) = \sum_{\mathcal{L} \in S^\prime}n_L $. The objective loss function over a model \textit{m} and a sample \textit{z} can be denoted as $\mathcal{L}(m,z)$.

Then in most prior FL work, the goal is to solve the following
\begin{mini*}|s|
{w}{f(w) = \sum_{m=1}^{M} p_m F_{m}(w)}
{}{}
\end{mini*}
where $p_m = \frac{n_m}{D}$ is the fraction of the total data worker, and thus, $\sum_{m}p_m = 1$. The local objective $F_m$ is typically
defined by the empirical loss over local data, $F_m (w) = \frac{1}{n_m} \sum_{j=1}^{n_m}\mathcal{L}_{j}(m,z)$. Here, \textit{w} is the model parameter that used for predicting loss over a sample data, and the goal is to find the optimal $w$ for which the loss should be minimized. Accordingly, we focus on utilizing a loss function that consider class imbalance problem in local data samples, and select a set of client worker's (fog system) models to aggregate that have certain level of accuracy (\ie mean intersection over union (mIoU) for semantic segmentation) to ensure the robustness of the global model. Hence, our new objective for this work would be as following:
\begin{mini*}|s|
  {w}{f(w) = \sum_{m=1}^{M} p_m F_{m}(w)}
  {}{}
  \addConstraint{mIoU(m)}{>=\gamma}
  \addConstraint{\theta}{>1}
\end{mini*}
Where, $\gamma$ is a dynamic threshold (initial value set to 50\% or 0.5) for checking the local trained model's mIoU with auxiliary test data, and $\theta$ is the user defined worker's weight with respect to oil spill class. Both of this parameters are used to select the relevant worker's model for aggregation into the global model that ensure the robustness, and consistency of the convergence for the aggregated model.

\begin{table}[]
\centering
\begin{tabular}{|l|l|}
\hline
Class       & Pixels  \\ \hline
Sea Surface & 797.7 M \\ \hline
Oil Spill   & 9.1 M   \\ \hline
Look-alike  & 50.4 M  \\ \hline
Ship        & 0.3 M   \\ \hline
Land        & 45.7 M  \\ \hline
\end{tabular}
\caption{Pixel distribution for each of the class in oil spill detection data set}
\label{tab:classImb}
\end{table}

\section{Federated Learning to Mitigate the Class Imbalance}\label{section:solution}
\begin{figure}[ht]
\centering
\includegraphics[width=0.60\textwidth]{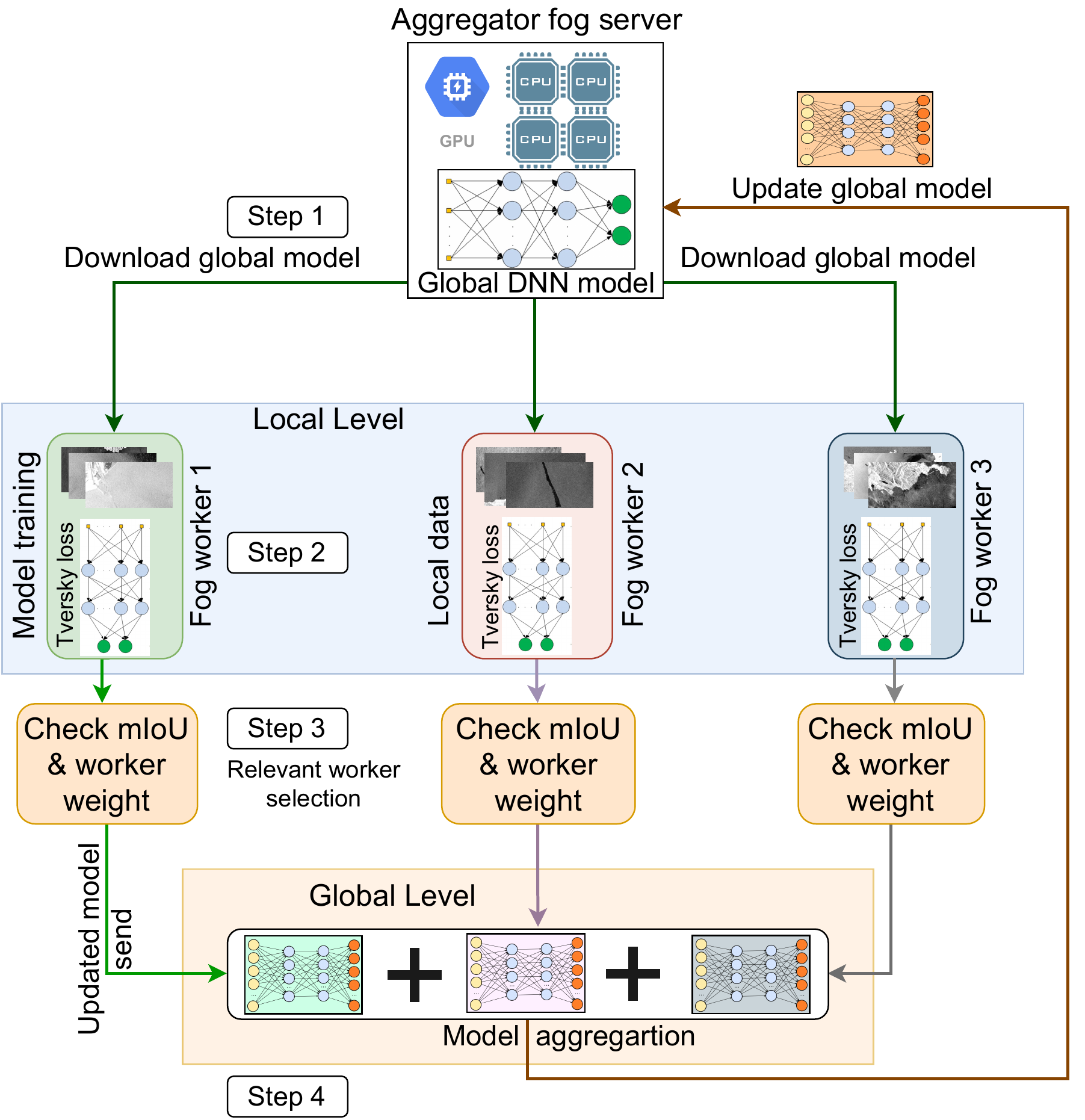}
\caption{Federated learning training considering class imbalance and global convergence. Tversky loss is used in the training considering class imbalance. After training of each epoch, mean intersection over union (mIoU) is checked with a dynamic threshold for global convergence.}
\label{fig:solFL}
\centering
\end{figure}

In a typical federated learning setup, the server (\eg fog device, cloud) stores a global ML model for training with local data of the participating workers (\ie fog systems). We use one of the popular semantic segmentation DNN models named as Unet \cite{ronneberger2015u} model for oil spill detection in the FL setup. The figure \ref{fig:solFL} represents a pictorial view of our solution. At first, some fog nodes agree to participate in the FL training, and they download the global model (\ie Unet) from the fog server presented in step 1 of figure \ref{fig:solFL}. Then, downloaded ML models are trained with their local data in step 2. Hence, we utilize \textit{tversky loss} function for local training that work efficiently for class imbalance issue proven by the research community \cite{abraham2019novel,salehi2017tversky,outeiral2022strategies,jadon2020survey,sau2022retinal}. In step 3, ML models are checked for relevant worker model selection. Finally, in step 4, selected workers updated models are aggregated (new model), and the previous global model is updated accordingly. This whole process is considered a federated round. The updated model is again downloaded by participating worker for the next federated round, and the training continues. The proposed solution is presented in algorithm \ref{fedbal}.

\SetKwComment{Comment}{/* }{ */}

\begin{algorithm}
\caption{The $K$ workers are indexed by $k$, $C$ is the initial worker selection percentage}\label{fedbal}
Initialize global model, $m_g$, test data, $D_{test}$, relevant Worker List, $r_f$\;
Set threshold, $t_h = 0.50$\;
 \For{each federated round f = 1,2,...}{
  $m \leftarrow \mathrm{max(C.K,1)}$\;
  $S_t \leftarrow (random~set~of~m~workers)$\;
  \For{each worker $k \in S_t$~\textbf{in parallel}}{
      $ClientUpdate(k,m_g)$\; 
    }
  $r_f =  selectionCriteria(S_t,D_{test},t_h)$\;
  $m_g = averageModels(m_g, r_f)$\;
  }
\end{algorithm}

Usually, the aggregator fog server provides the global model and aggregates the updates sent by the worker fogs. The \emph{FedBal} algorithm starts with initializing global model $m_g$, relevant worker list, $r_f$, and setting the threshold, $t_h$ value to 0.50. After that, the federated round continues as a for loop that is presented with variable $f$. Then $m$ number of workers are selected from $K$ participating worker, and assigned to selected worker list, $S_t$ for training (``ClientUpdate'' function) with their local data in the second for loop of the algorithm \ref{fedbal}. Finally, relevant workers are selected using function ``selectionCriteria'', and aggregate into the new global model, $m_g$ using ``averageModel'' function. The ``ClientUpdate'' function performs the training with the \textit{tversky loss} function and defined number of epochs to reduce the class imbalance at the local level. The proposed solution's global level is triggered in the ``selectionCriteria'' function, where trained worker models are evaluated according to their weight, $\theta$, and $mIoU$ value. The dynamic threshold mechanism also takes place in the ``selectionCriteria'' function to ensure the robustness of the global model. In this way, in every federated round,$f$, the global model updates and converges to a model robust against class imbalance with guaranteeing performance for our priority class, oil spill.  

\section{Experimental Setup}\label{section:expSetup}
The Federated learning setup can be synthesized by PyTorch's one of the popular library pysyft \cite{ziller2021pysyft}, and TensorFlow's federated learning library named as tff \cite{abadi2016tensorflow}. Due to pysyft's customization capability, we have selected pysyft as our development library. The oil spill detection is considered a semantic segmentation problem that typically uses real-world SAR image data sets (in this work, the data set is collected from MKLab \cite{krestenitis2019oil}, a research institute in Greece) for training a DNN model. To execute the DNN training operation, we used Google's Colab \cite{bisong2019google} run-time environment that provides a GPU platform with a high-speed ram of size 24 GB with storage of 128 GB.

The Colab provides Tesla P100, T4, or similar GPUs for the paid ``pro'' version. It also has the high-RAM option for faster execution while using GPU. We utilize pysyft's virtual worker's concept to synthesize fog devices. Our primary focus in this work is to reduce class imbalance issues and ensure a robust global model. Hence we concentrate on the computation part of FL and ignore the communication (\ie network) of conventional FL setup. Our federated learning setup can be utilized for any aggregation algorithms (\eg FedAvg, FedSGD, FedProx), and as such, we develop our codebase on top of these baseline algorithms. As our FL setup works on reducing the class imbalance, we named this setup ``FedBal''. In most of our experiments, we use 20 federated rounds where each round consists of 50 epochs. The reason behind choosing these values for the training parameters (\eg number of epoch, number of federated rounds) is to observe a significant difference among the aggregation algorithms. Finally, due to time constraints, we bound our experiments within 20 federated rounds of aggregation.

\section{Performance Evaluation}\label{section:performanceEval}
The federated learning setup is always beneficial for fog devices where data tends to be generated frequently. Hence, to understand the advantage of utilizing federated learning, we perform an experiment capturing the loss found in each epoch of training using federated learning and single machine training. The federated learning setup (a) could use more data as there are four workers perform the DNN model(Unet) training. On the other hand, a non-federated learning setup uses fewer data to train the model with a single fog device. Moreover, the uncertainty in federated learning setup is less severe than non-federated learning that we found in our initial experiment, (b). Considering the convergence of the training model, FL is also faster than non-fl. Although FL has better performance than typical machine learning, the class imbalance issue in the local data can make the global model's performance degradation. Hence, our local worker level solution utilizes the tversky loss function where $\alpha$ for penalizing false negative and $\beta$ for penalizing false positive parameters need to be tuned for better performance. The experiments with these parameters are provided in the following section.

\subsection{\textit{Tuning Loss Function}}
\begin{figure*}[!t]
\centering
\includegraphics[width=\textwidth]{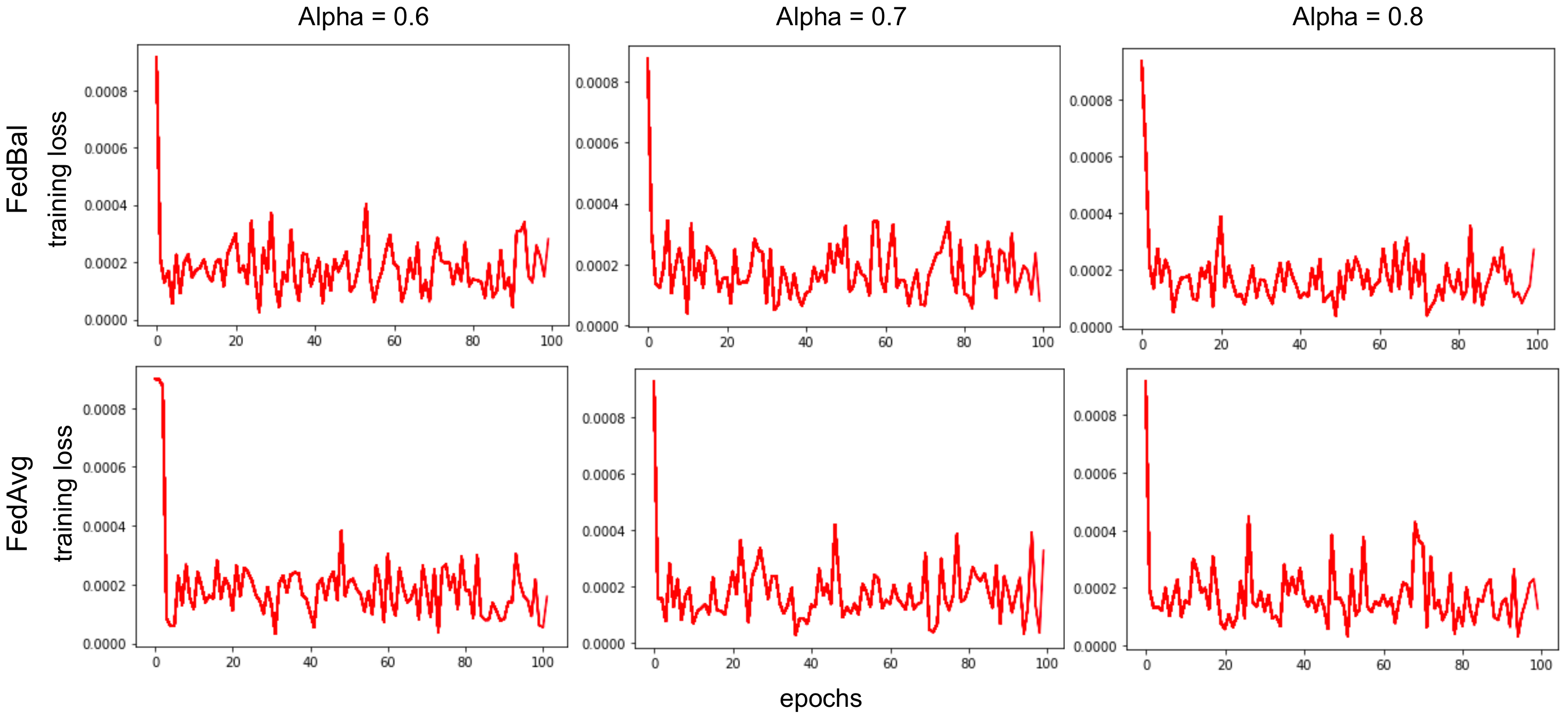}
\caption{Comparison of FedAvg and FedBal training loss utilizing tversky loss function. The alpha parameter of tversky index is changed from 0.6 to 0.8 (left to right) and the loss per epoch is captured for both FedAvg and FedBal algorithm.}
\label{fig:tverskyLoss}
\centering
\end{figure*}

~\\
To find the optimal Tversky loss function, we change the alpha parameter value from 0.6 to 0.8 and capture each training epoch's loss. The main goal is to find the optimum alpha value for which the loss will be minimal. The results of these experiments are demonstrated in figure \ref{fig:tverskyLoss}.

The figure \ref{fig:tverskyLoss} represents the training loss (\ie y-axis of the figure) for each epoch (\ie the x-axis of the figure) while using alpha values 0.6, 0.7, and 0.8 respectively within FedAvg, and FedBal FL setup. From figure \ref{fig:tverskyLoss}, we find that FedBal performed similar in comparison with FedAvg. It is also visible that for alpha value 0f 0.7, FedBal has minimum training loss. When we increase the alpha value to 0.8, the training loss does not decrease, which means we can penalize false negatives up to a certain point (\ie $\alpha = 0.7$). The reason is that while we are penalizing false negatives, the false positive predictions are ignored (\ie $\alpha + \beta = 1 $) as well. Hence for a higher value of alpha, we get less benefit by penalizing false-negative predictions. Therefore, we use $\alpha = 0.7$ for the rest of our experiments throughout this work.

\begin{figure}[h!]
\centering
\includegraphics[width=\textwidth]{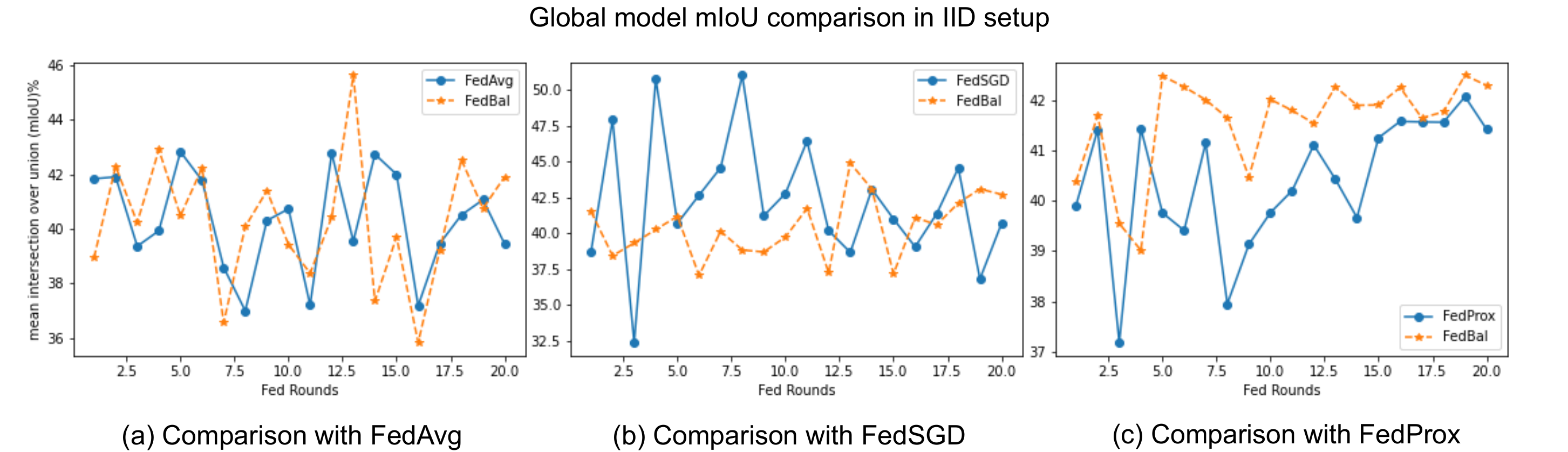}
\caption{Comparison of FedBal with FedAvg, FedSGD, and FedProx method's global model performance in IID setup.}
\label{fig:iidmiou}
\centering
\end{figure}

\subsection{\textit{The Impact of Using IID Data Distribution}}~\\
The benefit of a federated learning setup is reflected in the performance (\ie accuracy (mIoU)) of the global model after the aggregation step of FL is completed. Hence, we measure the mIoU of the global model after every communication or federated round of FedBal with FedAvg, FedSGD, and FedProx, respectively. Then, we plot the result as a line graph in figure \ref{fig:iidmiou}. For this experiment, we consider the data distribution among the FL workers is identical and independent distribution (a.k.a IID) which means every worker gets all the classes of images in their local data for training.

The x-axis of the figure \ref{fig:iidmiou} represents the federated rounds, whereas the y-axis presents the mIoU of the global model. From the figure \ref{fig:iidmiou} (a), it is visible that FedBal has outperformed FedAvg in most of the fed rounds. Considering FedSGD, in figure \ref{fig:iidmiou} (b), FedBal performed significantly well in the last few rounds. Although, in the initial rounds, FedSGD performed better than FedBal. Finally, from figure \ref{fig:iidmiou} (c), we find that comparing FedProx, our FedBal method performed significantly well. This improvement mainly comes from the utilization of left-out workers in the global model. In addition, the worker selection in FedBal considers the class imbalance issue and the priority class (\ie oil spill class) for aggregation into the global model. In contrast, other methods randomly select active workers for aggregation, leading to a less robust global model than FedBal.

\subsection{\textit{The Impact of Using non-IID Data Distribution}}
\begin{figure}[h!]
\centering
\includegraphics[width=0.55\textwidth]{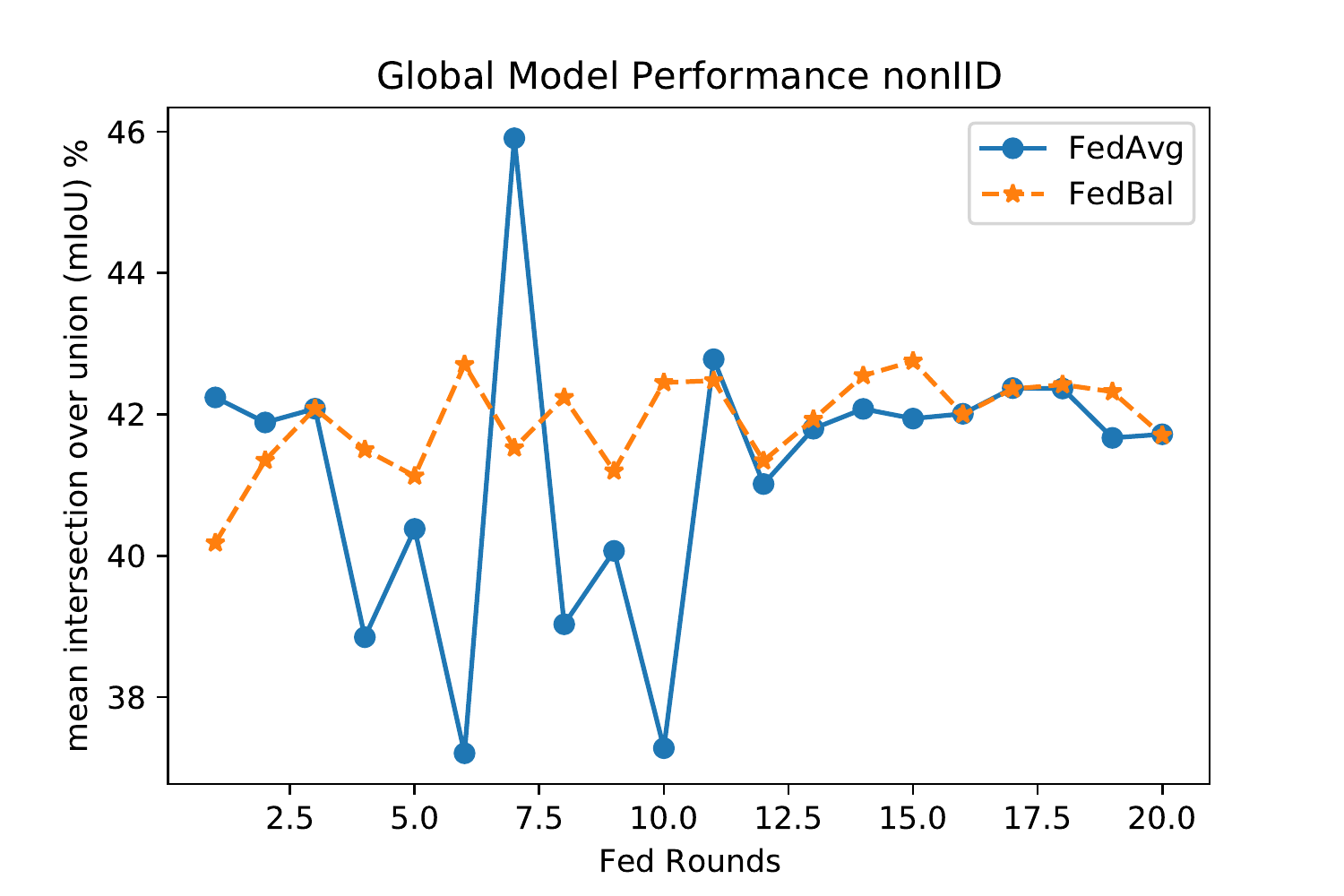}
\caption{The performance comparison of global models in terms of mIoU using FedAvg and FedBal methods. The data distribution is non-IID, the number of workers are 6, and in each fed round 50 epochs of training has been performed.}
\label{fig:noniid}
\centering
\end{figure}

~\\
In a real-world scenario, data distribution among FL workers is typically non-IID. That means every worker will get some fixed number of classes (not all the classes) for local training. Hence, we consider providing two classes for each worker, and these classes are different for every worker. Similar to our previous experiment, we measure the mIoU of the global models for FedAvg and FedBal algorithms in each federated round. The result is provided in figure \ref{fig:noniid} for 20 federated rounds with six federated fog workers.

The figure \ref{fig:noniid} reflects that FedBal has a consistent performance (mIoU) for 20 federated rounds then FedAvg. The FedBal method has less uncertainty (fewer spikes in orange line of figure \ref{fig:noniid}) across the federated rounds for selecting relevant workers in every federated round. Although FedBal has less significant performance improvement than FedAvg, the average mIoU of the global model of FedBal is higher than FedAvg. This consistent performance of FedBal represents the robustness of our method across the federated rounds.

\subsection{\textit{The Impact of Using non-IID and Unbalanced Data Distribution}}
\begin{figure}[h!]
\centering
\includegraphics[width=\textwidth]{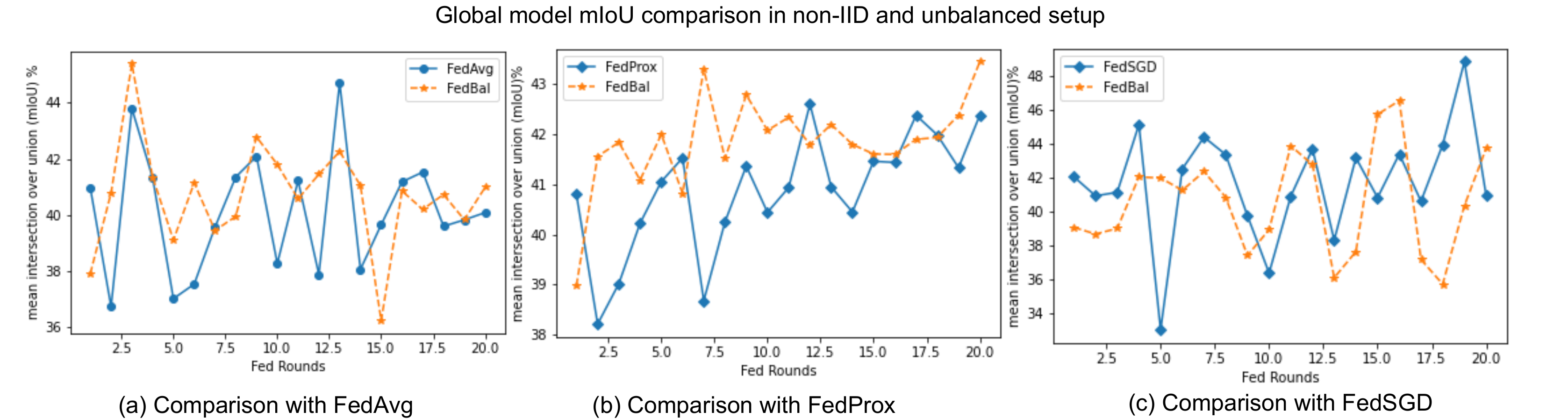}
\caption{Comparison of FedBal with FedAvg, FedProx, and FedSGD method's global model performance in non-IID and unbalanced data distribution.}
\label{fig:noniidunbalance}
\centering
\end{figure}

~\\
The non-IID and unbalance data distribution means each FL worker has a different number of classes. For instance, worker one can have two classes, whereas worker two can have only one class in its training data. Hence, we measure the mIoU of the global model and compare our method (FedBal) with the other three baseline methods named FedAvg, FedProx, and FedSGD, respectively. As our method is considered an improvement of any federated learning aggregation method (\eg FedAvg, FedProx, FedSGD), we compare the baselines separately in three different sub-figures. The result demonstrates as a line plot in the figure \ref{fig:noniidunbalance} where the x-axis represents the federated rounds, and the y-axis represents the mean intersection over union (mIoU) of the global model.

Figure \ref{fig:noniidunbalance} represents that FedBal outperforms FedAvg, FedProx, and FedSGD respectively in the final round. Although, in the 19th federated round, FedSGD and FedBal perform similarly. The performance improvement for FedBal is significant for FedProx, due to the utilization of the left out workers in the global model. It is also visible that baseline methods performance has severe uncertainty (more spikes than FedBal), whereas FedBal has comparatively consistent performance throughout the federated rounds. The main reason behind this consistency is the relevant worker selection in FedBal with a dynamic threshold mechanism that maintains a certain performance and provides a robust global model against class imbalance issues in the local data set.

\subsection{\textit{The Impact of Class Imbalance Intensity}}
\begin{figure}[h!]
\centering
\includegraphics[width=0.85\textwidth]{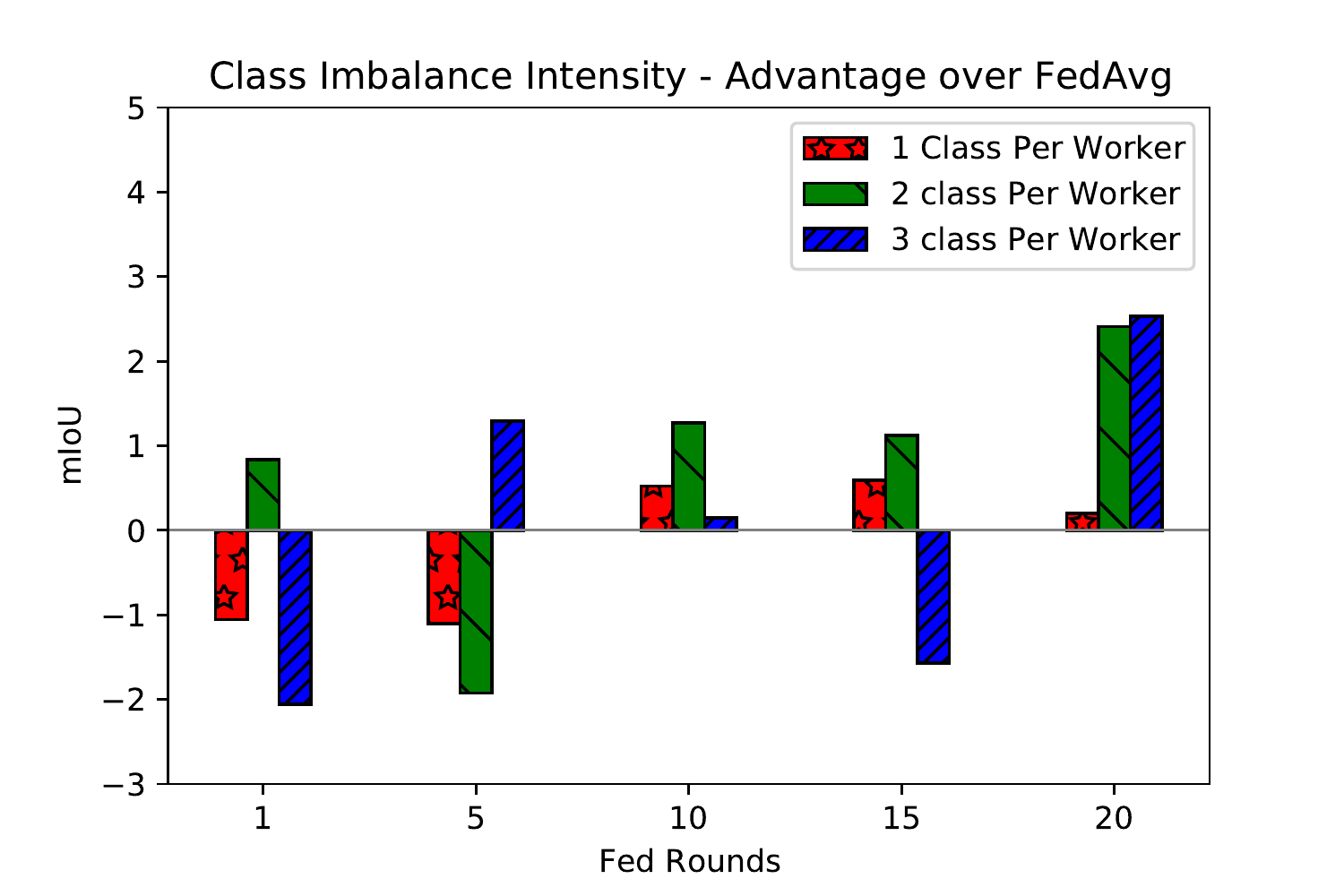}
\caption{Comparison of FedAvg, and FedBal method's global model performance in non-IID data distribution from high intensity(only 1 class per worker) to low intensity(3 classes per worker). The difference of mIoU of FedBal, and FedAvg is plotted as barchart for 3 case scenarios (1 class, 2 class, and 3 class).}
\label{fig:classimbalance}
\centering
\end{figure}

~\\
The class imbalance is a common phenomenon in the oil spill data set where the intensity of the imbalance can be severe within the FL setup. As such, we consider our solution, FedBal, to be performed consistently well than the baseline FedAvg algorithm in all levels of class imbalance intensity. To explore the class imbalance intensity, we distribute the classes from high imbalance to low imbalance using a non-IID setup and measure the mIoU of the global model for FedBal and FedAvg across the federated rounds of FL training. We estimate the difference of mIoU values for each federated round for three cases (one class, two classes, and three classes distribution) of imbalanced data distribution. Furthermore, then we plot a bar chart presented in figure \ref{fig:classimbalance} where positive values indicate FedBal's improvement over FedAvg, and negative values represent the opposite.

Figure \ref{fig:classimbalance} represents the advantage or disadvantage of FedBal over FedAvg algorithm across 20 federated rounds of training. For the first nine federated rounds, the improvement of FedBal over FedAvg is not significant. In the 10th federated round, the difference values are all positive, and in the final federated round (20th), we find the highest performance of FedBal over FedAvg. We also notice that for 2 class per worker distribution, FedBal constantly outperforms FedAvg. For high intense class distribution (only 1 class per worker), FedBal starts to perform well after the 10th round. In the final round, we find that for 3 class distribution FedBal has the most remarkable improvement. The main reason behind the less significant performance could be the dynamic threshold mechanism that starts with a good mIoU value (50\%) and dynamically change over federated rounds to increase the performance of the global model. After the 10th round, the threshold becomes stable with a sufficient number of relevant workers, and we see performance improvement for the last ten rounds of federated training.   

\begin{figure}[h!]
\centering
\includegraphics[width=0.55\textwidth]{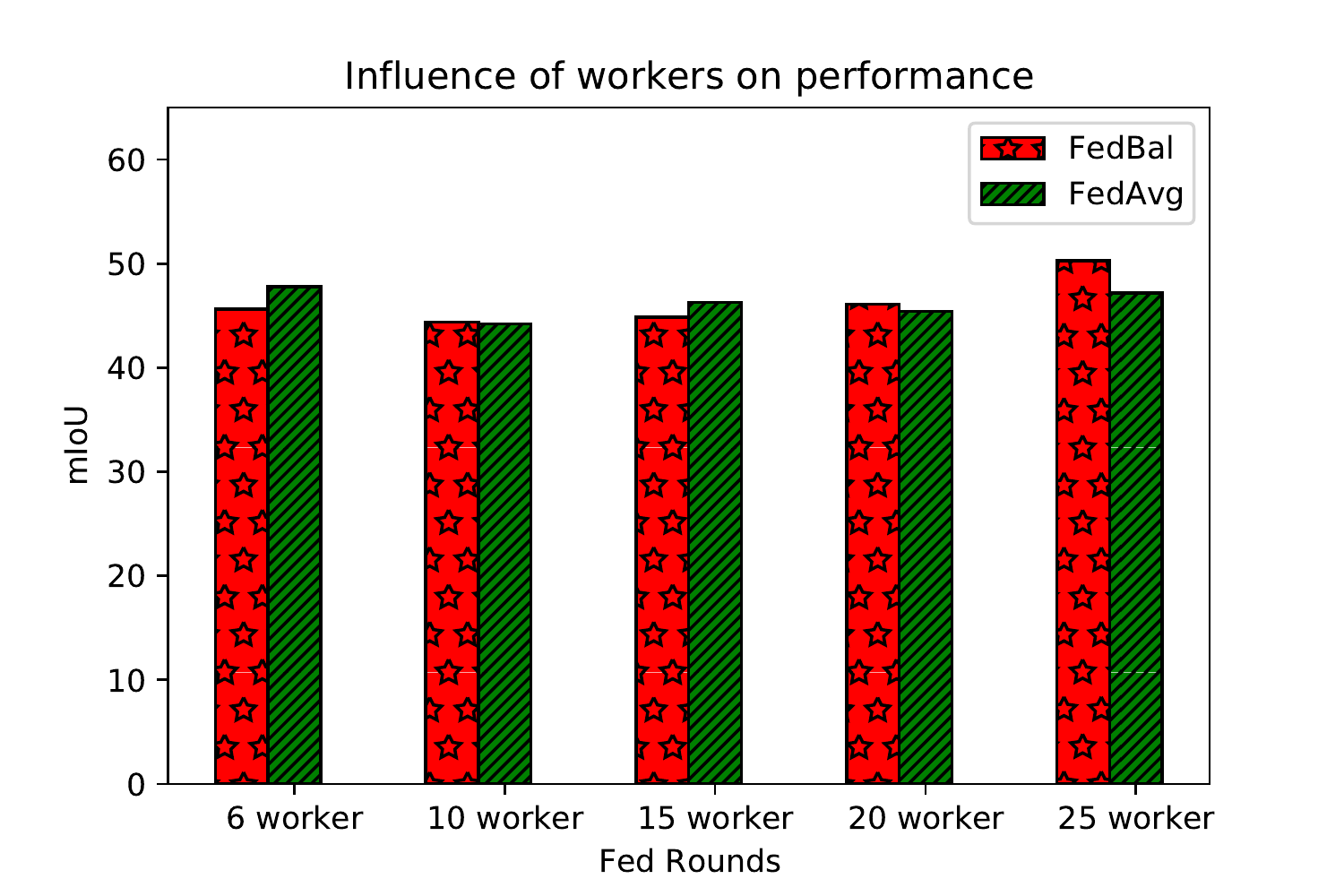}
\caption{The influence of federated worker on global models performance (mIoU) for FedBal, and FedAvg is measured by increasing the number of federated worker from 6 worker to 25 worker. For each case of worker pool 20 federated rounds of training are performed for both FedBal, and FedAvg method, and for each case maximum mIoU of both methods are considered for plotting as a barchart.}
\label{fig:numWorkes}
\centering
\end{figure}

\subsection{\textit{The Impact of Number of Workers on the Global Model}}~\\
To understand the influence of FL workers in our federated learning method, we measure the maximum mIoU of the global model for FedBal and FedAvg methods by gradually increasing workers from 6 to 25.  The main focus of this experiment is to compare the performance of our method, FedBal over FedAvg, while increasing the FL workers gradually. The result of this experiment is provided in the figure \ref{fig:numWorkes}.

The figure \ref{fig:numWorkes} presents that the performance improvement of FedBal is significant when we have an increased number of FL workers. It is visible that up to 15 workers FedBal does not show performance improvement then FedAvg. The reason is that FedBal selects relevant workers from the active workers' pool, and sometimes the relevant workers are very few, leading to a less improved global model. On the contrary, FedAvg always selects the same number of FL workers throughout the federated round, thus having better performance than the low number of FL workers pool. Therefore, an increased number of workers can significantly improve the global model's performance using the FedBal method.

\begin{figure}[h!]
\centering
\includegraphics[width=0.55\textwidth]{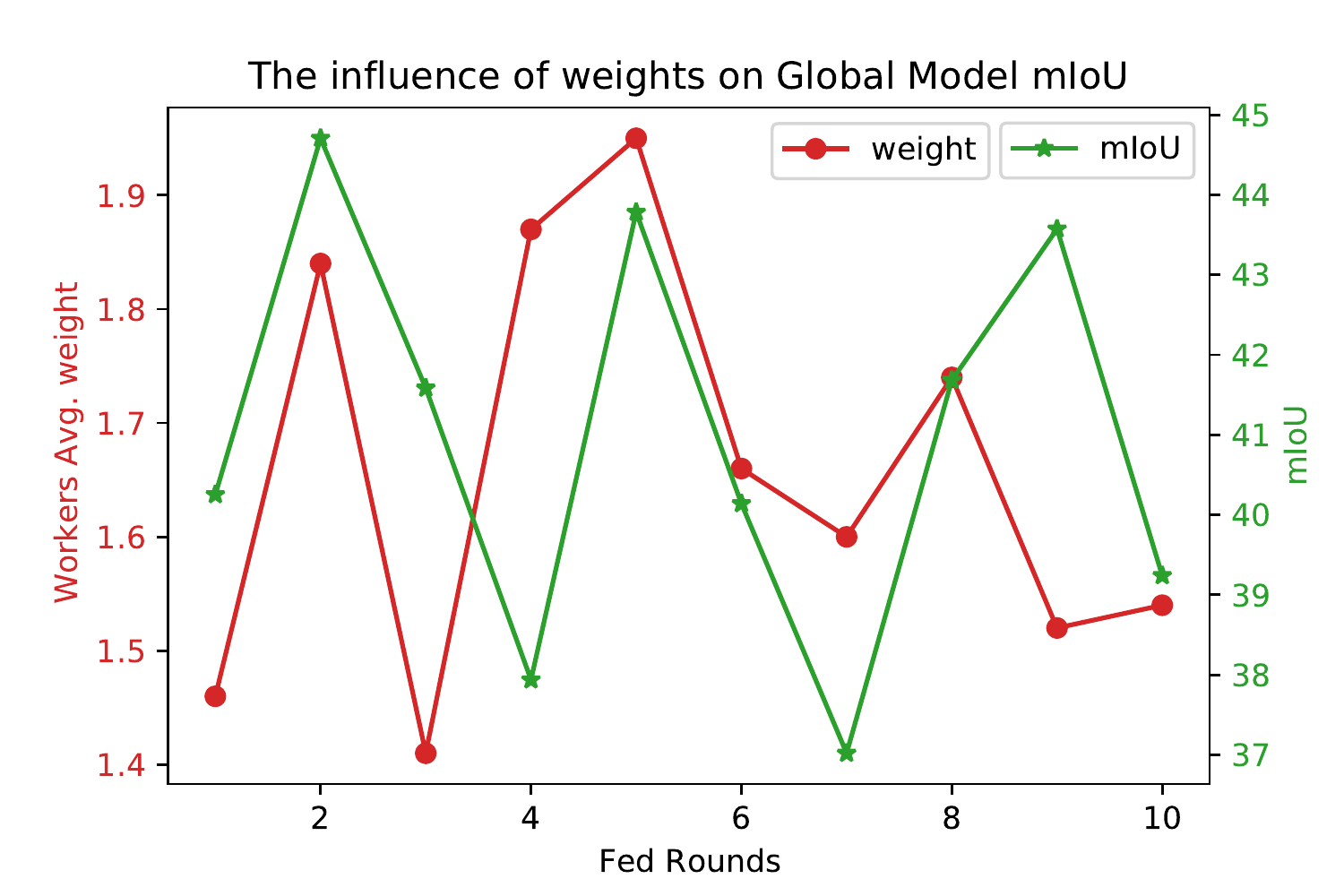}
\caption{The impact of workers weight (averaged on each of the federated round) on global model's mIoU.}
\label{fig:weightmIoU}
\centering
\end{figure}

\subsection{\textit{Verifying the Impact of Aggregation Scheme on Global Model}}\label{workerWeight}~\\
The selection technique of the FedBal algorithm considers the relevant worker by checking their weights defined by the priority class (oil spill class). Hence, workers with significant performance have a positive impact on the global model's mIoU. To explore the impact of the workers' weight in the global model, we measure the average workers' weight in each federated round and estimate the global model's mIoU after the aggregation. The result of this experiment is presented in figure \ref{fig:weightmIoU}, where the x-axis presents the federated round, the left y-axis presents the average workers' weight, and the right y-axis presents the global model's mIoU.

The figure \ref{fig:weightmIoU} presents that selected worker's weight has a positive impact on the global model. It is visible that with the increase of the average weight, the mIoU of the global model goes up, and with the decrease, the mIoU goes down. Therefore, the FedBal method's relevant workers are supported by the user-defined weight based on priority class, oil spill.

\section{Summary}\label{section:fedConclusion}
The federated learning technique has revolutionized distributed machine learning, especially considering data privacy and computational flexibility. With the emergence of IoT, Edge, and Fog computing, the data generation is getting faster and expensive when needed to be transferred utilizing network bandwidth. Hence federated learning can bring the ML model to the data generation sources that is less expensive and secure than cloud data centers. This case is more applicable than conventional centralized DNN model training considering ML support in remote areas. In addition, the data captured or collected in remote oil fields are sensitive, and privacy preservation is of significant importance for oil and gas companies. Although federated learning can overcome these challenges, the class imbalance issue can degrade the DNN model's performance. As such, we focus on reducing the effect of class imbalance at the local level while training the model, and the global level while aggregating the federated worker into the global model. At the local level, we use the tversky loss function with appropriate parameters (\eg $\alpha, \beta$) tuning to train each federated workers model considering the class imbalance issue. Then we assign each worker a weight, considering our priority class, oil spill. Finally, we check each federated worker's model mIoU with a predefined widely accepted mIoU value (50\% or 0.50) and dynamically change the threshold to ensure the robustness of the global model. 

In the empirical evaluation considering the global model's mIoU we find that for IID setup, FedBal has around 3\% performance improvement than FedAvg. For non-IID setup, we find similar performance in the final federated round (20th) compared to FedAvg. Although, FedBal's average performance for the non-IID setup is better than FedAvg. For non-IID and unbalanced setup, FedBal outperforms FedAvg, FedProx, and FedSGD respectively in the 20th federated round. Although, FedSGD has similar performance compared to FedBal, it has more uncertainty than FedBal (figure \ref{fig:noniidunbalance} (c)). In the class imbalance intensity, we find FedBal performs better than FedAvg in the final federated round (20th round) in three of the cases. Although, for high-class imbalance (only one class per worker) intensity FedBal has less significant improvement (0.25\%) whereas for low-class imbalance (three class per worker) intensity FedBal shows significant performance improvement (more than 2\%). The experiment with the increasing number of federated workers reflects that FedBal's performance can be improved (up to 2\%) with an increased number of federated workers. Due to time constraints and network vulnerability, we could not scale up the experiments, especially with the increased number of workers and class imbalance intensity. Finally, the impact of FL workers weight in FedBal method's global model reflects a positive relation that verifies the selection methods acceptability.
The ML training parameters (number of epochs per federated rounds, optimizer, batch size) can be tuned in a more granular way to explore the areas of improvement using the FedBal method that is considered as the future work of this research. Accordingly in future, we also plan to develop a custom loss function for the semantic segmentation field of deep learning and enhance our method (FedBal) as a service plugin that can be used on top of any federated learning algorithm to improve the robustness of the ML model. 
    \chapter{Threats and Side-Effects of Smart Solutions in Industry 4.0}\label{threatsSideEffect}

\section{Overview}
The convergence of new IoT technologies, cloud computing systems, improved wireless networks, and machine learning solutions have enabled smooth operations of large-scale cyber-physical industrial systems. Wireless connection, in particular, has altered operating paradigms to the point that most, if not all, production activity may now be managed remotely using a variety of sensors and actuators. Furthermore, these technologies have significantly increased the production and efficiency of various complex industrial operations. However, not everything about the digitization and smartness paradigm shift is positive! There are some disadvantages to consider as well—digital transformation and pervasive connection present weaknesses that criminals might use to launch cyber-attacks, thus jeopardizing industrial production, distribution, and even safety. As we have noticed in several recent instances, such as colonial pipeline \cite{kilovaty2023cybersecuring}, Amsterdam-Rotterdam-Antwerp (ARA) cyber-attack \cite{2022CyberattackARA}, and Norwegian energy company \cite{2022SomeCyberAttack}, malicious software systems (a.k.a. malware) have been able to take over the control of a system and block its regular operation until the intruders' demand has been fulfilled. Indeed, these recent cyber-attacks have proven that cyber-attacks can be as harmful as physical attacks in terms of both implications and severity.

Smart sectors, such as O\&G, have unique security challenges that can only be addressed through in-depth research and diagnostics of the entire system. However, specific security solutions for smart industries have yet to be available due to their high implementation complexity. This security vacuum has allowed countless cyberattacks to flourish in recent years, endangering people and communities worldwide. Therefore, it is imperative that, as part of the Industry 4.0 revolution, all-encompassing security solutions be investigated for smart industries due to the crucial nature of these sectors and the managerial and technical gaps between them.

As the oil and gas industry becomes increasingly complicated and digitized, we are considering researching key areas of smart O\& G that pose a security risk. The upstream, midstream and downstream deployment of a vast network of linked ``things" (IoT devices) presents a significant security risk for the oil and gas industry as a whole. Predictive maintenance and on-site worker safety are just two examples of the kinds of efficiency gains that may be made possible by processing the massive amounts of real-time, real-world data generated by smart sensors. There is a risk that the use of internet-connected devices might compromise the physical security and safety of O\&G infrastructure. For instance, interconnected cameras equipped with object-tracking capabilities, geofencing perimeter protection solutions, third-party infiltration, and other access control systems can cause security breaches in operational sites. Therefore, this thesis section focuses on the adverse outcomes of smart solutions for Industry 4.0 and the strategies for minimizing those outcomes.

\begin{figure}[h!]
\centering
\includegraphics[width=1.0\textwidth]{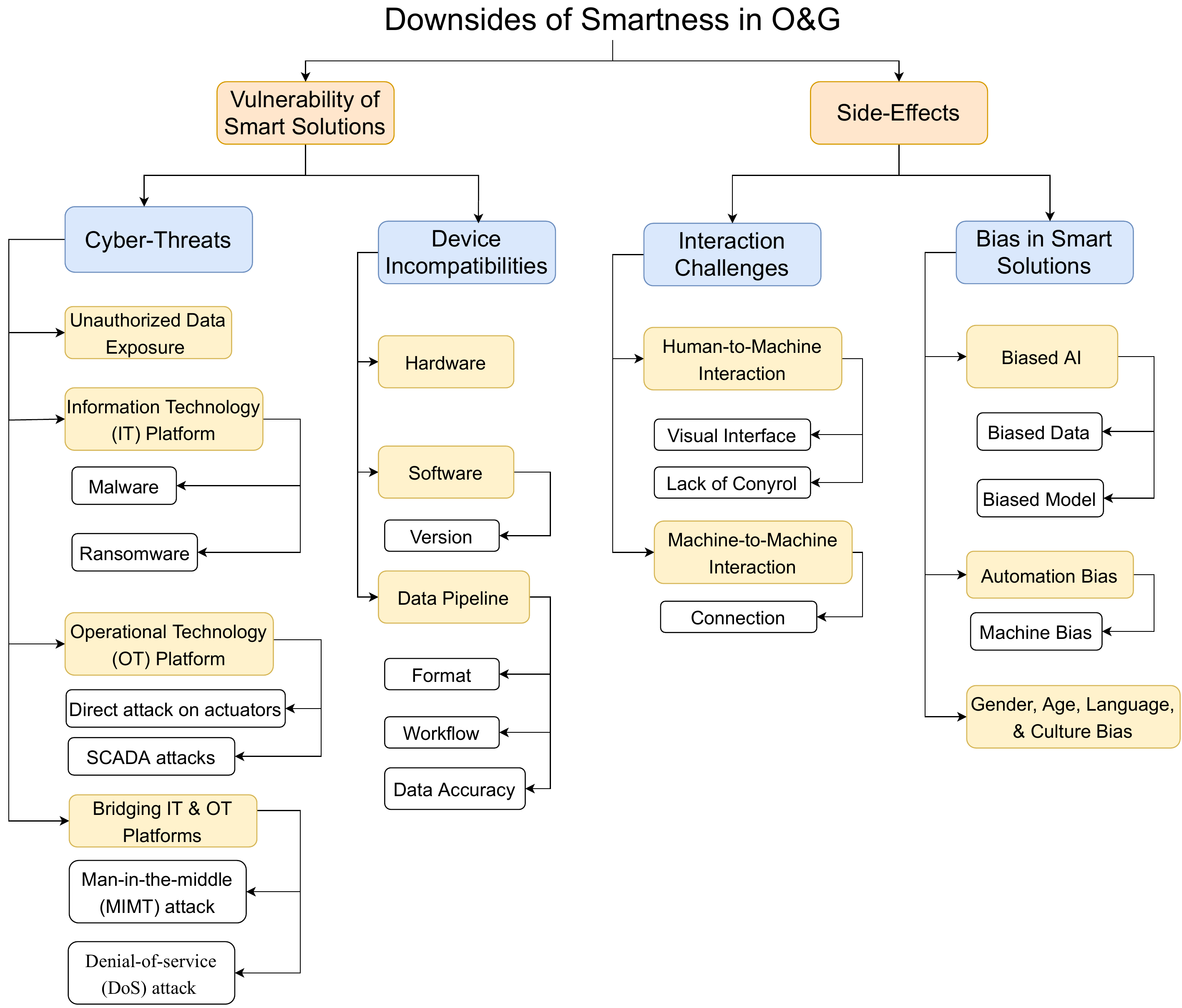}
\caption{A taxonomy reflecting the downsides of smart solutions implemented with advanced technology is organized using box flow-chart form. The main three levels are colored in orange, blue, and yellow. The white boxes represent different types (examples) of its parent box.}
\label{fig:threatTaxonomy}
\centering
\end{figure}

\section{Taxonomy of Cyber-Threats and Side-Effects in the Smart O\&G Industry}
To categorically explore various drawbacks of smart solutions in the O\&G industry, we develop a taxonomy that is presented in Figure \ref{fig:threatTaxonomy}. We separate the possible drawbacks of smart solutions into two groups in this taxonomy: vulnerabilities and side-effects. The vulnerability section investigates cyber-threats and challenges caused by device incompatibilities in a smart O\&G system, focusing on software, hardware, infrastructure, and data-related vulnerabilities in the oil and gas industry. On the other hand, the side-effect category focuses on difficulties that develop as a result of interactions with smart solutions (\eg human-machine and machine-machine interactions) and biases in a smart system.

Figure \ref{fig:threatTaxonomy} categorizes various drawbacks of smart solutions implemented or will be implemented in the near future. This taxonomy serves as the blueprint for this chapter, enabling readers to keep track of sophisticated smart solutions and their accompanying outcomes. Therefore, we will traverse major taxonomy sections in the following parts to comprehend the magnitude of smart solutions' drawbacks.

\section{Vulnerabilities caused by the Interplay of Informational and Operational technologies}
\begin{figure}[h!]
\centering
\includegraphics[width=1.0\textwidth]{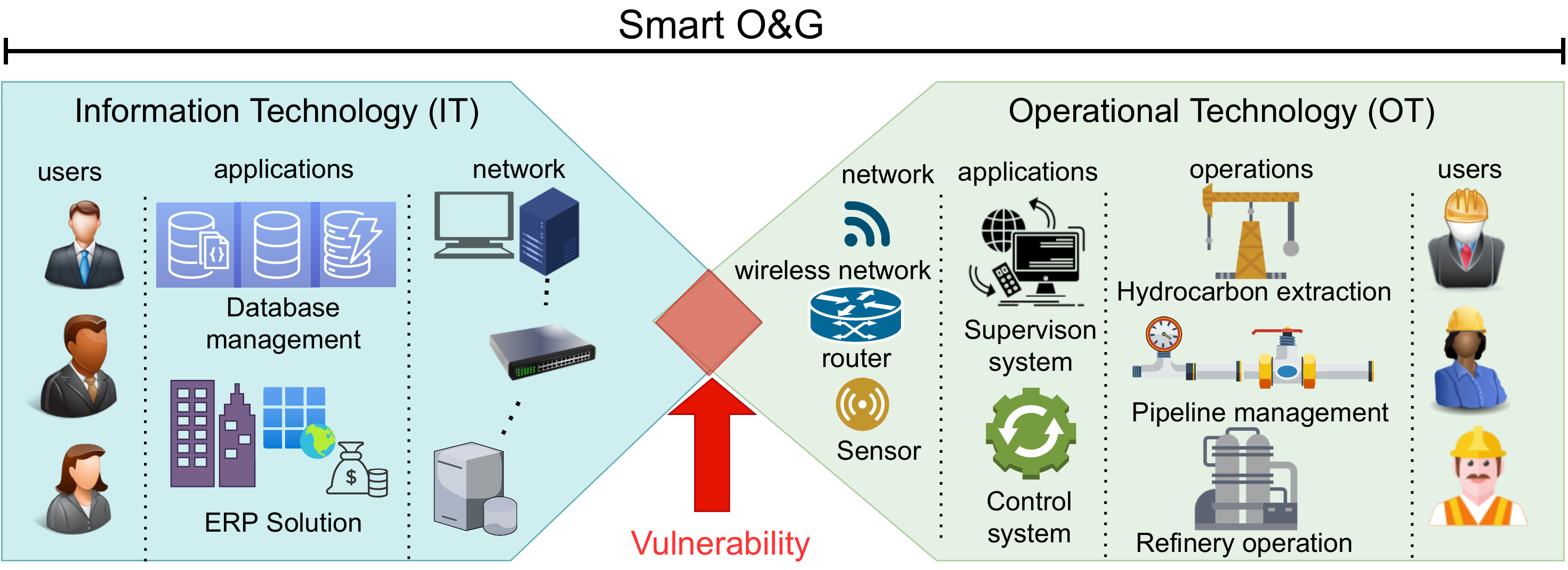}
\caption{Information technology (IT) and operational technology (OT) platforms of a smart oil and gas company that operates using different networks to run the entire operation of smart O\&G industry. The IT platform is significantly related to business applications and the financial side of O\&G, whereas the OT platform directly involves with oil or gas extraction and production operations. Both IT and OT platform is connected at some point which creates the sweet spot for cyber-attackers to penetrate into the whole system.}
\label{fig:itot}
\centering
\end{figure}

A smart oil and gas industry's technological operations are organized into two key technological platforms: information technology (IT) and operational technology (OT). Figure \ref{fig:itot} depicts an overview of an oil and gas company's IT and OT components. As seen in the diagram, the IT component is primarily concerned with the movement of data and information throughout the company. IT components frequently access outside networks due to their operational context, which is mainly business logic. In contrast, the OT component is involved with the operation of physical processes of oil and gas production and the machinery needed to carry them out. As a result, cyber thieves primarily target IT and OT platforms to meet their needs. Traditionally, the IT component has been more susceptible than the OT platforms because IT has numerous open windows (\eg operating systems, email servers, direct communication applications) that attackers may exploit.

On the other hand, OT platforms mainly deal with direct oil and gas production and processing activities with limited external access. Notably, the junction of IT and OT platforms is frequently a target for cyber attacks that system architects must effectively handle. Furthermore, smart IoT solutions based on sophisticated computing technologies are opening up access to OT platforms with the rise of IoT. As a result, we explore the extent of the vulnerabilities in these two platforms as well as their overlap.

The OT platform is comprised of technologies that are actively engaged in the production of petroleum end products. The activities include extraction, refining, pipelines, production, control, and monitoring systems. On the other hand, the oil and gas IT commodity primarily deals with finance, database administration, digital asset management, and other business operations using different computer platforms and communication protocols. In this case, the OT entity provides the petroleum end products, while the IT entity develops commercial prospects and financial policies by exploiting the OT entity's output. Therefore, compared to the OT entity, the contact with the outside network from the O\&G company's internal network is substantially greater for the IT entity. Because of this relationship, a petroleum company might become a victim of ransomware and other cyber-attacks.

OT was traditionally an ``air-gapped" environment, which was not linked to public networks or other digital technologies. For decades, traditional OT has depended on computers to monitor or modify a system's physical state, such as employing SCADA systems to monitor and control equipment to increase operational efficiency. Traditional OT security largely comprises simple physical tasks, such as ensuring that a machine performs the same operation correctly and that an assembly line continues to run. Nonetheless, the emergence of Industry 4.0 in recent years has altered the conventional OT environment. Companies have started to deploy new digital solutions in their networks to boost automation via the addition of ``smart devices" that can gather data more effectively and have network access. The IT and OT systems were integrated as a consequence of this connection and to process/analyze the OT data as it was generated. Although this technological paradigm change (referred to as \emph{IT-OT Convergence} \cite{ITotconvergence,zhang2022productivity}) has generated new possibilities and unlocked new use cases, it has also offered scope for cybersecurity vulnerabilities. For example, Colonial Pipeline's assault \cite{cyberThreatColonial} demonstrates how poor password management may harm the country's largest gasoline pipeline. The hackers found the password for an old but still working VPN account. In light of this threat, oil and gas companies should establish strict cybersecurity safeguards, including employee training. STUXNET \cite{knapp2014industrial} was the first specialized hack into industrial control system (ICS)  to attract considerable attention, although not being the first cyberattack against an industrial environment. STUXNET is a computer worm that is accused of creating havoc on Iran's nuclear programme, damaging more than 20\% of the country's nuclear centrifuges. Since then, cyber-attacks on industrial organizations have progressively risen, affecting a wide range of industries, including power grids (Industroyer), energy (Black Energy), petrochemicals (Havex), and oil and gas (Havex) (TRISIS). Hackers are hacking into industrial networks, among other things, to shut down machines, demand ransom, and steal data. 

\section{Cyber Threats in Smart Oil and Gas Industry}
The challenge with the oil and gas industry is that its systems need to be designed with network connections in mind. For instance, plants were never designed to be network-connected. However, they are today as a result of the developing digital revolution. This may create a dangerous scenario since a cyberattack on such a system can damage operations and cause loss of life. In terms of cybersecurity, the O\&G industry lags behind other industries. Even though cybersecurity is vital to the company's sustainability, many companies still need to spend more on robust systems. The remainder of this section discusses some security problems the O\&G industry confronts.

\subsection{\textit{Vulnerabilities of Sensitive Data}}~\\ 
When stored on industrial IoT devices (sensors and actuators), sensitive information must be protected by rigorous security protocols. As a result, oil and gas companies now routinely examine private information gathered from a wide range of sources. Here are some examples of such data sources:
\begin{itemize}
    \item Historical oil \& gas exploration, delivery, and pricing data
    \item Demographic data
    \item Response data from job postings
    \item Web browsing patterns (on informational websites)
    \item Social Media
    \item Traditional enterprise data from operational systems
    \item Data from sensors during oil and gas drilling exploration, production, transportation, and refining
\end{itemize}
The aforementioned are examples of highly confidential information for any private corporation. Various confidential information belonging to one company might be precious to a company's competitors in the oil and gas industry due to the intense rivalry in this sector. As a result, hackers with questionable ethics increasingly focus on gaining access to these sensitive records.

\subsection{\textit{Vulnerabilities of Smart Systems}}~\\
In earlier chapters, we covered smart solutions that empower Industry 4.0. Although these are intriguing and future technologies that might assist the oil and gas sector as a whole, their weaknesses should also be acknowledged. The following are some of the ways a smart solution might fail or be compromised:

\noindent\textbf{Inherent bias in a machine learning method:} The quality of the training dataset is crucial to the success of any machine learning model. A biased dataset is one that has been selected in such a way that some types of examples are given more weight than others. For example, suppose the photo dataset used to train the model for pipeline leakage detection by drones mostly covers bright weather settings. In that case, the trained model will perform badly in rainy or snowy weather. Predictive maintenance may also be used when a model is taught to work with a certain brand of equipment under certain conditions. Consequently, the model failed to generalize previously observed data correctly. Therefore, it would need to improve in accuracy before it could be used as a predictive maintenance model.

\noindent\textbf{Uncertainty exists in the machine learning model:} Machine learning models are susceptible due to their inherent ambiguity. However, it is feasible that the model may provide false-positive or false-negative findings, which might have disastrous repercussions. For example, if a refinery's smart fire detection system overlooks a fire, it might cause severe damage quite rapidly.

\noindent\textbf{Failure in the workflow of a smart application:} Smart solutions are usually composed of many parts that work together to build a directed acyclic graph or DAG. Face detection on an oil rig, for example, entails capturing videos, removing frames, and then analyzing each frame individually. Interrupting such a smart application cycle at any point might cause the whole application to fail, making it vulnerable. Similarly, if the command is not sent to the actuator, the whole workflow may be deactivated, resulting in a loss of control over the system.

\subsection{\textit{Malware and Vulnerability of Information Technology (IT)}}~\\
Malware, an abbreviation for ``malicious software," refers to any invasive program created by cyber criminals (also referred to as ``hackers") to steal data and damage or destroy computers and computer systems. Examples of malware include viruses, worms, trojans, spyware, adware, and ransomware. Recent malware attacks have resulted in massive data leaks. Therefore, the malicious actor(s) must be identified swiftly to remove malware. Among many forms of malwares, we discuss four major types in the following paragraphs.


\textit{\textbf{Virus:}} In order to infect other computers, \textit{viruses} often attach themselves to files that can run macros. The virus will remain latent inside the downloaded file until it is opened. Viruses are malicious programmes that interfere with normal system functioning. This means that infections may interrupt operations and lead to lost data.

\textit{\textbf{Worm:}} A \textit{worm} is a piece of malicious software that can quickly copy itself and infect any system on a network. In contrast to viruses, worms may spread without the help of any host software. For example, a worm may infect a device by a file download or a network connection, then rapidly replicate and spread. Worms, like viruses, may drastically impair a device's functionality and lead to data loss.

\textbf{\textit{Trojan:}} Often, \textit{Trojan} malware may mask as seemingly valuable pieces of software. However, once downloaded, the Trojan virus may access the user's private information and make changes, prevent access, or even erase it. The device's functionality may suffer severely as a result. In contrast to common viruses and worms, Trojan viruses are not programmed to multiply.

\textbf{\textit{Spyware:}} \textit{Spyware} is malicious software that works surreptitiously on a computer and feeds data back to an outside source. Spyware is especially hazardous since it affects device performance, targets sensitive data, and allows would-be attackers remote access. Spyware often targets financial or personal data. A key-logger, for example, is a kind of spyware that records users' keystrokes in order to steal passwords and other confidential information.

\begin{figure}[h!]
\centering
\includegraphics[width=1.0\textwidth]{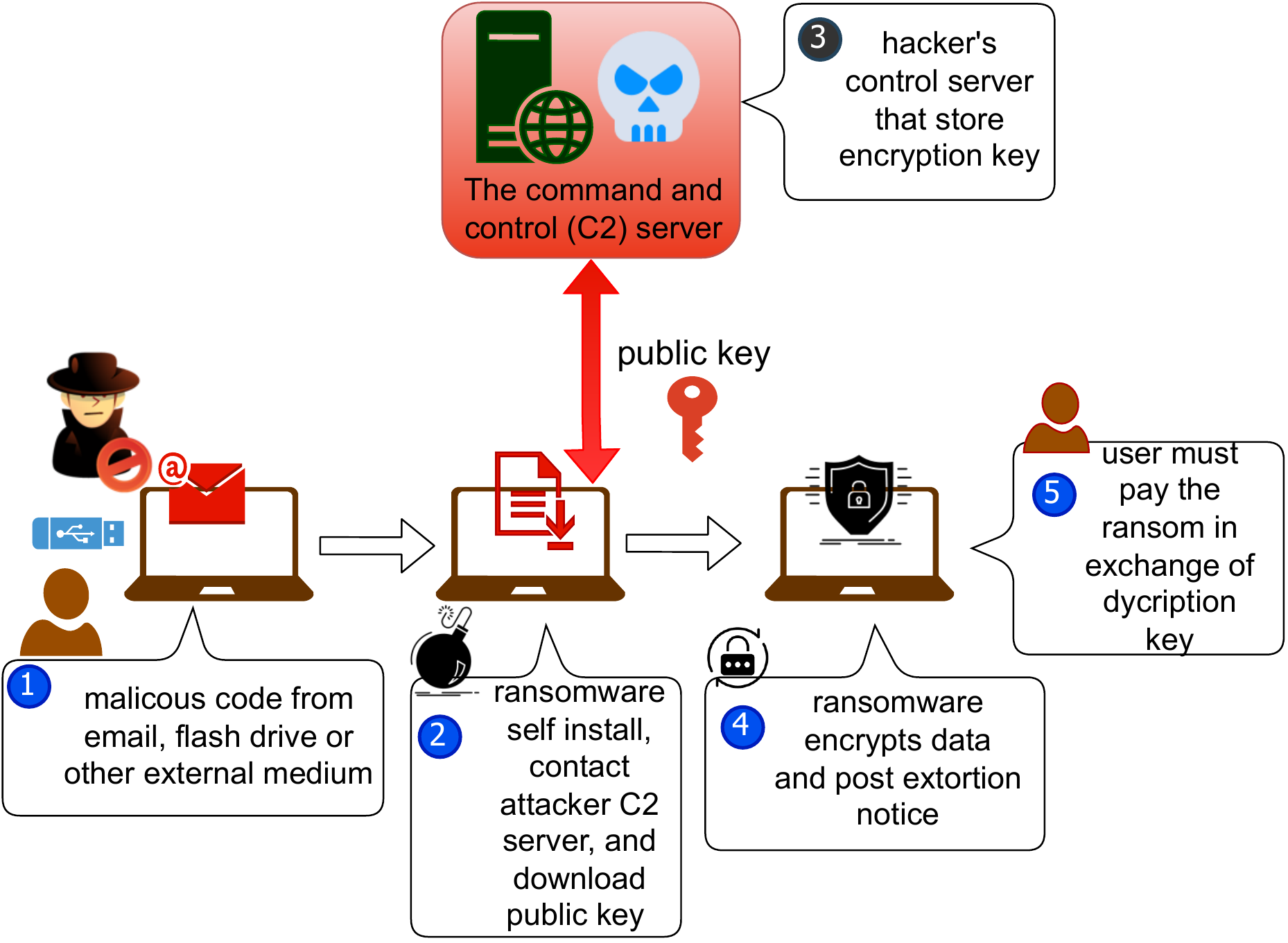}
\caption{The anatomy of ransomware from start to end. The ransomware client enters the IT platform through malicious email or other external mediums. The client then communicate with hacker's command and control server to download the encryption key. The user's data encrypted by the ransomware client, and finally extortion notice is sent.}
\label{fig:ransomware}
\centering
\end{figure}

\textbf{\textit{Ransomware:}} \textit{Ransomware} is a kind of malicious software that infiltrates a system, encrypts its data so that the user cannot access it, and then demands payment in exchange for decrypting the data. The use of ransomware is often associated with a phishing scheme. Figure \ref{fig:ransomware} depicts the stages of an actual ransomware attack. As can be seen, in these assaults, the victim downloads the ransomware by accidentally clicking on a spoofed link. The attacker then encrypts the targeted data using a cryptographic key that is known only to the attacker. Finally, in exchange for money, the hacker will release the information. We then go on to analyze this threat in further depth because of its rising prevalence over the last several years, especially in the oil and gas sector.

\subsubsection{Ransomware attack incidents. }
During a targeted cyberattack, a single virus may be used for a variety of reasons, including data theft, spread, and penetration. The threat actor's goal is to maintain persistence inside the victim's network. Therefore, they have to constantly communicate with and update their virus. Using the DNS protocol, a process known as DNS tunneling \cite{jan2021investigation} transmits information between malware and the controller. Additionally, email and cloud services have greatly expanded the scope of modern-day communication, which creates a wide door for ransomware criminals. 

Historically, hackers used spam botnets to infiltrate as many systems as possible and propagate ransomware. reference mandal2020digital. Although ransomware has always been a huge problem for everyone with digital files, it has become an even bigger problem as criminals have begun to specifically target businesses in assaults that may have devastating effects on operations. The following are examples of some of the most notable ransomware attacks:

\textbf{BitPaymer19: } BitPaymer19 \cite{mandal2020digital} is a particularly deadly type of ransomware that recently attacked a U.S. firm providing oil well drilling services. BitPaymer actors often employ phishing emails to infect their victim with first malware before moving laterally throughout the network to compromise other sensitive data. When IT personnel are unavailable (\eg on weekends and holidays), the ransomware attacks are 

\textbf{APT33: }One well-known actor group's primary concentration is on the oil sector and its supply networks. Organizations in the energy sector with linkages to petrochemical manufacturing and the aviation industry, where APT33 is involved in both military and commercial capacities, have been targeted. APT33 has also hit energy companies in Europe and Asia. From October 2018 through December 2018 and into 2019, a Powerton C\&C server was hosted on the C\&C timesync.com website and communicated with a small number of IP addresses belonging to oil corporations. Over the course of three weeks in late November and early December of 2019, a database server run by a European oil company in India spoke with a Powerton C\&C server used by APT33. In the fall of 2018, it was discovered that a significant UK-based corporation offering specialized services to oil refineries and petrochemical plants might have been penetrated by APT33.

Email phishing was APT33's most common method of infiltration. For many years, this scam has relied on the same bait: an email that seems legitimate but is a spear phishing attempt to offer a job. Other campaigns were directed against the recruiting procedure in the aviation and oil industries \cite{APT33}. Additionally, a link to the malicious ``.hta" file is provided in the email. To further entrench themselves in the target's network, APT33 may use the PowerShell script downloaded with the ``.hta" file to download further malware.

\section{Incompatible IoT Devices}
Among the smart O\&G sector's most common vulnerabilities is the use of incompatible Internet of Things (IoT) devices, as seen in Figure \ref{fig:threatTaxonomy}. In fact, automated systems that take data from a wide variety of Internet of Things (IoT) devices and sensors, a process that data using machine learning or statistical models, and then implement their decisions via a variety of actuation operations are the real engines behind a smart industry like oil and gas. In reality, the sensors and other linked IoT devices are created and purchased by a wide variety of companies, making them inherently heterogeneous. This diversity may cause incompatibilities and can be exploited by cyber-attackers or lead to gaps in service during times of crisis. For effective data transmission and offering real-time communication during emergency scenarios (for example, poisonous gas detection), it is crucial to configure linked and suitable IoT devices that can interact seamlessly.

Since acquiring uniform and completely compatible IoT equipment is difficult, if not impossible, researchers are looking at other solutions, such as the development of standard protocols that would enable effective communication across all industrial IoT devices—because of this, leading IT firms are collaborating to create a single protocol (called matter protocol \cite{conde2019iot}) that will be compatible with any and all Internet of Things (IoT) gadgets. The issue is a new protocol for inter-network communication between smart homes that are being backed by the Connectivity Standards Alliance, which includes tech giants like Apple, Google, Amazon, and others. The problem is the lack of a standard, IP-based communication protocol that is based on tried and true technologies to construct safe and secure IoT ecosystems.

In smart O\&G and other smart industrial contexts, we might examine three different kinds and degrees of incompatibility. In the following sections, we will discuss the incompatibilities that exist: those at the hardware, the software, and the data pipeline.

\subsection{\textit{Hardware-level Incompatibility}}~\\
Commonly available products are increasingly being utilized to replace specialized equipment in the oil and gas sector. They are more susceptible to security problems than traditional process control systems because of their adaptability. Because they are so widely used and deployed, the attack surface is widened significantly. There are several methods to assault an oil field. For instance, a smart real-time video camera may be employed to keep an eye on a potentially dangerous region for anomaly detection. Still, an unauthorized user might be able to use the control system to open a valve that lets poisonous or explosive gas escape. Sensors, actuators, cameras, and their supporting hardware might be protected against this kind of assault if they all used the same protocol to check for vulnerabilities and flag any unusual activity.

\subsection{\textit{Software-level Incompatibility}}~\\
Compatibility issues at the software level might arise from the usage of outdated or unsupported software, which can lead to system failure. Furthermore, there is a risk that malicious viruses will be introduced into internal systems through third-party software. But antiquated software must be updated to work with modern hardware and applications. Systems are more likely to be attacked if they haven't been kept up-to-date or are using enhancements that weren't made for their operating system.

Companies in the oil and gas industry often purchase digital items on the assumption that they are secure and can be integrated into the more extensive system. However, it is common practice for them to verify that everything else in the system is compatible with the new component. Also, oil and gas companies may not always have the resources available to verify incompatibility at the software level. That's how cybercriminals get into oil and gas firms' private networks via vulnerabilities in ``smart" technology.

As a result of the Internet of Things (IoT) smartness, business leaders in the petroleum supply chain must come up with novel approaches to preventing cybersecurity concerns. In recent hacks, vulnerabilities in the software were used. In 2017, a cyberattack known as NotPetya hit a variety of institutions, including a single electricity provider, banks, public transportation networks, and a large international container shipping firm. Interestingly, the virus propagated via Ukrainian companies' updated accounting software. When the infection spread to other computers, it caused crashes; in this case, the cybercriminals had infected customer-ready, certified software with spyware known as ``SolarWinds" (2021). In both cases, hackers used vulnerabilities in software to get into connected vendors' systems. In addition, they put in place loopholes that might be used to steal IP financial data or propagate malware among user machines.

\section{Blockchain to Overcome Cyber-Threats in Smart O\&G}
Blockchain technology, which has recently risen to prominence as the foundation of cryptocurrencies such as BitCoin and Etherium, is an effective security method. As the data is stored, it is linked in a series of blocks, and the hash value of the preceding block is kept in each block. Since the hash value of a tampered data block would no longer be consistent with that of the succeeding block, the attack could be traced. Several subsystems of Industry 4.0 and smart O\&G are now using blockchain technology.

\subsection{\textit{Blockchain-based Control Systems (SCADA)}}
\begin{figure}[h!]
\centering
\includegraphics[width=1.0\textwidth]{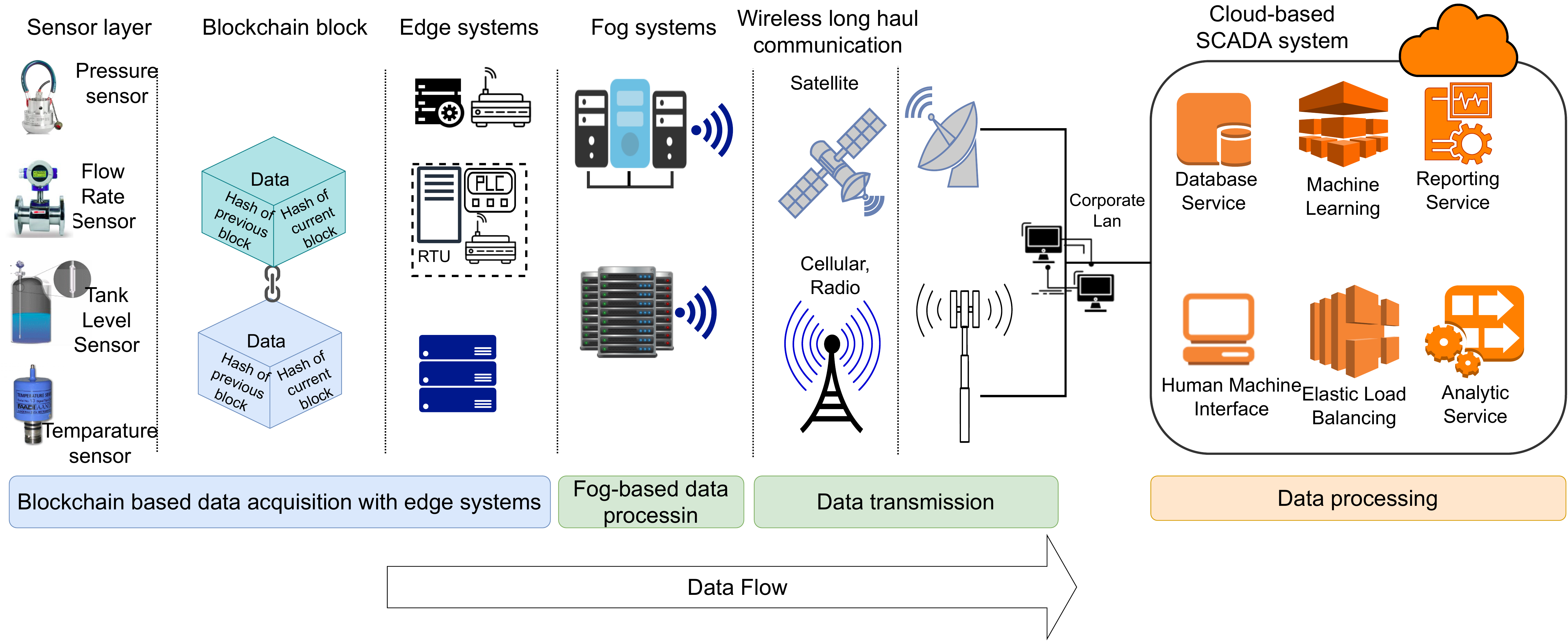}
\caption{Blockchain based data transmission within end-to-end  SCADA system of an oil and gas company. Blockchain enable encryption while transmitting the data for processing that increase the data security even data is hijacked while transmitting.}
\label{fig:blockchain}
\centering
\end{figure}

~\\
The Industry 4.0 movement has transformed the role of IT and OT in the modern industry. The SCADA systems that gather information from the smart IoT devices and send it to the servers where it is analyzed constitute the backbone of most OT platforms. However, this kind of data collection is inherently unsafe and unreliable, providing an opening for hackers. For this reason, edge and fog computing-based blockchain security procedures have been suggested to safeguard SCADA systems' data collection transactions. The gathered sensor data are encrypted in data blocks before being processed on a cloud-based SCADA system, and a high-level overview of this method is shown graphically in Figure \ref{fig:blockchain}. Data hashes from the previous and current blocks are stored in each block. Then, the Data Aggregator (DA) and all relay servers participate in the block verification procedure. The servers will answer many times for the purpose of verification. Upon consensus that the block is legitimate, the DA will forward the request to all participating servers. The DA adds the new block to the blockchain and then successfully sends the updated blockchain to the command center. Both the mining node selection technique and a more secure consensus process that is compatible with Industry 4.0 have been suggested in a recent paper \cite{hossain2020porch}, and these are discussed in the following sections.

\subsubsection{Consensus mechanism. } Simply put, a consensus mechanism is a process by which validators/miners verify the authenticity of freshly released blocks before adding them to the blockchain, hence preserving the integrity of the network. Various blockchain networks have spent considerable time and energy developing various consensus techniques. Both public and private blockchains use some consensus mechanisms. Proof-of-Work (PoW), Proof-of-Stake (PoS), Delegated Proof-of-Stake (DPoS), Practical Byzantine Fault Tolerance (PBFT), Proof-of-Authority (PoA), and RAFT \cite{ongaro2014search} are all examples of popular consensus techniques. Both the benefits and drawbacks of each consensus technique are distinct. PoW, for instance, is unjust to new entrants since it has a large processing expense and favors the richest validators. On the other hand, DPoS is less robust and decentralized. Due to its lack of anonymity, PBFT has restricted to permission (non-public) blockchains \cite{mollah2020blockchain}. 

\subsubsection{Mining node selection. } A machine that participates in a blockchain network by hosting blockchain software and facilitating data transfer is called a ``node." Nodes in a network might be anything from a laptop to a phone to a router. ``mining nodes" are the nodes that participate in the processing and verification of blockchain transactions. Any participant in the blockchain network may choose to take part in the mining process. As a term, ``mining" refers to the activity of adding new transactions to a blockchain. Figure \ref{fig:blockchain} shows the Data Aggregator (DA) edge server collecting data, processing it, and coordinating mining node selection and verification. If you want to save as much time and processing power as possible, place the DA on the same private network as the relay servers. Because of this, the fog servers used in the pre-processing stage of the data shown in Figure \ref{fig:blockchain} are hosted inside the DAs' own private network

In \cite{hossain2020porch}, the authors provide a specialized method for selecting mining nodes. To begin, the DA server initiates a data request to the relay servers. Once the DA collects all of the readings from the various relays, it will produce a random number and send it out across the network. To count how many times a random number appears, relay servers hash their data and compare the results. At this point, the DA's server statistics are identical to every other server's. Ultimately, every server casts a vote for the one that has made the most random appearances during the process. The DA server will choose the relay with the highest count as the mining node for the current cycle if all other relay servers agree. On the other hand, let's pretend that a large number of relay hosts have the same highest count or that they all have 0. The DA here selects the mining node at random using a cryptographically sound process \cite{zhang2019security}.

\subsection{\textit{Blockchain to Enable Trust Across Industrial IoT}}~\\
The problem of trust is one of the barriers to the security of the industrial Internet of Things (IIoT). The conventional Public Key Infrastructure (PKI) design, which is built on a single root of trust, does not operate well in this heterogeneous dispersed IoT environment, which may be subject to several administrative domains. Therefore, a distributed trust model that can be constructed on top of current trust domains and produce end-to-end trust across IoT devices without depending on a single root of trust is necessary for this sort of scenario. As a result, establishing a credit-based Blockchain with an integrated reputation system might be beneficial \cite{di2018blockchain}. 

Another potential use of blockchain in the oil and gas sector is the storage of credentials required to operate safety-critical industrial machinery. For example, employee and contractor qualifications, such as H2S training, first aid, and welding, may be securely recorded and preserved on a company's blockchain network. By storing such information in a blockchain network \cite{Abdul2021Council}, all members may perform verification of credentials and standard operating procedures at any time.

\section{Risks of Smart Solutions in industrial IoT}
As technology improves and more industries and products are connected to the internet, it is important to understand the risks of industrial IoT installations. Any business that wants to use IoT in manufacturing or industry or connect existing technologies for automated and remote monitoring should consider the advantages and disadvantages. In the next section, we'll discuss about the possible inadequate performances of smart solutions.

\subsection{\textit{Human-Machine Interaction Issues}}~\\
The industrial IoT has come a long way; machines can now process data from connected devices automatically. In addition, various automated sensors and actuators (like video cameras, smart glasses, and automatic valves with audio input/output) are in place to help or replace the human worker in order to make sure that production runs smoothly and/or that workers are safe when using different machines to do their jobs.

\begin{figure}[h!]
\centering
\includegraphics[width=0.8\textwidth]{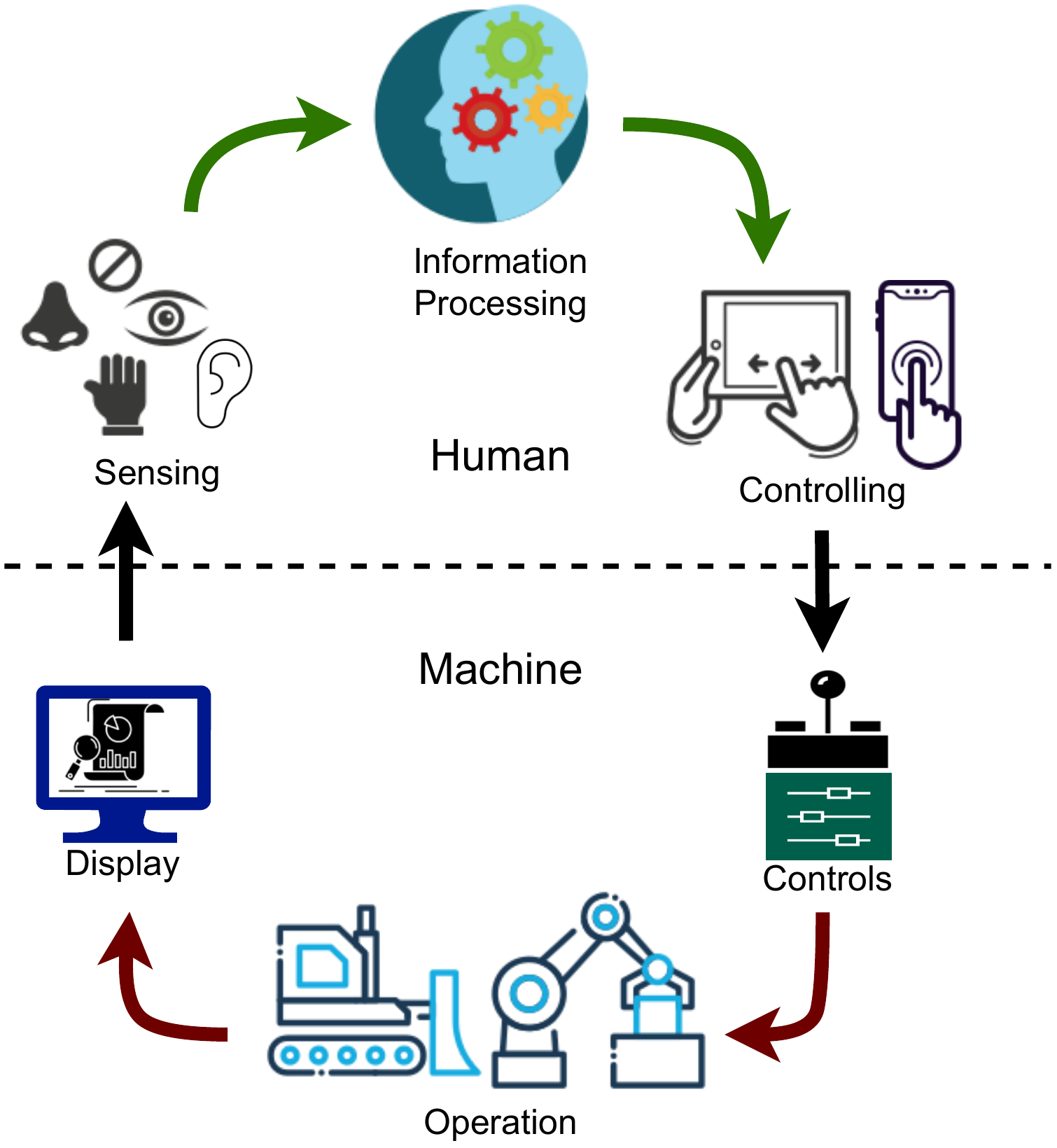}
\caption{Human-machine interaction workflow from sensing to control operation.}
\label{fig:h2missue1}
\centering
\end{figure}

Figure \ref{fig:h2missue1} shows how the human and machine sides of human-machine interaction work together. As this picture shows, people use different senses (like sight, smell, and hearing) to look at the machine's results. So, a human worker uses information processing to run or control the machines. Then, machines do their jobs, and the results are shown to people so they can figure out what they mean. The whole process of how people and machines work together is called the human-machine interaction process. 

Production and safety on the job site may be jeopardized if the interdependent machines fail to operate as intended or are not user-friendly. To perform vital industrial tasks or, more crucially, to "emphasize overrule" the choice of a smart system, human interactions are often required beyond those with a computational interface through input/output devices. Take the case of a drone or ROV sent to a politically sensitive location (like a border region) to conduct autonomous oil and gas surveys. However, inefficiently or a glitch in its algorithm may cause it to survey regions beyond the designated zone and prevent the operator from navigating the survey route. Unforeseen repercussions on the political or military front may result from such a glitch in human-machine interaction. 

\subsection{\textit{Machine-to-Machine Interaction Issues}}~\\
Machines communicate with one another in networked autonomous systems to complete various activities. In these systems, an automated sequence of actions is carried out using multiple devices; if anything goes wrong, it could be due to (A) the devices producing misleading output (for instance, automated valve shutdown with wrong anomaly detection or automatic door closing that traps onsite workers with false alarm), or (B) incompatibility across devices. Accidents or catastrophes may arise due to machine-to-machine interface issues in a production setting with fault-intolerant operations.

\begin{figure}[h!]
\centering
\includegraphics[width=1.0\textwidth]{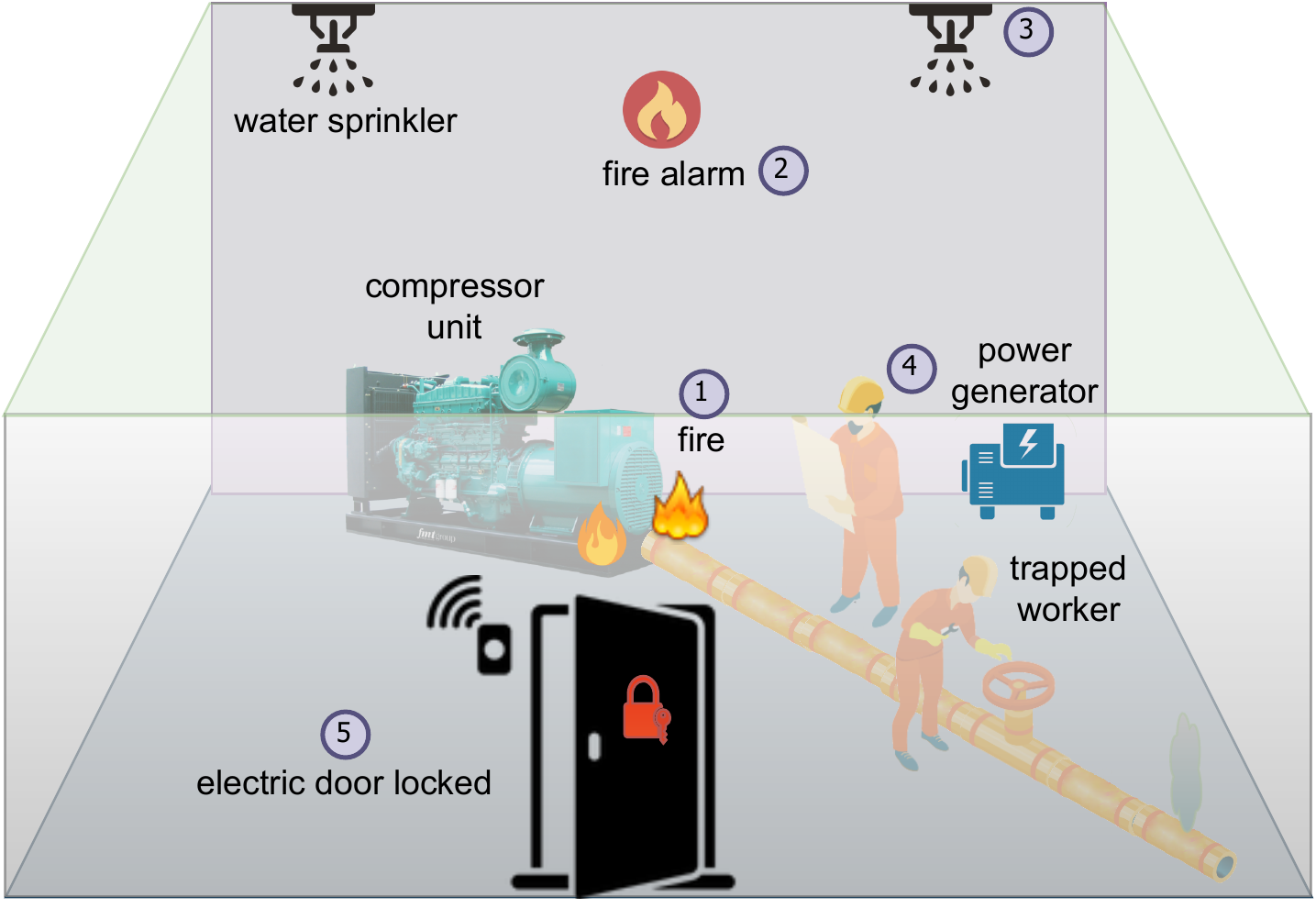}
\caption{A small fire breakout accident occurs in a closed oil production area in a compressor unit. The fire alarm generates, and water sprinkler starts to sprinkle water that causes power failure in power generator that made the electric door locked.  Unfortunately, workers were working on pipeline maintenance, and were trapped inside the facility due to door closure. Here, machine to machine interaction cause the safety issue of the onsite worker.}
\label{fig:m2missue1}
\centering
\end{figure}

Figure \ref{fig:m2missue1} depicts one scenario illustrating the repercussions of the machine-to-machine interaction problem. Consider pipeline maintenance and communication with numerous pieces of equipment in a production scenario. Consider a pipeline linked to a machine (for example, a distillation unit) in an enclosed space that requires repair. As a result, maintenance staff works within the enclosed space when a fire danger occurs. As a precaution, the gas sensor detects smoke (Step 2 in Figure \ref{fig:m2missue1}), activates the water sprinkler (Step 3), and sends an alert to the controller. The electricity generator shuts down due to sprinkler water (step 4). To safeguard the safety of the outside employees, the controller instantly sends the automatic door to shut (Step 5) while disregarding the workers within the area. In this scenario, the controller cannot recognize personnel within the facility and executes a safety action for outside workers, putting the workers inside at risk. We can see that machine-to-machine interaction concerns may sometimes lead to scenarios that must be solved by evaluating a smart solution for the O\&G industry.

\section{Bias in Smart Industry}
Human intervention at different stages of software development may introduce a wide range of biases that might undermine the quality of otherwise intelligent software solutions \cite{Grundy2020We}. Accidents and potentially dangerous situations have occurred as a result of prejudices in the past. Many types of bias exist, including those based on age, race, sexual orientation, disability, and other demographics. In addition, many onsite team members (\eg workers, engineers, and coordinators) rely on software tools and simulations that are prone to the above mentioned flaws. Our discussion here will center on the different kinds of bias and the damage they might do to the smart O\&G business and beyond.

\subsection{\textit{Biases Caused by the Artificial Intelligence (AI) Solutions}}~\\
Even though AI systems have shown to be revolutionary in several contexts, they are prone to the following two types of bias:
\begin{itemize}
    \item There are gaps in the training dataset that cause the model to underperform on certain inputs.
    \item The model's biases are the same as those found in the original dataset used for training.
\end{itemize}

An absence of training datasets is one source of AI bias, as shown by the commercial face recognition system. The lack of dark-skinned women \cite{lohr2018facial} in the training dataset is the root cause of the face recognition system's discrepancy between its 99.9 percent accuracy with white males and its 35.0 percent accuracy with women of colour, as determined by the researchers. The issue, however, is that ``Big Data" does not necessarily provide valid and trustworthy models. For instance, social media is a well-established mine for massive datasets. Conclusions obtained from Twitter data should be treated with caution since just 24\% of internet teenagers utilize the platform, as reported by \cite{Madden2013teens}. 

An unfair model does not necessarily perform poorly on a demographic subset. Even if the model is correct, it is still unjust. The dataset is skewed in this situation, and the model repeats or amplifies the inherited bias. Natural Language Processing (NLP) models, for example, are often trained on a vast corpus of human-written text (\eg article news). However, word embeddings trained on Google News articles have been observed to reflect female/male gender stereotypes. The models, for example, replied that a father is a doctor while the mother is a nurse, or that a ``man" is a ``computer programmer" while a ``woman" is a ``homemaker." This kind of bias occurs when a model is trained on skewed data owing to unjust procedures or structures \cite{bolukbasi2016man}.

Another example of AI bias is Yelp's restaurant review system. Restaurants may pay Yelp to promote their locations on the Yelp platform, but this inevitably influences how many people see adverts for a particular restaurant and, as a result, who decides to dine there. As a result, Yelp evaluations may be unjustly slanted in favor of more prominent eateries. 

\subsection{\textit{Automation Bias in Smart Solutions of Industry 4.0}}~\\
One of the most respected psychologists in the field, Linda J. Skitka of the University of Illinois at Chicago, defined automation bias as ``a specific class of errors individuals tend to make in highly automated decision-making scenarios when many decisions are handled by automated aids (such as computers, IoT devices, and smartphones) and the human actor is primarily present to monitor ongoing tasks.''

A bias toward using automated assistance and decision support systems is often known as ``automation bias." When the Enbridge pipeline ruptured  \cite{wesley2017complacency} on July 26, 2010, sending enormous amounts of crude oil into the Kalamazoo River and Talmadge Creek, automation bias was a major factor. Both complacency and a leaning toward automation were shown to have played significant factors in the Enbridge oil pipeline disaster. Therefore, businesses, governments, and regulators must account for automation bias while designing systems to reduce the potential for careless errors. ``Automation bias" is humans' propensity to favour actions requiring the least amount of mental effort. Similar thinking applies to the underlying principle of AI and automation: learning from massive amounts of data. Such calculations imply that future conditions will mostly stay the same. Another factor to consider is the possibility that faulty training data may lead to faulty learning \cite{automationdatabrick} that is implicitly related to AI bias.

\subsection{\textit{Gender Bias in O\&G Industry}}~\\
According to a study report \cite{rick2017untapped}, the oil and gas sector is confronting a skilled personnel scarcity, while gender prejudice is exacerbating the problem by excluding female workers from recruiting. The research contains interviews with various male and female workers from around the globe and an analysis of their comments. Indeed, the oil and gas industry has a reputation for being controlled by males. However, while some oil and gas businesses work hard to achieve gender parity and worker diversity, others are allowing the gender gap to widen. Although many businesses strive to include gender equality in their policies, actions, and procedures, they still face challenges such as gender imbalance and various types of implicit prejudice.

\subsection{\textit{Cognitive Bias in Smart O\&G Solutions}}~\\
Cognitive biases, a newly discovered notion, are mental faults in human thinking and information processing that may result in inaccurate or irrational assessments or decisions. Amos Tversky and Daniel Kahneman first proposed it in a 1974 article for Science Magazine (Tversky \& Kahneman, 1974 \cite{tversky1974judgment}). Since then, a great deal of literature has been produced on cognitive biases and how they impact human thoughts and actions.

According to a common understanding of cognitive bias, it is a mental flaw that results in incorrect interpretation of external data and impairs the logic and precision of choices and verdicts. Biases are unconsciously occurring, automatic processes that speed up and improve decision-making effectiveness. There are several factors that might contribute to cognitive biases, including public influence and emotions. There has been a growing awareness of the threats cognitive bias may bring to operational safety during the last several years. Biases like deviance, normalization, and group thinking, for instance, are now widely accepted. Additionally, the Deepwater Horizon \cite{prause2017rig} investigation in 2010 brought cognitive bias to the public's attention, at least among those working in the offshore drilling industry. Consequently, the International Association of Oil and Gas Producers (IOGP) has brought attention to how crucial these cognitive impairments are to safety. Therefore, it is high time that cognitive bias should be addressed while building smart, automated solutions that require human decisions for complex industrial operations.

\section{Summary}
Oil and gas operations have seen dramatic changes as a result of the digital Industry 4.0 revolution, which has made extensive use of cutting-edge computer hardware and software. However, with these developments come opportunities for cyber criminals to improve their efficiency in locating vulnerabilities in either IT or OT systems, or in the hybrids that exist between the two. Another possible entry point for cyber criminals is provided by the heterogeneity and incompatibility of smart technologies, as well as the connection difficulty between them. Problems with human-machine and machine-to-machine interactions, as well as incompatibilities between technologies acquired through time, are among the most significant obstacles to the widespread adoption of smart technologies in the legacy and smart oil and gas sectors. Though Industry 4.0 has been a boon to the oil and gas sector, business executives and professionals working in the sector should be wary of its smarts being misapplied. In the last several years, we've learned the hard way that blindly installing or embracing smart technology may open the door to a wide variety of risks. Researchers and practitioners must bear in mind these drawbacks while deciding whether or not to use smart technology. In this regard, as a part of this dissertation, we published a book \cite{hussain2022iot} on the scope of IoT technologies with the rise of the Industry 4.0 revolution that addresses a detailed analysis of smart solutions and their drawbacks in the context of smart O\&G Industry. 
    \chapter{Conclusion and Future Research Directions}
\label{section:thesiscon}
The ever-growing IoT and smart devices (\eg smart gateway, sensors, controllers, actuators) produce a substantial amount of raw data that need to be stored, pre-processed, and analyzed to bring out potential insights that can make the industrial systems more efficient. Accordingly, various Industry 4.0 latency-sensitive applications operate based on machine learning (ML) and utilize the generated sensor data to achieve automation and other industrial activities. Hence, the cloud computing platform has been offering services \cite{rathod2022cloud} to perform various operations on the ever-growing data generated in the industrial sectors. Privacy, centrality, and expenses have been significant constraints to utilizing cloud data centers effectively. As such, edge and fog computing bring the computational services \cite{elhadad2022fog,cheng2022contract,sarkar2022deep} near the end-users closer to the data sources. However, edge devices may support limited computing demands due to resource limitations. In contrast, the fog system can be a preferable option to meet computing needs due to its availability of computational resources and more robust middleware compared to edge systems. Because, fog systems are heterogeneous and the heterogeneity is one factor that introduces stochasticity in the execution time of Industry 4.0 applications that affects the completion times of these applications. To develop a robust solution for Industry 4.0, it is necessary to study the execution time behaviour of various ML-based applications in heterogeneous fog systems. As such, we perform statistical analysis of ML-based Industry 4.0 applications to understand the execution time pattern of these applications. In addition, we introduce real-world Industry 4.0 smart application execution traces in fog computing systems that can be beneficial for the future research works. 

Even though fog systems have more computing resources than edge systems, the surge in computing demands at disasters can reduce performance. Therefore, in this dissertation, we propose federating fog computing systems (owned by private companies) from nearby sites to support such scenarios. Furthermore, the fog federation concept can be practical with system administrators' efficient resource allocation mechanisms adopted by research works related to load-balancing methods. A real-world Industry 4.0 application execution traces on fog computing platforms can be crucial for devising effective resource allocation methods. As a result, we utilize our prior workload trace to devise a statistical resource allocation method across federated fog systems for Industry 4.0 latency-sensitive applications. 

In addition, the heterogeneous software methodologies (\eg monolithic, micro-service) of Industry 4.0 applications can affect the execution plan of a fog federation due to their diverse latency constraints, resulting in decreasing system performance. Hence, the decomposition of micro-service applications with effective resource allocation methods can maintain the systems' performance in oversubscribed situations (\eg accidents, and disasters). Accordingly, the industrial computing platform (\ie federated fog system) should be cognizant of stochastic execution behaviour, software structure, and latency requirements of micro-service workflow applications. We propose a resource allocation method based on probability estimation that partition micro-service workflows across the federated fog computing systems to support their latency requirements. Furthermore, the concept of federating fog resources raises data security and privacy concerns for private fog systems participating in the federation due to having sensitive company data stored or processed in these fog systems. Thus, we propose a data privacy preserving solution that works based on federated learning method for training ML-based Industry 4.0 application across federated fog systems.  

\section{Discussion}
In this dissertation, our main objective was to investigate and develop effective resource allocation solutions using modern distributed computing systems for remote Industry 4.0. As such, we first explore and identify various smart computing aspects in remote offshore industries (\eg oil and gas, minerals, sustainable energy) where computing demand is significantly high and conventional computing systems are inefficient due to latency constraints. Hence, privately owned fog systems in remote areas can support industrial computing demands. Hence, we identify stochastic execution time behaviours of latency-sensitive tasks executing in heterogeneous fog systems. As such, we explore the execution time behaviour of various ML-based applications in heterogeneous execution platforms (\eg amazon web service, chameleon). Consequently, we introduce a real-world workload of execution time in heterogeneous computing resources. Furthermore, in remote industries, the surge in computing demand can decrease the fog systems' performance at disaster times by not completing latency-sensitive task requests on time. Accordingly, we propose federating nearby fog systems in remote industries and forming a fog federation to support surge computing demands. Thus, we enable the federation concept and develop a statistical resource allocation method using prior synthesized real-world application workload considering an oversubscribed situation. Hence we evaluate our proposed solution for monolithic applications widely used in Industry 4.0. After that, we investigate smart micro-service applications' internal structure to understand the impact of the decomposition on application workflow completion. We suggest a probabilistic workflow partitioning method along with the previously proposed resource allocation method that improves the fog federation's performance and ensures safety in remote Industry 4.0. Finally, we address the data privacy issue for sharing privately owned fog systems in developing accurate ML models for Industry 4.0. Hence, we explore the federated learning techniques across the fog federation that ensure data privacy for privately owned fog systems. In this context, we address the class imbalance issue in a federated learning setup that can reduce the robustness of the global model. Therefore, we propose a federated learning method that is robust against the class imbalance issue.

In chapter~\ref{section:performanceAnalysis}, we analyze and estimate the performance of DNN-based applications in heterogeneous cloud and fog resources (\eg amazon, chameleon). Here, we identify stochastic execution behaviours of various Industry 4.0 applications. Thus we explore, and model the inference execution behaviours of various Industry 4.0 smart applications utilizing different statistical tools from two distinct perspectives, namely application-centric and resource-centric, respectively. Furthermore, we introduce an execution time workload of four different DNN-based applications for Industry 4.0 with the intent of developing robust resource allocation methods across federated fog systems.

In chapter~\ref{section:loadBalance}, we explore the usability and benefits of fog federation that can be formed to support emergencies such as disasters (\eg fire explosions, oil spills). As an example of a smart industry, we consider remote smart oil fields with multiple oil extraction sites in close vicinity, each with fog computing systems to support its local computing demands. Although in case of an emergency like an oil spill, the computing demands can rise due to the coordination of multiple activities (\eg drone inspection of oil spill, video camera images, sensors data processing) to support the situation. Hence we propose a probabilistic resource allocation method for monolithic latency-sensitive applications that effectively selects a relevant fog system from the federation by utilizing our prior workload. As the resource allocation method is aware of the receiving applications stochastic execution behaviours from our prior work, it ensures the robustness of the fog federation by completing majority of the receiving workload on time.

In chapter~\ref{section:microservice}, we explore modern software architecture (\ie micro-service) of Industry 4.0 applications to create an efficient execution strategy over fog federation. In contrast, we identified legacy applications with monolithic software architecture are still exists in various industrial sector. Therefore, to support the computational demands in remote industry the execution platform (federated fog system) should be aware of software architectures of the Industry 4.0 applications.
Hence, for micro-services we consider the idea of using an application breakdown strategy to increase the chance of finishing the execution on time. Furthermore, for monolithic applications and individual micro-services we utilizes our prior knowledge of stochastic execution behaviour to efficiently allocate fog resources across the federated fog systems. As a result, we propose a statistical micro-service partitioning and resource allocation method that considers the underlying software architecture and the stochastic execution latencies of Industry 4.0 applications.

In chapter~\ref{section:fedLearn}, we explore the data security and privacy aspects of fog federation while training ML-based applications in remote Industry 4.0. In this case, we investigate the federated learning techniques utilizing fog federation to train a ML-based oil spill detection application that provide data security to privately owned fog systems of the federation. Accordingly, we identified low occurrence events in training data (\ie class imbalance) can reduce the accuracy of the ML-model that can be detrimental in emergency situations. Here, we propose a customized federated learning technique, considering the class imbalance issue across fog federation to increase the safety measures of remote Industry 4.0. 

In chapter~\ref{threatsSideEffect}, we investigate the downsides and side-effects of smart solutions developed with the integration of various applications in the industrial sectors. Hence, we introduce a taxonomy of cyber threats and side-effects of smart solutions in the context of the O\&G industry that structurally address the unsafe areas of these smart solutions. Accordingly, various vulnerable areas, including both software and hardware components, machine-human interaction issues, and different forms of biases in smart solutions, are addressed with efficient resilience methods that would help system architects or industrial researchers to develop robust smart solutions for Industry 4.0.

In conclusion, we explore and investigate the stochastic execution behaviour of various Industry 4.0 applications and introduce a real-world execution workload that has been utilized in our resource allocation research works. Then, we explore the federation concept using privately owned fog systems for various computing demands of Industry 4.0. Especially in oversubscribed situations like disasters, the federation could be more efficient if the load is adequately balanced. Hence we develop a load-balancing method to make the federation robust in emergencies that we consider the system administration level of our research track. Then we dive into the application level by investigating various software architectures (\eg monolithic, micro-service) of Industry 4.0 applications. Hence, we identify micro-service workflow applications can be decomposed to improve the application workflow completion on time. Accordingly, we propose a probabilistic micro-service partitioning and resource allocation method that can enhance the performance of the fog federation. Then, we explore the data security and privacy aspects of federated fog systems while training ML-based Industry 4.0 applications. Finally, in the end, we identify various pitfalls of smart solutions that need to be appropriately addressed to develop efficient and robust smart solutions for Industry 4.0 applications.  

\section{Future Research Directions}
Based on our findings during the development of the resource allocation, micro-service workflow partitioning, and secure resource-sharing solutions, we identify some of the expansion areas that can improve the robustness and safety of Industry 4.0. There are several points where the work could be expanded.

\subsection{\textit{Resource Allocation Using Reinforcement Learning for Industry 4.0 Applications across Federated Fog System}}~\\
In this dissertation, we suggest a statistical application completion time estimation method across the fog federation system to allocate Industry 4.0 applications into a relevant fog system. Our estimation of task completion success could be coarse that sometimes leads to the deadline miss of a receiving task. We think that a resource allocation method operating based on the reinforcement learning technique \cite{ye2019deep,zhang2022federated} can improve the quality of service for the federated fog system. The field of reinforcement learning (RL) has emerged as an important subset of machine learning because it enables autonomous agents to make sound decisions in response to changing conditions in their environment. Hence, uncertain execution behaviour can be addressed effectively using RL technique. In this scenario, RL might be used to allocate resources \cite{lakhan2022efficient} for offloading and executing tasks in a federated fog computing system, leading to better overall performance.

\subsection{\textit{Data Locality-Aware Resource Allocation Across Federated Fog System}}~\\
The proliferation of IoT devices has coincided with a surge in network traffic that may overwhelm the potential of the current network. Furthermore, data privacy and latency are important concerns when these devices analyze sensor or user information. Therefore, conventional methods such as cloud computing don't work. Although, advanced computing platform like fog computing can fill this need. Understanding how the initial input data's localization impacts on fog platforms' performance is critical to developing reasonable load balancing and resource allocation solutions \cite{stavrinides2020orchestration,inides2022data}. As a result, if several data-intensive applications with deadline restrictions arrive dynamically, performance evaluation of a heterogeneous federated fog environment is required. For example, the applications may need data from the IoT layer or from local fog resources (\eg sensor data that have already been transferred to the fog layer or data processed by prior applications). In this scenario, examining the influence of input data localization on system performance across federated fog systems with varying data placement probability might influence the federation's resource allocation efficiency. Therefore, we consider exploring the impact of data localization on resource allocation methods across federated fog systems in the oversubscribed situation for remote Industry 4.0.


\subsection{\textit{Dynamic Fault-Tolerant Federated Fog Systems for Industry 4.0 Operation}}~\\
The fog computing systems provide low latency to the end users being close to the data sources. In contrast, the distributed characteristics of fog aid in processing vast amounts of sensor-generated data of Industry 4.0. Hence, federating fog computing resources can support latency-sensitive tasks and data processing services. However, fog systems have various uncertainties (\eg transient failures \cite{al2022transient}, network and power outage) that need to be considered when supporting surge computing demands. Especially in an emergency, those uncertainties can lead to unsuccessful task completion causing significant damage to the environment and even human lives. Hence, the resource management system for fog federation should be aware of the uncertainties and provide efficient fault-tolerant \cite{saeed2021fault,wang2018adaptive,zhang2021fault} solution to ensure successful completion of the receiving latency-sensitive tasks on time. Accordingly, the federation management system should consider providing a service that continuously monitors the fog resources and then sends the signal to all the participant fogs about the neighbouring fog's state. In addition, already ongoing service execution can get disrupted or fail due to the fog system's internal issues (\eg software, hardware). In this case, every fog system should have a method (\eg re-execution, offloading the failed task to another fog) to ensure successful completion of receiving task's execution. Therefore, a fault-tolerant federated fog system is crucial for supporting surge computing demands in emergency or disaster situations, enabling the system's robustness and leading to a safe Industry 4.0.

Similar to super cloud \cite{jia2015supercloud}, fog systems provide various virtual services \cite{li2017virtual} like application deployment, multi-tenancy, interoperability, and service migration across fog federation. Hence, to enable a fog federation that can avail various fog virtual services need to support fault-tolerant characteristics for efficient utilization of fog services. In addition, in oversubscribed situation, the fault-tolerant service needs to address the scalability of fog federation, ensuring the system's robustness in a dynamic condition. Therefore, to achieve all these requirements, we are considering performing research works in future on developing a fog system (super fog) that provides fault-tolerant fog services across the federated fog systems to improve the efficiency of the federation.

The popularity of fog systems with heterogeneous resources and dynamic fog federation \cite{hammoud2022dynamic} concept has created the demand for developing the fog-friendly application that requires proper investigation of the application stack and fog resources. However, building this type of application is time-consuming and requires overcoming some major obstacles. The first is to support the dynamic nature of the fog network; the second is to manage the context-dependent qualities of application logic; the third is to cope with the system's massive size. As a result, we must consider how to decompose and deploy applications to a geographically dispersed fog federation utilizing current software components that may be altered and reused to participate in fog applications \cite{giang2020developing}. Hence abstracting the application layer from the execution layer can be the primary objective to solve the heterogeneity challenge of the fog systems. Thus, we would like to perform research on developing fog-friendly applications for dynamic federated fog systems that are cognizant of super fog systems' characteristics.



\subsection{\textit{The Cognitive Aspects of Human-Machine Interaction for Smart Industry 4.0 Solutions}}~\\
Industry 4.0, an industrial technology paradigm shift, mandates new ways in which human and machine (\eg robots, drones) will work together. The introduction of ever-more-advanced sensors and collaborating machines raises important questions about the influence on safety in the highly technical and inventive scenario of Industry 4.0, defined by a succession of enabling technologies and a strong interconnectedness of resources between human and machine. On the one hand, advanced software tools and machines facilitate human work (human-machine cooperation). On the contrary, it must communicate and share data with other intelligent devices (human-machine interaction) \cite{ansari2018rethinking}. Since the advent of ``smart" technologies, both the environment in which these innovations are deployed and the responsibilities of front-line human workers have changed. In complex industrial operations or at disaster times, the human-machine interaction can be challenging that is significantly related to cognitive aspects \cite{chacon2020developing} of the human workers. When some tasks need specialized human abilities, there is genuine ``collaboration" between humans and machines. In today's modern industries, workers' interactions with ``smart machines" can make their jobs easier by making their tasks more automated and less prone to human error. In contrast, it makes the workplace more complicated by increasing information and communication flow between different systems. For example, using sensors and cutting-edge technology, we can collect the information we need to make accurate forecasts about the health of industrial machinery and carry out precise treatments. As humans must handle the massive amounts of data (big data) that need to be gathered, analyzed, and understood, the cognitive interaction effort of the machine operator rises from the skill level to the knowledge level \cite{madonna2019evolution}. Therefore, we would like to address various cognitive aspects of human-machine interaction issues in Industry 4.0 and develop smart solutions for human workers to aid in a complex industrial scenario.

\subsection{\textit{Fog Computing and Advanced Analytics for Human-Machine Interaction in Industrial Sector}}~\\
The advancement in computing technology with industrial revolution has transformed the industrial operations using automation, robotics, artificial intelligence and other modern smart solutions. Although, various complex industrial operations (\eg machine maintenance, oil well drilling operation, manufacturing machines) need human intervention and interaction \cite{gorecky2014human,kumar2022human}  to ensure precision and accuracy of the operation. Hence, human operator that communicate with machine sometimes need to process machine generated data to efficiently communicate with machines \cite{nawaz2022intelligent,shang2022interactive}. In this case human operators can use a mobile device with them to process the data or visualize the data that is processed a nearby computing systems. Hence, fog computing can be a potential candidate to support the computing demands of human-machine interactions \cite{wang2019enabling,na2022human,ahmed2020assessment}. Furthermore, fog computing utilizing various advanced analytics (\eg machine learning, deep neural network, reinforcement learning) on machine generated data can provide useful insights to the human operators that can improve human-machine interactions.     
\bibliographystyle{IEEEtran}
\bibliography{myBib}


\closingpages
\begin{biosketch}
   \indent \Author~received his
    Bachelor of  Science in computer science and engineering in the fall of 2011 from Military Institute of Science and Technology, Bangladesh. He immediately began his career in February of 2012, as a programmer in one of the software company named ``ERA Infotech Ltd''. After nine months of his first job, he joined one of the top multinational software company ``Samsung R\&D Institute" as a software engineer. \Author~worked in Samsung for around 4 year and 5 months, and decided to enrich his academic knowledge by pursuing higher education. As such \Author~started his Ph.D. journey in computer science in the fall of 2017 at the University of Louisiana at Lafayette. While pursuing his Ph.D. degree \Author~ received his M.Sc. degree in computer science in the spring of 2019. His research interests are: cloud computing, resource allocation in a fog federation, task scheduling, machine learning, DNN-based applications for Industry 4.0, and federated learning. Currently, \Author~is working as a Software Engineer in one of the software company named ``TryCycle Data Systems'', providing software solutions in healthcare sector of Canada \& United States.
\end{biosketch}

\end{document}